\documentclass[aps,prd,eqsecnum,11pt,showpacs,preprintnumbers,nofootinbib]{revtex4-1}
\usepackage{amssymb,amsmath,bm,color}
\usepackage{graphicx}
\allowdisplaybreaks[1]

\def\nbox#1#2{\vcenter{\hrule \hbox{\vrule height#2in
\kern#1in \vrule} \hrule}}
\def\sq{\,\raise.5pt\hbox{$\nbox{.09}{.09}$}\,}
\def\sqb{\,\raise.5pt\hbox{$\overline{\nbox{.09}{.09}}$}\,}
\def\sech{{\rm sech}}

\newcommand{\bea}{\begin{eqnarray}}
\newcommand{\eea}{\end{eqnarray}}
\newcommand{\be}{\begin{equation}}
\newcommand{\ee}{\end{equation}}
\newcommand{\bes}{\begin{subequations}}
\newcommand{\ees}{\end{subequations}}
\newcommand{\nn}{\nonumber \\}
\newcommand{\ups}{\upsilon}
\newcommand{\lam}{\lambda}
\newcommand{\lag}{\langle}
\newcommand{\rag}{\rangle}

\begin{document}

\preprint{LA-UR-13-26338}

\title{\hspace{1cm}\\  \Large
On the Instability of Global de Sitter Space to Particle Creation
\vspace{9mm}}

\author{Paul R. Anderson}
\altaffiliation{\tt anderson@wfu.edu}
\affiliation{Department of Physics, \\
Wake Forest University, \\
Winston-Salem, NC 27109, USA}
\author{Emil Mottola}
\altaffiliation{\tt emil@lanl.gov}
\affiliation{Theoretical Division, MS B285\\
Los Alamos National Laboratory\\
Los Alamos, NM 87545 USA}

\begin{abstract}
\vspace{9mm}\noindent

We show that global de Sitter space is unstable to particle creation, even for a massive
free field theory with no self-interactions. The $O(4,1)$ de Sitter invariant state is a definite
phase coherent superposition of particle and anti-particle solutions in both the asymptotic
past and future, and therefore is not a true vacuum state. In the closely related case
of particle creation by a constant, uniform electric field, a time symmetric state analogous
to the de Sitter invariant one is constructed, which is also not a stable vacuum state.
We provide the general framework necessary to describe the particle creation process, 
the mean particle number, and dynamical quantities such as the energy-momentum tensor 
and current of the created particles in both the de Sitter and electric field backgrounds in real 
time, establishing the connection to kinetic theory. We compute the  energy-momentum tensor 
for adiabatic vacuum states in de Sitter space initialized at early times in global ${\mathbb S}^3$ 
sections, and show that particle creation in the contracting phase results in exponentially large 
energy densities at later times, necessitating an inclusion of their backreaction effects, and leading 
to large deviation of the spacetime from global de Sitter space before the expanding phase can begin.

\end{abstract}

\maketitle

\section{Introduction}
\label{Sec:Intro}

The problem of vacuum zero-point energy and its effects on the curvature of space through
Einstein's equations has been present in quantum theory since its inception, and was first recognized
by Pauli \cite{Pau}. Largely ignored and bypassed during the steady stream of successes of
quantum mechanics and then quantum field theory (QFT) over a remarkable range of scales
and conditions for five decades, the role of vacuum energy was raised to prominence by
cosmological models of inflation. Inflation postulates a large vacuum energy density to drive
exponential expansion of the universe, and invokes quantum fluctuations in the de Sitter epoch
as the primordial seeds of density fluctuations that give rise both the observed cosmic microwave
background anisotropies, and the formation of all observed structure in the universe \cite{Infl}.
The problem of quantum vacuum energy and the origin of structure are both strong motivations
for the study of QFT in de Sitter space.

Further motivation comes from the discovery of dark energy in $1998$ by measurements of the redshifts
of distant type SNI supernovae \cite{SNI}. This has led to the realization that cosmological vacuum energy
may be more than $70\%$ of the energy density in the universe and be responsible for its accelerated Hubble
expansion today. If correct, this implies that de Sitter space is actually a better approximation than flat
Minkowski space to the geometry of the present observable universe. Accounting for the value of the apparent vacuum
energy density today and elucidating its true nature and possible dynamics is widely viewed as one of the
most important challenges at the intersection of quantum physics and gravitation theory, with
direct relevance for observational cosmology.

Being a maximally symmetric solution of Einstein's equations with positive cosmological constant, which itself
can be regarded as the energy of the vacuum, de Sitter space is the simplest setting for posing questions
about the interplay of QFT, gravity, and cosmology. Although global de Sitter space is clearly an idealization,
it is an important one amenable to an exact analysis of fundamental issues. Progress toward a consistent
theory of quantum vacuum energy and its gravitational effects, and the formation of structure in the universe
almost certainly requires  a thorough understanding of quantum effects in de Sitter space. One of the most
basic of quantum effects that arise in curved spacetimes is the spontaneous creation of particles from the
vacuum \cite{Parker,Zel,ParFul,BirBun,BirDav}. This process converts vacuum energy to ordinary matter
and radiation, and therefore can lead to the dynamical relaxation of vacuum energy with time \cite{PartCreatdS,Fluc}.
An introduction to these quantum effects, summary of the earlier literature and the prospects for a dynamical theory
of vacuum energy based on conversion of vacuum energy to particles may be found in \cite{NJP}.
Since that review, there has been further interesting work on various aspects of QFT in de Sitter space,
particularly on quantum infrared and interaction effects \cite{Poly,AkhBui}.

Because of the mathematical appeal of maximal symmetry, much of both the earlier and more recent
research assumes the stability of the de Sitter invariant state obtained by continuation from the Euclidean
${\mathbb S}^4$. This state of maximal $O(4,1)$ symmetry, known as the Chernikov-Tagirov or Bunch-Davies 
(CTBD) state \cite{Nacht,CherTag,BunDav}, is also the one most often considered in inflationary models \cite{Infl}.
However the CTBD state raises a number of questions that seem still to require clarification.
It is known that a freely falling detector in de Sitter space in the CTBD state will detect a
non-zero, thermal distribution of particles at the Hawking-de Sitter temperature \cite{GibHaw}.
This shows at the outset that the Euclidean CTBD state is not a `vacuum' state in the usual
sense familiar from Minkowski spacetime. In flat space the separation into positive and negative
energies, hence particle and anti-particle solutions of any Lorentz invariant wave equation is itself Lorentz
invariant. Minimizing the energy of the Hamiltonian in flat space in any inertial frame also
produces a vacuum state that is Lorentz invariant. In de Sitter space neither of these statements is true.
Arbitrarily chosen positive and negative energy solutions are transformed into each other by the
action of the de Sitter group generators \cite{Nacht}.  As a result, a de Sitter invariant separation
of particles from anti-particles is not possible, and the notion of a stable vacuum which
minimizes a positive Hamiltonian, upon which QFT in flat space depends, does not exist.

While interactions are clearly important for the subsequent evolution, the absence of
a minimum energy vacuum state free of particle excitations, as well as the spontaneous
particle creation rate  \cite{PartCreatdS} makes quite dubious the assumption of
a stable vacuum state in global de Sitter space, even at the level of free field
theory for massive fields. Progress in describing particle interactions also requires
a physically correct definition of particles in curved space. The question of
the existence or not of a stable vacuum and particle production should be settled
unambiguously first for massive fields since it is apparent that the effects of light fields
and gravitons are even more subtle \cite{Ford,AntMot,Wood}. Finally this basic question
of vacuum stability to particle creation is clearly relevant to the fundamental problem of
vacuum energy, and the ultimate fate of the cosmological term in a full quantum theory.

The status of the vacuum in de Sitter space is very much analogous to that in other background field
problems of QFT, and in particular to the example of a charged quantum field in the presence of a
constant, uniform electric field. Such an idealized static background field is completely invariant
under time reversal and time translations. In this case, as first shown by Schwinger \cite{Schw},
there can be little doubt that the vacuum is unstable to the spontaneous creation of charged
particle/anti-particle pairs. This spontaneous process breaks the time reversal and translational 
symmetry of the background. Mathematically it is implemented by the $m^2 \rightarrow m^2 - i0^+$ 
prescription in the Feynman propagator, which distinguishes positive and negative frequency
solutions as particles and anti-particles respectively. It is this crucial distinction that 
provides the consistent definintion of the `vacuum' and `particle' concepts 
for QFT in background fields, including also gravitational fields \cite{DeWitt,Rumpf}.

The analogy with charged particle creation in an external uniform electric field is one that we shall develop
in parallel with the de Sitter case, since it is quite illustrative. Although this problem has a long
history \cite{Schw,Nar,Nik,NarNik,FradGitShv,KESCM,GavGit,QVlas}, the features directly relevant
to de Sitter space are worth emphasizing. In particular, one can find a completely time symmetric
state in a constant, uniform electric field background which has exactly zero decay rate by time
reversal symmetry, and is the close analog of the maximally symmetric CTBD state in de Sitter space.
The construction of such a time symmetric state, however appealing it may be mathematically,
is an artificial coherent superposition of particle and anti-particle waves, which assures neither its nature
as a vacuum state, nor its stability. The spontaneous particle creation process in an electric field leads
to an electric current that grows linearly with time and whose backreaction on the classical background
electric field must eventually be taken into account \cite{KESCM,QVlas}.

Our main purpose in this paper is to present a detailed time dependent description of
particle creation in de Sitter space, extending and deepening previous analyses of its instability
to quantum particle creation \cite{PartCreatdS}. The standard Feynman-Schwinger prescription of 
particle excitations moving forward in time with negative energy modes interpreted as anti-particles 
propagating backwards in time provides the framework for defining both an exact asymptotic definition of {\it in}
and {\it out} vacuum states in the remote past and future of de Sitter space, as well as an
approximate instantaneous adiabatic particle number at intermediate times. With this definition
it becomes evident that particles are created spontaneously, and the overlap $|\lag in | out\rag|^2$
provides the vacuum decay probability, just as it does in the electric field analog.
By investigating the particle creation process in real time and computing the corresponding 
energy-momentum of the particles, the vacuum instability of global maximally extended 
de Sitter space to particle creation, the breaking of both time reversal and global de Sitter 
symmetries, and the necessity to include backreaction of the particles on the geometry become 
clear. The nature of the CTBD state as a particular coherent squeezed state combination 
of particle and anti-particle excitations is also clarified.

In the case of a uniform electric field, the Feynman-Schwinger prescription is known also to
be equivalent to an adiabatic prescription of switching the electric field on and off again smoothly
on a time scale $T$, evaluating the particles present in the final field free {\it out} region starting
with the well-defined Minkowski vacuum in the initial field free {\it in} region \cite{Nar,NarNik,GavGit}.
We present evidence that for the analogous gentle enough switching on of the de Sitter phase from 
an initially static Einstein universe phase, for large values of $T$ and for modes with small enough 
momenta, the initial state produced for these modes is the asymptotic {\it in} state defined previously 
in eternal de Sitter space. Particle creation takes place for these modes after adiabatic switching 
on of de Sitter space and the particle spectrum produced is the same as in global de Sitter space. 

The paper is organized as follows. In the next section we begin with the preliminaries of defining a
(non-self-interacting) scalar quantum field theory in de Sitter space in order to fix notation.
In Sec. \ref{Sec:Scat} we formulate the problem of particle creation in de Sitter space in terms
of a one-dimensional time-independent scattering problem, review the construction of the {\it in}
and {\it out} vacuum states, the non-trivial Bogoliubov transformation between them, and the
de Sitter decay rate this implies. We also show that the de Sitter invariant CTBD state
is a definite phase coherent squeezed superposition of particle and anti-particle solutions
in both the past and the future and not a true vacuum state. In Sec. \ref{Sec:ConstantE} we
digress to consider the case of particle creation in a constant, uniform electric field, showing
the close analogy to the de Sitter case, including also the construction of the time symmetric
state analogous to the CTBD state. In Sec. \ref{Sec:Adb} we provide the general framework necessary to
interpolate between the asymptotic {\it in} and {\it out} states and describe the particle
creation process, the adiabatic particle number, and physical quantities such as the current and
energy-momentum tensor of the created particles in real time. In Sec. \ref{Sec:AdbdS} we apply
this adiabatic framework to global de Sitter space with the spatial ${\mathbb S}^3$ sections, showing
how the particle creation process through semi-classical creation `events' may be described in real time.
In Sec. \ref{Sec:Flat} we show how these methods may be applied equally well in the Poincar\'e
coordinates of de Sitter space with flat spatial sections most often used in cosmology.
In Sec. \ref{Sec:SET} we compute the energy-momentum tensor for vacuum states set at early initial
times in the global ${\mathbb S}^3$ sections, and show that particle creation in the contracting
phase leads to exponentially large energy densities and pressures which necessitate an inclusion
of their backreaction, breaking of de Sitter symmetry and large deviation of the spacetime from global
de Sitter space before the expanding phase even begins. In Sec. \ref{Sec:Switch} we present the numerical results
for the adiabatic turning on and off of de Sitter curvature on a time scale $T$ starting from a static space
and back, showing that the {\it in} state is produced in this way and particle production proceeds
as in global de Sitter space.

Sec. \ref{Sec:Conc} contains a Summary and Discussion of our results. There is
one Appendix which contains reference formulae for de Sitter geometry and coordinates,
included for completeness. The reader interested primarily in the main results in de Sitter 
space only may proceed from Sec. \ref{Sec:Scat} directly to Secs. \ref{Sec:AdbdS}-\ref{Sec:SET} 
and the Summary and Discussion, with reference to the other sections and Appendix as needed. 
This paper may be read in conjunction with the closely related paper \cite{AndMotDSVacua}, where the 
instability of the CTBD state to  perturbations, and their relation to the conformal trace anomaly 
are considered.

\section{Wave Equation and de Sitter Invariant State}

We consider in this paper a scalar field ${\bf \Phi}$  with mass $m$ and conformal curvature coupling $\xi = \frac{1}{6}$ 
which in an arbitrary curved spacetime satisfies the free wave equation
\be
(-\sq + M^2) {\bf \Phi} \equiv \left[ - \frac{1}{\sqrt{-g}} \frac{\partial}{\partial x^a}
\left(\sqrt{-g}\,g^{ab}\frac{\partial}{\partial x^b}\right) + M^2\right] {\bf \Phi} = 0
\label{waveq}
\ee
with effective mass $M^2 \equiv m^2 + \xi R$. In cosmological spacetimes with $\mathbb{S}^3$ closed spatial sections 
the Robertson-Walker line element is
\be
ds^2 = -d\tau^2 + a^2 d\Sigma^2 \;.
\label{RW}
\ee
Here $d\Sigma^2 = d\hat N\cdot d\hat N$ denotes the line element on $\mathbb{S}^3$ and $\hat{N}$, 
defined in Eq.\eqref{Ndef}, is a unit vector on $\mathbb{S}^3$. Specializing to de Sitter spacetime and 
defining the dimensionless cosmological time $u\equiv H\tau$, the scale factor
is $a(u) = H^{-1}\cosh u$, the Ricci scalar $R=12H^2$ is a constant, and the effective mass is
\be
M^2 = m^2 + 12\,\xi H^2 = m^2 + 2H^2
\ee
for $\xi= \frac{1}{6}$. The wave equation (\ref{waveq}) is separable in coordinates (\ref{RW}) with solutions of the form 
${\bf \Phi} = y_k (u)\, Y_{klm}(\hat N)$ and $Y_{klm_l} (\hat N)$ a spherical harmonic on $\mathbb{S}^3$ given
explicitly in \cite{AndMotDSVacua}. The equation for $y_k$ is
\be
\left[\frac{d^2}{du^2} + 3 \tanh u\, \frac{d}{du} + (k^2 -1)\, \sech^2 u
+ \left(\frac{9}{4} +\gamma^2\right) \right]\, y_k = 0
\label{modeq}
\ee
with the dimensionless parameter $\gamma$ defined by
\be
\gamma \equiv \sqrt{\frac{M^2}{H^2}-\frac{9}{4}} = \sqrt{\frac{m^2}{H^2} - \frac{1}{4}}
\label{gamdef}
\ee 
with the latter expression valid for conformal coupling. The range of the integers $k= 1, 2, \dots$ is taken to 
be strictly positive, so that the constant harmonic function on $\mathbb{S}^3$ corresponds to $k=1$, 
conforming to the notation of \cite{EMomTen}-\cite{Attract}. In some works the sign under the square root 
in (\ref{gamdef}) is reversed and the quantity $\nu = i\gamma$ is defined, which is real for 
$M^2 \le \frac{9}{4}H^2$ \cite{BirDav}. In this paper we shall be interested mainly in the massive 
case $M^2 > \frac{9}{4}H^2$ (the {\it principal series}) where $\gamma$ is real and positive, and for simplicity, 
in the case of conformal coupling $\xi = \frac{1}{6}$, so that $m^2 > \frac{1}{4}H^2$.

With the  change of variables $z = (1 -i \sinh u)/2$, the mode equation (\ref{modeq}) can be recast in the form
of the hypergeometric equation. The fundamental complex solution $y_k= \ups_{k\gamma}(u)$ may be taken to be
\be
\ups_{k\gamma}(u) \equiv H c_{k\gamma} \ (\sech u)^{k+1} \, (1 - i \sinh u)^k\,
F \left(\frac{1}{2} + i\gamma, \frac{1}{2} - i\gamma ; k+1 ; \frac{1 - i \sinh u}{2}\right)
\label{BDmode}
\ee
where $F \equiv \,_2F_1$ is the Gauss hypergeometric function and
\be
c_{k\gamma} \equiv \frac{1}{k!} \left[ \frac{ \Gamma\left(k + \frac{1}{2} + i\gamma\right)
\Gamma \left( k  + \frac{1}{2} - i\gamma\right)}{2}\right]^{\frac{1}{2}} =
 \frac{ \big|\Gamma\left(k + \frac{1}{2} + i\gamma\right)\!\big |}{\sqrt 2\ \Gamma(k+1)\ }  
\label{Cdef}
\ee
is a real normalization constant, fixed so that $\ups_{k\gamma}$ satisfies the Wronskian condition
\be
iH a^3(u) \left[\ups_{k\gamma}^* \frac{d}{du}\ups_{k\gamma} - \ups_{k\gamma}  \frac{d}{du}\ups_{k\gamma}^*\right] = 1 
\label{Wronups}
\ee
for all $k$. Note that under time reversal  $u \rightarrow -u$ the mode function  (\ref{BDmode}) goes
to its complex conjugate
\be
\ups_{k\gamma}(-u) = \ups_{k\gamma}^* (u) 
\label{timerev}
\ee
for all $M^2 > 0$.

The scalar field operator ${\bf \Phi}$ can be expressed as a sum over the fundamental solutions
\be
{\bf \Phi} (u,\hat N) = \sum_{k=1}^{\infty}\sum_{l=0}^{k-1}\sum_{m_l=-l}^{l}
\left\{ a^{\ups}_{klm_l}\, \ups_{k\gamma}(u)\, Y_{klm_l} (\hat N) +  a^{\ups\,\dagger}_{klm_l}\, \ups^*_{k\gamma}(u)
Y_{klm_l}^*\, (\hat N)\right\}  \;,
\label{Phiop}
\ee
with the Fock space operator coefficients $a^{\ups}_{klm_l}$ satisfying the commutation relations
\be
\Big[a^{\ups}_{klm_l}, a^{\ups\,\dagger}_{k'l'm'_l}\Big] = \delta_{kk'}\delta_{ll'}\delta_{m_lm'_l}\;.
\label{commutator}
\ee
With (\ref{Wronups}), (\ref{commutator}) and the standard unit normalization of harmonic functions
on the unit sphere
\be
\int_{\mathbb{S}^3} d^3\Sigma \ Y_{k'l'm_l'}^* \, Y_{klm_l} =  \delta_{k'k}\delta_{l'l}\delta_{m'_lm_l}  
\ee
the canonical equal time field commutation relation
\be
\Big[{\bf \Phi} (u, \hat N) , {\bf \Pi} (u, \hat N')\Big] = i\, \delta_{\Sigma}(\hat N, \hat N') 
\label{cancom}
\ee
is satisfied, where ${\bf \Pi} = \sqrt{-g}\, \dot{\bf \Phi} = H a^3 \frac{\partial{\bf \Phi}}{\partial u}$ is the field
momentum operator conjugate to ${\bf \Phi}$, the overdot denotes differentiation $H \partial/\partial u$ 
and $ \delta_{\Sigma}(\hat N, \hat N')$ denotes the delta function on the unit $\mathbb{S}^3$
with respect to the canonical round metric $d\Sigma^2$.

The Chernikov-Tagirov  or Bunch-Davies (CTBD) state $\vert \ups\rag$ \cite{Nacht,CherTag,BunDav} is defined by
\be
a^{\ups}_{klm_l} \, \vert \ups\rag = 0 \qquad \forall\quad  k, l, m_l 
\label{dS}
\ee
and is invariant under the full $O(4,1)$ isometry group of the complete de Sitter manifold (\ref{dSman})-(\ref{flat5}),
including under the discrete inversion symmetry of all coordinates in the embedding space,
$X^A \rightarrow - X^A$ (\ref{inversion}), which is not continuously connected to the identity.
This de Sitter invariant state $\vert \ups\rag$ is the one usually discussed in the literature, and
the Green functions in this state are those obtained by analytic continuation to de Sitter spacetime
from the Euclidean $\mathbb{S}^4$ with full $O(5)$ symmetry. The existence of an $O(4,1)$ invariant
symmetric state does not imply that this state is a stable vacuum. In this and a closely related
paper \cite{AndMotDSVacua}, we shall present the evidence that it is not.

\section{de Sitter Scattering Potential, In and Out States, and Decay Rate}
\label{Sec:Scat}

The solutions (\ref{BDmode}) and de Sitter invariant state $|\ups\rag$ are defined once and for all,
globally in de Sitter space without any reference to a separation between positive and negative
frequencies, which is axiomatic in flat spacetime to discriminate between particle and anti-particle states,
and necessary to define a stable vacuum which is a minimum of a positive definite Hamiltonian.
Whereas in flat spacetime such a separation into particle and anti-particle solutions of the
wave equation is Lorentz invariant, an $O(4,1)$ transformation will generally rotate the solution
$\ups_{k\gamma}$ into a linear combination of the $\ups_{k\gamma}$ and $\ups_{k\gamma}^*$
in de Sitter spacetime \cite{Nacht}, making a clean separation into particle and anti-particle
waves on symmetry grounds alone impossible. Corresponding to this mixing of positive and
negative frequency modes is the absence of a unique definition of `particles,'  or a positive
definite Hamiltonian operator to be minimized in de Sitter space. These important differences
with flat space are responsible for the non-trivial features of quantum fields and the quantum
vacuum in de Sitter space.

To appreciate the sharp distinction from flat space, it is useful to eliminate the factors of $a^3$
in the Wronskian condition (\ref{Wronups}) by defining the mode functions
\be
f_k= a^{\frac{3}{2}}\,y_k  
\label{fdef}
\ee
which satisfy the equation of a time dependent harmonic oscillator
\be
\frac{d^2}{d\tau^2}\, f_k + \Omega_k^2 \,f_k = 0  
\label{oscmode}
\ee
in each $k$ mode, with the time dependent frequency given by
\be
\Omega_k^2 \equiv \omega_k^2  + \left(\xi -\tfrac{1}{6}\right) R
- \frac{\dot h}{2} - \frac{h^2}{4}\,,\qquad \omega_k^2 \equiv  \frac{k^2}{a^2} + m^2 \;.
\label{defOmega}
\ee
Here $h\equiv \dot a/ a$ in a general RW spacetime with line element (\ref{RW}). Specializing
to de Sitter space and again using $u=H\tau$, the time dependent harmonic oscillator frequency is
\be
\Omega_k^2\Big\vert_{dS} = H^2\left[ \left(k^2 - \tfrac{1}{4}\right) \sech^2 u + \gamma^2\right] 
\label{OmegdS}
\ee
so that we may rewrite (\ref{oscmode}) in dimensionless form as a stationary state scattering problem
\be
\left[-\frac{d^2 }{du^2}  + {\cal U}_k(u) \right] f_{k\gamma} = \gamma^2\,  f_{k\gamma}  
\label{scat}
\ee
in the one-dimensional effective scattering potential
\be
{\cal U}_k(u) \equiv - \left(k^2 - \tfrac{1}{4}\right) \sech^2 u\;.
\label{scpot}
\ee
Here the `energy' $\gamma$ is defined by (\ref{gamdef}) and is positive for the fields with $\xi = \frac{1}{6}$ and 
$m^2 >\frac{1}{4}H^2$ considered in this paper.

Since the scattering potential (\ref{scpot}) is negative definite, and approaches zero exponentially
as $|u| \rightarrow \infty$, the solutions of (\ref{scat}) for $\gamma^2 >0$ describe {\it over the barrier}
scattering and are everywhere oscillatory. The vanishing of the potential at large $|u|$ implies well-defined
free asymptotic solutions as $u \rightarrow \mp \infty$, behaving like $e^{\pm i\gamma u}$. Because of the
scattering by the potential, a positive frequency wave $e^{-i\gamma u}$ incident from the left (the past
as $u\rightarrow -\infty$) will be partially transmitted to a positive frequency $e^{-i\gamma u}$ wave to the
right (the future as $u\rightarrow +\infty$) and partially reflected to a negative frequency $e^{i\gamma u}$
wave to the left. Potential scattering of this kind  and mixing of positive and negative energy solutions
clearly does not occur in static spacetimes such as Minkowski spacetime.

Now the crucial point is that these asymptotic pure frequency scattering solutions correspond exactly
to the Feynman prescription of positive energy solutions as particles propagating forward
in time and negative energy solutions as anti-particles propagating backward in time \cite{Feyn,Rumpf}.
This leads to the covariant definition of the Feynman propagator as the boundary value of
a function defined in the complex $m^2$ plane with the $m^2 - i 0^+$ prescription specifying the
limit in which the real axis is approached and pole contributions evaluated. This definition
is easily generalized to non-vanishing background fields and curved spacetimes by the same
generally covariant $m^2 - i 0^+$ prescription, and is then completely equivalent to the Schwinger-DeWitt
proper time method of defining the propagator and effective action functional in such situations \cite{Schw,DeWitt}.
This gives a rigorous definition of particles and anti-particles whenever the solutions of the time
dependent mode equation (\ref{oscmode}) behave as pure oscillating exponential functions in the
asymptotic past and the asymptotic future.  This definition is physically based on the corresponding
definitions in Minkowski space \cite{Rumpf}, free of any assumptions of analytic continuation
from Euclidean time or ${\mathbb S}^4$, and generally quite different from that prescription.
It is also the Feynman-Schwinger $m^2 - i 0^+$ definition of the propagator, and {\it only}
that definition that satisfies the composition rule for amplitudes defined by a single path integral \cite{FeynPath}.
Finally, in Sec. \ref{Sec:Switch} we provide evidence that this definition of asymptotic particle and anti-particle
solutions to the wave equation is the also the unique one produced by adiabatically switching the background 
gravitational field on and off.

Let us therefore denote by $f_{k\gamma (+)} (u)$ the properly normalized positive frequency solution
of (\ref{scat}) which behaves as $e^{-i\gamma u}$ as $u \rightarrow -\infty$ (the particle {\it in} solution),
and by $f_{k\gamma}^{(+)} (u)$ the properly normalized positive frequency solution of (\ref{scat})
which behaves as $e^{-i\gamma u}$ as $u \rightarrow +\infty$ (the particle {\it out} solution).
The corresponding negative frequency (or anti-particle) solutions $f_{k\gamma (-)} (u)$ and
$f_{k\gamma}^{(-)} (u)$ which behave as $e^{i\gamma u}$ as $u \rightarrow \mp\infty$ respectively
are obtained from these by complex conjugation. Moreover since the potential ${\cal U}_k(u)$ is real
and even under $u \rightarrow -u$, we have
\bes
\bea
f_{k\gamma (-)} (u) &= & [ f_{k\gamma (+)} (u)]^* = f_{k\gamma}^{(+)} (-u)    \\
f_{k\gamma}^{(-)} (u) &=& [f_{k\gamma}^{(+)} (u)]^* = f_{k\gamma (+)} (-u)  
\eea
\label{posneg}\ees
by which any one of the four solutions determines the other three. The proper normalization condition
for each set of modes analogous to (\ref{Wronups}) is
\be
iH \left(f_{k\gamma (+)}^* \frac{d}{du} f_{k\gamma (+)}
 - f_{k\gamma (+)} \frac{d}{du} f_{k\gamma (+)}^*\right) = 1
= iH \left(f_{k\gamma}^{(+)*} \frac{d}{du}  f_{k\gamma}^{(+)}
- f_{k\gamma}^{(+)}  \frac{d}{du}  f_{k\gamma}^{(+)*} \right)\;.
\label{scatnorm}
\ee
The CTBD mode function (\ref{BDmode}), which we may write in terms of a Legendre function \cite{Bate}
\be
F_{k\gamma}(u) \equiv a^{\frac{3}{2}} \ups_{k\gamma}(u) =
e^{-\frac{ik \pi}{2} {\rm sgn} (u)} \,\frac{\big\vert\,\Gamma\left(k + \frac{1}{2} + i \gamma\right)\!\big\vert}
{\sqrt{2H}}\, (\cosh u)^{\frac{1}{2}}\,  P^{-k}_{-\frac{1}{2} + i \gamma} (i \sinh u)  
\label{BDosc}
\ee
satisfies (\ref{scat}) and the normalization (\ref{scatnorm}) by virtue of (\ref{Cdef}), (\ref{Wronups}) and
(\ref{fdef}). The dependence of the phase on the sign of $u$ enters to compensate for the discontinuity of the Legendre
function  $P^{-k}_{-\frac{1}{2} + i \gamma}(\zeta)$ as conventionally defined with a branch cut along the real axis from
$\zeta = -\infty$ to $\zeta = +1$ \cite{Bate}, so that $F_{k\gamma}(u)$ is in fact continuous as $u$ crosses zero.

Since the {\it in}, the {\it out}, and the CTBD mode functions together with their
complex conjugates are each a complete set of solutions to (\ref{scat}), which preserve the
Wronskian relation (\ref{scatnorm}), they are expressible in terms of each other
by means of a Bogoliubov transformation. Specifically, the {\it in} mode functions are
expressible in terms of $F_{k\gamma}$ and $F_{k\gamma}^*$ by
\be
\left( \begin{array}{c} f_{k\gamma (+)} \\  f_{k\gamma (-)} \end{array}\right) =
\left( \begin{array}{cc} A_{k\gamma}^{in} & B_{k\gamma}^{in}\\
B^{in\, *}_{k\gamma} & A^{in\, *}_{k\gamma}\end{array}\right)\
\left( \begin{array}{c} F_{k\gamma}\\  F_{k\gamma}^* \end{array}\right)  
\label{BoginF}
\ee
and likewise for the {\it out} mode functions
\be
\left( \begin{array}{c} f_{k\gamma}^{(+)} \\  f_{k\gamma}^{(-)} \end{array}\right) =
\left( \begin{array}{cc} A_{k\gamma}^{out} & B_{k\gamma}^{out}\\
B^{out\, *}_{k\gamma} & A^{out\,*}_{k\gamma}\end{array}\right)\
\left( \begin{array}{c} F_{k\gamma}\\  F_{k\gamma}^* \end{array}\right)  \;.
\label{BogoutF}
\ee
Each set of the (strictly time independent) $A_{k \gamma}$ and $B_{k \gamma}$
Bogoliubov coefficients satisfies the relation
\be
|A_{k\gamma}|^2 - |B_{k\gamma}|^2 = 1\,.
\label{ABcond}
\ee
By using (\ref{posneg}) and $F_{k\gamma}(-u) = F_{k\gamma}^*(u)$ we immediately infer the relations
\be
A_{k\gamma}^{out} = A^{in\, *}_{k\gamma} \qquad {\rm and} \qquad B_{k\gamma}^{out} = B^{in\, *}_{k\gamma} 
\label{ABinout}
\ee
between the {\it in} and {\it out} Bogoliubov coefficients.
Furthermore, by inverting (\ref{BogoutF}) and substituting the result in (\ref{BoginF}) we obtain
\be
\left( \begin{array}{c} f_{k\gamma (+)} \\  f_{k\gamma (-)} \end{array}\right) =
\left( \begin{array}{cc} A_{k\gamma}^{in} & B_{k\gamma}^{in}   \\
B^{in\, *}_{k\gamma} & A^{in\, *}_{k\gamma}\end{array}\right)\!
\left( \begin{array}{cc} A_{k\gamma}^{out\, *} & -B_{k\gamma}^{out}\\
-B^{out\, *}_{k\gamma} & A^{out}_{k\gamma}\end{array}\right)\!
\left( \begin{array}{c} f_{k\gamma}^{(+)} \\  f_{k\gamma}^{(-)} \end{array}\right)
= \left( \begin{array}{cc} A_{k\gamma}^{tot} & B_{k\gamma}^{tot}\\
B^{tot\,*}_{k\gamma} & A^{tot \,*}_{k\gamma}\end{array}\right)\!
\left( \begin{array}{c} f_{k\gamma}^{(+)} \\  f_{k\gamma}^{(-)} \end{array}\right)  
\vspace{1mm}
\label{Boginout}
\ee
which with (\ref{ABinout}) gives
\vspace{-2mm}
\bes
\bea
&&A_{k\gamma}^{tot} = (A_{k\gamma}^{in})^2 - (B_{k\gamma}^{in})^2   \\
&&B_{k\gamma}^{tot} = A^{in\, *}_{k\gamma}\,B_{k\gamma}^{in} -  A_{k\gamma}^{in}\,B^{in\, *}_{k\gamma} 
\eea
\label{ABtot}\ees
for the coefficients of the total Bogoliubov transformation relating the {\it in} and {\it out} bases.

To find the Bogoliubov coefficients explicitly we construct the de Sitter scattering solutions (\ref{posneg}).
From the asymptotic form of the Legendre functions for large arguments \cite{Bate}, the pure positive frequency
solutions of (\ref{scat}) as $u \rightarrow \mp \infty$ are Legendre functions of  the second kind,
$Q^{-k}_{-\frac{1}{2} \pm i\gamma}$. Fixing the normalization by (\ref{scatnorm})
these exact {\it in} and {\it out} solutions of (\ref{scat}) may be taken to be
\bes
\bea
\hspace{-1cm} f_{k\gamma (+)}\Big\vert_{u<0}&=&
\frac{e^{-\frac{\pi\gamma}{2}}}{\big|\Gamma\left(\frac{1}{2} -k + i\gamma\right)\! \big|}
\left[ \frac{\cosh u}{H\sinh (\pi \gamma)}\right]^{\frac{1}{2}}
Q^{-k}_{-\frac{1}{2} - i\gamma} (i \sinh u) \label{modein}   \\
\hspace{-1cm} f_{k\gamma}^{(+)}\Big\vert_{u>0} &=&
\frac{e^{-\frac{\pi\gamma}{2}}}{\big|\Gamma\left(\frac{1}{2} -k + i\gamma\right)\! \big|}
\left[ \frac{\cosh u}{H\sinh (\pi \gamma)}\right]^{\frac{1}{2}}
Q^{-k}_{-\frac{1}{2} + i\gamma} (i \sinh u)   \label{modeout}
\eea
\ees
in the indicated regions of $u$, which have the required asymptotic behaviors \cite{Gutz}
\bes
\bea
f_{k\gamma (+)} \ &\raisebox{-1.5ex}{$\stackrel{\textstyle\longrightarrow}{\scriptstyle u \rightarrow -\infty}$}&\
\frac{\ (-)^k}{\sqrt{2H\gamma}} \, e^{\frac{i \pi}{4}}\, e^{-i\eta_{k\gamma}}\, e^{-i\gamma u} \\
f_{k\gamma}^{(+)} \ &\raisebox{-1.5ex}{$\stackrel{\textstyle\longrightarrow}{\scriptstyle u \rightarrow  \infty}$}
&\ \frac{\ (-)^k}{\sqrt{2H\gamma}} \, e^{-\frac{i \pi}{4}}\, e^{i\eta_{k\gamma}}\, e^{-i\gamma u} 
\eea
\label{inoutasym}\ees
respectively, and where the phase $\eta_{k\gamma}$ here is defined by
\be
\eta_{k\gamma} \equiv {\rm arg}\,\Big\{\Gamma(1-i \gamma)\,\Gamma\left(k+ \tfrac{1}{2} + i \gamma\right)\Big\} \;.
\label{thetadef}
\ee
Then by using Eq.\, (\ref{BDosc}) and Eq.\, 3.3.1 (11) of \cite{Bate} relating the Legendre functions of the second
kind to those of the first kind, we obtain
\bes
\bea
f_{k\gamma (+)} &=& \frac{1}{\sqrt{2 \sinh (\pi \gamma)}}\,
\left(i\,e^{-\frac{ik \pi}{2}} e^{\frac{\pi \gamma}{2}}\, F_{k\gamma}
+ e^{\frac{ik\pi}{2}} e^{-\frac{\pi \gamma}{2}}\, F_{k\gamma}  ^*\right) \label{intoBD}\\
f_{k\gamma}^{(+)} &=& \frac{1}{\sqrt{2 \sinh (\pi \gamma)}}\,
\left(-i\,e^{\frac{ik\pi}{2}} e^{\frac{\pi \gamma}{2}}\, F_{k\gamma}
+ e^{-\frac{ik\pi}{2}} e^{-\frac{\pi \gamma}{2}}\, F_{k\gamma}  ^*\right)\;.
\eea
\label{inouttoBD}\ees
which are valid for all $u$. Making use of the definitions (\ref{BoginF}) and (\ref{BogoutF}), we may read off
the Bogoliubov coefficients
\vspace{-2mm}
\bes
\bea
&&A_{k\gamma}^{in} = \frac{i}{\sqrt{2 \sinh (\pi \gamma)}} \,e^{-\frac{ik\pi}{2}}\, e^{\frac{\pi \gamma}{2}}
= A_{k\gamma}^{out\,*}  \\
&&B_{k\gamma}^{in} = \frac{1}{\sqrt{2 \sinh (\pi \gamma)}}\,e^{\frac{ik\pi}{2}}\, e^{-\frac{\pi \gamma}{2}}
= B_{k\gamma}^{out\,*}   \label{Boutvalue}\eea
\label{ABinvalues}\ees
relating the {\it in} and {\it out} scattering solutions of (\ref{scat}) to the fundamental CTBD solution in de Sitter space.

Notice that in inverting (\ref{BoginF}) or (\ref{BogoutF}) with (\ref{ABinvalues}), these relations imply that the
CTBD solution (\ref{BDosc}) is a very particular phase coherent superposition of positive and negative frequency
solutions at both $u \rightarrow \pm \infty$ \cite{Rumpf81}. Hence the $O(4,1)$ invariant state $|\ups\rag$ they define
through (\ref{dS}) contains particles in both the {\it in} and {\it out} bases and is not a particle vacuum state in either limit.
A direct consequence of this is that the $O(4,1)$ invariant propagator function constructed from the CTBD modes and
obtained also by analytic continuation from the Euclidean ${\mathbb S}^4$ manifold does not obey
the composition rule of a Feynman propagator function \cite{Poly}.

Clearly the quantization of the scalar field $\Phi$ may be formally carried out in either the {\it in} or {\it out} bases
and the corresponding Fock space operators introduced as in (\ref{Phiop})-(\ref{commutator}) for the CTBD basis.
Since there is scattering in the de Sitter potential (\ref{scat}) and the {\it in} and {\it out} states are related by a
non-trivial Bogoliubov transformation (\ref{ABtot}), which from (\ref{ABtot}) and (\ref{ABinvalues}) has coefficients
\bes
\bea
&&A_{k\gamma}^{tot} = (-)^{k -1}\, \coth (\pi \gamma)  \\
&&B_{k\gamma}^{tot} = i \,(-)^{k - 1}\, {\rm csch} (\pi \gamma) 
\eea
\label{ABtotval}\ees
the vacuum state $| in\rag$ defined by absence of positive frequency particle excitations at early
times is different from the corresponding vacuum state $|out\rag$ defined by the absence of positive
frequency particle excitations at late times.  Equivalently the early time $| in\rag$ state contains particle
excitations relative to the late time $| out\rag$ vacuum. The mean number density of particles of the
{\it out} basis in the vacuum state defined by the {\it in} basis is
\vspace{-1mm}
\be
|B_{k\gamma}^{tot}|^2 ={\rm csch}^2 (\pi \gamma) 
\label{Boutpart}
\ee
in the mode labeled by $(klm_l)$. Also
\vspace{-1mm}
\be
w_{\gamma} \equiv \left\vert\frac{B_{k\gamma}^{tot}}{A_{k\gamma}^{tot}}\right\vert^2 = \sech^2 (\pi \gamma) 
\label{wgam}
\ee
is the relative probability of creating a particle/anti-particle pair in this mode. Note that both (\ref{Boutpart})
and (\ref{wgam}) are independent of $(klm_l)$, depending only upon the mass of the field and its coupling 
to the scalar curvature. Equivalent results were found in earlier work \cite{PartCreatdS} with a different 
choice of the arbitrary phases for the scattering solutions and Bogoliubov coefficients.

The overlap between the {\it in} and {\it out} bases yields the probability that no particles are created, or
that the vacuum remains the vacuum, and is given by
\be
|\lag out | in \rag |^2 = \prod_{klm_l}\, (1- w_{\gamma})
= \exp \bigg\{\sum_{klm_l} \ln \left[\tanh^2 (\pi \gamma)\right]\bigg\}\,.
\label{vacprob}
\ee
Because the summand in the last expression is independent of $(klm_l)$, the sum is quite divergent.
This is an expression of the fact that in the infinite time limits $u \rightarrow \pm \infty$ the overlap
between the $|in\rag$ and $|out\rag$ states vanishes, and one should ask instead about the decay
rate per unit volume. This can be extracted from (\ref{vacprob}) by the following physical considerations,
which we justify more rigorously in Sec. \ref{Sec:AdbdS}. First the sums in (\ref{vacprob}) are regulated
by introducing a cutoff in the principal quantum number at $k_{max} = K$, so that
\be
\sum_{k=1}^K \sum_{l = 0}^{k-1} \sum_{m_l = - l}^{l} \, 1 = \frac{K(K+1)(2K+1)}{6} \rightarrow \frac{K^3}{3}  
\label{sumK}
\ee
for $K \gg 1$. Then one recognizes that the cutoff in the mode sum corresponds to a time
dependent cutoff in physical momenta at
\be
P_K(u) = \frac{K}{a} = \frac{KH}{\cosh u}  \;.
\label{Pcutoff}
\ee
Alternatively, for a fixed physical momentum cutoff $P_K$, an increase in time by $\Delta u$ results 
in an increase in $K$ such that
\be
\frac{\Delta K}{K} = \frac{\Delta (\cosh u)}{\cosh u} \rightarrow {\rm sgn}(u)\, \Delta u = \vert \Delta u\vert 
\label{DKDu}
\ee
as $K$ and $|u| \rightarrow \infty$. Thus the $K$ cutoff in the sum (\ref{sumK}) may
be traded for a cutoff in the time interval $u$ according to
\vspace{-1mm}
\be
\ln K \leftrightarrow |u| + {\rm const.} 
\label{lnKu}
\ee
where the constant is dependent upon the finite fixed $P_K$ and is unimportant in the limit $K, |u| \rightarrow \infty$.
Since the four-volume enclosed by the change of $u$ is
\be
\Delta V_4 = \int d^4 x \sqrt{-g}\Big\vert_{u}^{u + \Delta u} = \frac{2 \pi^2}{H^4} \,\cosh^3 u |\Delta u| \rightarrow
\frac{\pi^2}{4H^4} \, e^{3|u|} \frac{\Delta K}{K} \rightarrow \frac{\pi^2}{4H^4} \, K^2\Delta K  
\label{DelV4}
\ee
in this limit, the change in the sum in the exponent of (\ref{vacprob}) as the cutoff $K$ is changed, {\it viz.}
\be
\Delta \sum_{klm_l} \ln \left[\tanh^2 (\pi \gamma)\right] = -2 \ln \left[\coth (\pi \gamma)\right] \,K^2 \Delta K
\rightarrow - \frac{8H^2}{\pi^2} \ln \left[\coth (\pi \gamma)\right] \,\Delta V_4  
\label{Dsum}
\ee
may be regarded as giving rise to the finite decay rate per unit four-volume according to
\be
|\lag out | in \rag |^2_{\ \,V_4}  =  \exp\, (- \Gamma\, V_4)  
\label{vacpersist}
\ee
as $V_4 \rightarrow \infty$, with
\be
\Gamma = \frac{8 H^4}{\pi^2} \, \ln \big[\coth (\pi \gamma)\big]  
\label{Decayrate}
\ee
the decay rate of the vacuum {\it in} state due to particle creation in de Sitter space \cite{PartCreatdS}.
For $m \gg H$ the decay rate goes to zero exponentially
\be
\Gamma \rightarrow \frac{16 H^4}{\pi^2}\, e^{- 2 \pi m/H}\qquad {\rm for} \qquad m \gg H   
\ee
while the divergence of (\ref{Decayrate}) at $\gamma = 0$ indicates that the case of light masses must be treated differently.

The argument leading from (\ref{vacprob}) to (\ref{Decayrate}) will be justified in Sec. \ref{Sec:AdbdS}
by a more careful procedure based on an analysis of the real time particle creation process in de Sitter space.
This requires evolving the system from a finite initial time to a finite final time and defining
time dependent adiabatic vacuum states which interpolate smoothly between the $|in\rag$
and $|out\rag$ states, so that the infinite time limit is taken only at the end. The analysis of
particle creation in real time introduces the momentum dependence that is absent
from the asymptotic Bogoliubov coefficients (\ref{ABtotval}) in infinite times and which
justifies the replacement (\ref{DKDu}). It will also enable consideration of the stress tensor
of the created particles and their backreaction on the classical geometry. Before
embarking upon that more complete treatment of the particle creation process in de Sitter
space, we review the analogous case of particle creation in a constant uniform electric field,
which shares many of the same features, and for which the implication of an instability is clear.

\section{In/Out States and Decay Rate of a Constant Uniform Electric Field}
\label{Sec:ConstantE}

The case of a charged quantum field in the background of a constant uniform
electric field has many similarities with the de Sitter case. Although this case has
been considered by many authors \cite{Schw,Nar,Nik,NarNik,FradGitShv,KESCM,GavGit,QVlas},
the aspects relevant to the de Sitter case are worth emphasizing, including
the existence of a time symmetric state analogous to the CTBD state in de Sitter space,
which apparently has not received previous attention.

Treating the electric field as a classical background field analogous to the classical
gravitational field of de Sitter space,  the wave equation of a
non-self-interacting complex scalar field ${\bf \Phi}$ is
\be
\left[-(\partial_\mu -i eA_\mu) (\partial^\mu-i eA^\mu) + m^2\right]{\bf \Phi} = 0
\label{eomE}
\ee
in the background electromagnetic potential $A_{\mu}(x)$. Analogous to choosing global time
dependent coordinates (\ref{RW}) or (\ref{hypermet}) in de Sitter space, one may choose the time
dependent gauge
\be
A_z = -Et\,,\qquad A_t=A_x=A_y = 0
\label{Egauge}
\ee
in which to describe a fixed constant and uniform electric field in the $\bf \hat z$ direction.
Then the solutions of the field equation (\ref{eomE}) may be separated in Fourier modes
${\bf \Phi} \sim e^{i {\bf k\cdot x}} f_{\bf k}(t)$ with
\be
\left[ \frac{d^2}{dt^2} + (k_z + eEt)^2 + k_{\perp}^2 + m^2 \right] f_{\bf k}(t) = 0  \;.
\label{Emodeq}
\ee
This is again the form of a time dependent harmonic oscillator analogous to
(\ref{oscmode}), with the frequency function now given by
\be
\omega_{\bf k}(t) \equiv \left[ (k_z+eEt)^2 + k_{\perp}^2 + m^2\right]^{\frac{1}{2}} = \sqrt{2|eE|}\,\sqrt{\frac{u^2}{4\,} + \lam}
\label{OmegaE}
\ee
instead of (\ref{defOmega}) of the de Sitter case. We have defined here the dimensionless variables
\be
u \equiv \sqrt{\frac{2}{|eE|}} \ (k_z + eEt)\,,\qquad \lam \equiv \frac{k_{\perp}^2 + m^2}{2 \, |eE|\,} >  0\,.
\label{ulamdef}
\ee
Without loss of generality we can take the sign of $eE$ to be positive. With $f_{\bf k}(t) \rightarrow  f_{\lam}(u)$,
the wave equation (\ref{Emodeq}) then becomes
\be
\left[ \frac{d^2}{du^2} + \frac{u^2}{4\,} + \lam \right] f_{\lam}(u) = 0
\label{yEmode}
\ee
whose solutions may be expressed in terms of confluent hypergeometric functions $_1F_1(a;c;z)$
or parabolic cylinder functions \cite{Bate}
\be
D_{-i\lam - \frac{1}{2}}( e^{\frac{i\pi}{4}} u)\,, \quad D_{-i\lam - \frac{1}{2}}(-e^{\frac{i\pi}{4}} u)\,, \quad
D_{i\lam - \frac{1}{2}}( e^{-\frac{i\pi}{4}} u)\,, \quad D_{i\lam - \frac{1}{2}}( -e^{-\frac{i\pi}{4}} u)\,.
\label{pcylsolns}
\ee
Any two of the solutions (\ref{pcylsolns}) are linearly independent for general real $\lambda$.

As in the de Sitter case Eq. (\ref{yEmode}) may viewed as a one-dimensional stationary state scattering problem
for the Schr\"odinger equation in the inverted harmonic oscillator potential $- u^2/4$, independent
of $k$ in this case, with `energy' $\lam$ (the analog of $\gamma^2$). We again have over the barrier scattering
in a potential that is even under $u \rightarrow -u$, with no turning points on the real $u$ axis and the solutions
(\ref{pcylsolns}) are everywhere oscillatory for positive $\lam$. Although the potential $- u^2/4$
grows without bound as $|u| \rightarrow \infty$, pure positive frequency {\it in} and {\it out}  particle modes
can be defined by the requirement that they behave as $(2 \omega_{\bf k})^{- \frac{1}{2}}e^{- i\Theta_{\lam}(u)}$,
where the adiabatic phase $\Theta_{\lam}(u)$ is defined by
\bea
&&\Theta_{\lam}(u) \equiv \int_{t(u=0)}^{t(u)} \, dt\,\omega_{\bf k}(t) = \frac{1}{2} \int_0^u du\,\sqrt{u^2+ 4\lam}
\, = \frac{u}{4} \sqrt{u^2 + 4 \lam} \,+\, \lam \ln \left( \frac{u +  \sqrt{u^2 + 4 \lam}}{2 \sqrt\lam}\right)\nn
&&\qquad \qquad \rightarrow\ {\rm sgn}(u) \left\{\frac{u^2}{4} + \frac{\lam}{2} \left[ \ln \left(\frac{u^2}{\lam}\right) + 1 \right]\right\}
+ {\cal O} \left(\frac{\lam^2}{u^2}\right)
\label{Ephase}
\eea
as $|u| \rightarrow \infty$. The fact that the phase (\ref{Ephase}) has a well-defined asymptotic form with small corrections
means that well-defined positive and negative frequency mode functions exist in the limit of large $|u|$,
although the potential (\ref{Egauge}) does not vanish in this limit. Examining the asymptotic form of the various
parabolic cylinder functions (\ref{pcylsolns}) one easily finds the exact solutions of (\ref{yEmode}) which behave
as pure positive frequency adiabatic solutions of (\ref{Emodeq}) or (\ref{yEmode}), {\it viz.}
\bes
\bea
f_{\lam \,(+)}(u)
&=&(2eE)^{-\frac{1}{4}}\, e^{-\frac{\pi \lam}{4}}\, e^{i\eta_{\lam}}\, D_{- \frac{1}{2} + i\lam}( -e^{-\frac{i\pi}{4}} u)\\
f_{\lam}^{(+)}(u)
&=& (2eE)^{-\frac{1}{4}}\, e^{-\frac{\pi \lam}{4}}\, e^{-i\eta_{\lam}} \,D_{- \frac{1}{2} -i\lam}( e^{\frac{i\pi}{4}} u)
\eea
\label{inoutyEmodes}\ees
and which satisfy the Wronskian normalization condition
\be
i\left(f_{\lam\, (+)}^*\frac{d}{dt}f_{\lam\, (+)} - f_{\lam\, (+)} \frac{d}{dt} f_{\lam\, (+)}^*\right) = 1
= i \left(f_{\lam}^{(+)\,*} \frac{d}{dt}f_{\lam}^{(+)} - f_{\lam}^{(+)} \frac{d}{dt}f_{\lam}^{(+)\,*}\right)
\label{normE}
\ee
analogous to (\ref{scatnorm}). These {\it in} and {\it out} scattering solutions are chosen to have the
simple pure positive frequency asymptotic behaviors
\bes
\bea
&&f_{\lam\, (+)} \ \raisebox{-1.5ex}{$\stackrel{\textstyle\longrightarrow}{\scriptstyle u \rightarrow -\infty}$}
\ (2 \omega_{\bf k})^{-\frac{1}{2}}\, e^{- i \Theta_{\lam}(u)} \label{inadb}\\
&&f_{\lam}^{(+)} \ \ \,\raisebox{-1.5ex}{$\stackrel{\textstyle\longrightarrow}{\scriptstyle u \rightarrow +\infty}$}
\ (2 \omega_{\bf k})^{-\frac{1}{2}}\, e^{- i \Theta_{\lam}(u)} \label{outadb}
\eea
\label{inoutadb}\ees
provided the arbitrary constant phase $\eta_{\lam}$ in (\ref{inoutyEmodes}) is taken to be
\be
\eta_{\lam} \equiv \frac{\lam}{2} - \frac{\lam}{2}\,\ln \lam - \frac{\pi}{8}\,.
\label{phaselam}
\ee
The {\it in} and {\it out} particle mode solutions (\ref{inoutyEmodes}) and the corresponding complex
conjugate anti-particle mode solutions are a set of four solutions of (\ref{yEmode}) which are related
to each other by the precise analog of (\ref{posneg}) in the de Sitter case. Here we have
chosen to incorporate the phase $\eta_{\lam}$ into the definition of the modes (\ref{inoutyEmodes})
rather than have it appear in the asymptotic forms (\ref{inoutadb}), as the analogous phase
$\eta_{k\gamma}$ does in (\ref{inoutasym}) of the previous section.

Now an additional point of correspondence is the existence of a $u$-time
symmetric solution to (\ref{yEmode}) analogous to the CTBD mode solution (\ref{BDmode})
or (\ref{BDosc}) in de Sitter space, and a corresponding maximally symmetric state of the
charged quantum field in a constant, uniform electric field background. That such a mode
solution to (\ref{yEmode}) obeying
\be
\ups_{\lam}(-u) = \ups_{\lam}^*(u)
\label{symmode}
\ee
exists is clear from the $u \rightarrow -u$ symmetry of the real scattering potential $-u^2/4$.
Since there is no expansion factor $a(u)$ in this case, this symmetric function is also the
analog of $F_{k\gamma}(u)$ (\ref{BDosc}) in the de Sitter case. It is most conveniently expressed
in terms of the confluent hypergeometric function defined by the confluent hypergeometric series
\be
\Phi(a,c;z) \equiv  \,_1F_1(a;c;z) = \sum_{n=0}^{\infty} \frac{(a)_n}{(c)_n}\, \frac{z^n}{n!}\,,\qquad
(a)_n \equiv \frac{\Gamma(a+n)}{\Gamma(a)}
\label{confseries}
\ee
or the integral representation
\be
\Phi(a,c;z) = \frac{\Gamma(c)}{\Gamma(a) \Gamma(c-a)} \int_0^1dx \, e^{xz}  \, x^{a-1} \, (1-x)^{c-a-1}
\,,\qquad {\rm Re}\, c > {\rm Re}\, a > 0
\label{confint}
\ee
in the form
\be
\ups_{\lam}(u)= 2^{-\frac{1}{2}}(k_{\perp}^2 + m^2)^{-\frac{1}{4}}\, e^{-\frac{iu^2}{4}}\,
\left[\Phi\left( \frac{1}{4} + \frac{i\lam}{2}, \frac{1}{2}; \frac{iu^2}{2}\right)
- i u\, \lam^{\frac{1}{2}}\,\Phi\left( \frac{3}{4} + \frac{i\lam}{2}, \frac{3}{2}; \frac{iu^2}{2}\right)\right]
\label{Esymodes}
\ee
which is correctly normalized by (\ref{normE}), and satisfies (\ref{symmode}) by use of the
Kummer transformation of the function $\Phi(a,c;z)$, {\it c.f.} Eq. 6.3 (7) of \cite{Bate}.
By making use of the value $\Phi(a,c;0) =1$ from (\ref{confseries}) or (\ref{confint}), we find
\bes
\bea
&&\ups_{\lam}(0) = 2^{-\frac{1}{2}}(k_{\perp}^2 + m^2)^{-\frac{1}{4}} = \frac{1}{\sqrt{2\omega_{\bf k}}}\bigg\vert_{u=0}\\
&&\frac{\partial  \ups_{\lam}}{\partial t} \bigg\vert_{u=0} = -i\sqrt{eE\lam}\, (k_{\perp}^2 + m^2)^{-\frac{1}{4}} =
\frac{-i\, \omega_{\bf k}}{\sqrt{2\omega_{\bf k}}}\bigg\vert_{u=0}
\eea
\label{upsinit}\ees
so that the symmetric solution $\ups_{\lam}$ matches the adiabatic positive frequency form
$(2 \omega_{\bf k})^{- \frac{1}{2}}e^{- i\Theta_{\lam}(u)}$ at the {\it symmetric} point of the potential
$u=0$, halfway in between the asymptotic limits $u \rightarrow \pm \infty$.
The solution of (\ref{yEmode}) with these properties is unique.

The existence of such a time reversal invariant solution to (\ref{yEmode}) implies the existence
of a maximally symmetric state constructed along the lines of the maximally $O(4,1)$ invariant
invariant state (\ref{dS}) in the de Sitter background. The existence of this state of maximal
symmetry does not imply that it is the stable ground state of either the de Sitter or electric field
backgrounds. In the electric field case this is well known and the decay rate of the electric field
into particle/anti-particle pairs \cite{Schw} is becoming close to being experimentally
verified in the near future \cite{Laser}. That result is easily recovered in the present formalism
by calculations exactly parallel to those of the de Sitter case in the last section.

First the Bogoliubov transformation analogous to (\ref{BoginF}) relating the $\it in$ state mode
function to the symmetric one $\ups_{\lam}(u)$ and its complex conjugate are determined
from the relation between the parabolic cylinder function in $f_{\lam \,(\pm)}(u)$ and the
confluent hypergeometric functions, {\it c.f.} Eqs. 6.9.2 (31) and 6.5 (7) of \cite{Bate}, which give
\be
f_{\lam\, (+)}(u) = A^{in}_{\lam}\, \ups_{\lam} (u) + B_{\lam}^{in} \ups_{\lam}^*
\label{insym}
\ee
with
\bes
\bea
&&A^{in}_{\lam}= \sqrt{\frac{\pi}{2}}\, 2^{\frac{i\lam}{2}}\,e^{i\eta_{\lam}}\,e^{-\frac{\pi\lam}{4}}
\left[ \left(\frac{\lam}{2}\right)^{\frac{1}{4}} \frac{1}{\Gamma\left(\frac{3}{4} - \frac{i\lam}{2}\right)}
+ \left(\frac{2}{\lam}\right)^{\frac{1}{4}}\frac{ e^{\frac{i\pi}{4}}}{\Gamma\left(\frac{1}{4} - \frac{i\lam}{2}\right)}\right]\\
&&B^{in}_{\lam} =  \sqrt{\frac{\pi}{2}}\, 2^{\frac{i\lam}{2}}\, e^{i\eta_{\lam}}\,e^{-\frac{\pi\lam}{4}}
\left[ \left(\frac{\lam}{2}\right)^{\frac{1}{4}} \frac{1}{\Gamma\left(\frac{3}{4} - \frac{i\lam}{2}\right)}
- \left(\frac{2}{\lam}\right)^{\frac{1}{4}}\frac{ e^{\frac{i\pi}{4}}} {\Gamma\left(\frac{1}{4} - \frac{i\lam}{2}\right)}\right]
\eea
\label{ABinE}\ees
Because the $\it in$ and $\it out$ mode functions satisfy the same relations as (\ref{posneg}),
and have the same relation to the symmetric mode function $\ups_{\lam}(u)$
as the corresponding $\it in$ and $\it out$ mode functions have to the CTBD mode
function (\ref{BDosc}) in the de Sitter case, the Bogoliubov coefficients defined by the
analogs of (\ref{BoginF})-(\ref{ABtot}), and the coefficients of the total Bogoliubov
transformation from $\it in$ to $\it out$ states in the electric field case are given by
the same relations as (\ref{ABtot}), {\it viz.}
\bes
\bea
&&A_{\lam}^{tot} = (A_{\lam}^{in})^2 - (B_{\lam}^{in})^2 =
\frac{\sqrt{2\pi}}{\Gamma\left(\frac{1}{2} - i \lam\right)}\,
e^{ - \frac{\pi \lam}{2}}\,e^{i \lam - i \lam \ln \lam}\,,\\
&&B_{\lam}^{tot} = A^{in\, *}_{\lam}\,B_{\lam}^{in} -  A_{\lam}^{in}\,B_{\lam}^{in\, *}
= -i\, e^{-\pi \lam}\,.\label{BtotE}
\eea
\label{ABtotE}\ees
Thus the number density of {\it out} particles at late times in the mode labeled
by ${\bf k}$ or $(k_z, k_{\perp})$ if the system is prepared in the {\it in} vacuum is
\be
|B_{\lam}^{tot}|^2 = e^{-2\pi \lam} = \exp \left[- \frac{\pi (k_{\perp}^2 + m^2)}{eE}\right]
\label{Blampart}
\ee
and the relative probability of finding a particle/anti-particle charged pair in the mode
characterized by $(k_z, {\bf k_{\perp}})$ in the $| in\rag$ `vacuum' is
\be
w_{\lam} \equiv \left\vert\frac{B_{\lam}^{tot}}{A_{\lam}^{tot}}\right\vert^2 = \frac{1}{e^{2\pi\lam} + 1}
\label{wlam}
\ee
which is independent of $k_z$. The vacuum overlap or vacuum persistence probability is given then
by the analog of (\ref{vacprob}),
\be
|\lag out | in \rag |^2 = \prod_{\bf k} (1-w_{\lam}) =  \exp \Big\{ -\sum_{\bf k} \,\ln\, (1 + e^{-2\pi \lam})\Big\}\,
\label{vacprobE}
\ee
Taking the infinite volume limit and converting the sum into an integral according to
\be
\sum_{\bf k} \rightarrow  \frac{V}{(2\pi)^3} \int dk_z \int d^2 \bf k_{\perp}
\label{sumint}
\ee
we see that the exponent in (\ref{vacprobE}) diverges both in $V$ and because the integrand
is independent of $k_z$. Thus we should again define the decay rate  by dividing the exponent
in the vacuum persistence probability (\ref{vacprobE}) by the four-volume $VT$,  before taking
the infinite time limit $T \rightarrow \infty$.  Recognizing that the physical (kinetic) longitudinal
momentum of the particle in mode $k_z$ is $p = k_z + eEt$, for a fixed large $p=P$ cutoff
we have
\be
dk_z = -eE\, dt\,.
\label{dkdt}
\ee
Thus the positive integral over $k_z$ in (\ref{sumint}) may be replaced by $eET$, $T$ being the
total elapsed time over which the electric field acts to create pairs. In this way we obtain
from (\ref{vacprobE})-(\ref{dkdt}) the vacuum decay rate per unit three-volume $V$ per unit time $T$
to be
\bea
\Gamma &=& \frac{eE}{(2\pi)^3} \int d^2 {\bf k}_{\perp}\, \ln (1 + e^{-2\pi \lam})\nn
&=& \frac{eE}{(2\pi)^3} \int_0^{\infty}\pi\,  dk^2_{\perp}\, \sum_{n=1}^{\infty} \frac{(-)^{n+1}\hspace{-6mm}}{n}
\hspace{5mm}
\exp\left[-\frac{\pi n (k_{\perp}^2 + m^2)}{eE}\right]\nn
&=& \frac{(eE)^2}{(2\pi)^3}\sum_{n=1}^{\infty} \frac{(-)^{n+1}\hspace{-6mm}}{n^2} \
\hspace{4mm}\exp\left(-\frac{\pi n  m^2}{eE}\right)
\label{Erate}
\eea
which is Schwinger's result for scalar QED. (Schwinger actually obtained the
result for fermionic QED in which the alternating sign in the sum over $n$ is absent \cite{Schw}).

Thus the definition of the {\it in} and {\it out} states which are purely positive frequency as
$t \rightarrow \mp \infty$ respectively according to (\ref{inoutadb}) gives a non-trivial particle creation
rate and imaginary part of the one-loop effective action which agrees with \cite{Schw}, notwithstanding
the existence of a fully time symmetric state with mode functions (\ref{Esymodes}). Clearly a non-zero
imaginary part and decay rate breaks the time reversal symmetry of the background. Mathematically
this is of course a result of initial boundary conditions on the vacuum, implemented in the present
treatment by the definition of positive frequency solutions at early and late times, or in Schwinger's
proper time original treatment by the $m^2 \rightarrow m^2 - i0^+$ prescription of avoiding a pole.
As in the de Sitter case, the time symmetric modes (\ref{Esymodes}) can be defined and have the
maximal symmetry of the background $\bf E$ field. They do {\it not} describe a true vacuum state,
but rather a specific coherent superposition of particles and anti-particles with respect to
either the {\it in} or {\it out} vacuum states, `halfway between.' The time symmetric state defined
by the solution (\ref{Esymodes}) is a very curious state indeed, corresponding to the rather unphysical
boundary condition of each pair creation event being exactly balanced by its time reversed pair annihilation
event, these pairs having been precisely arranged to come from great distances at early times in order to
effect just such a cancellation.

We note also that taking the strict asymptotic states (\ref{inoutyEmodes}) in a constant
uniform electric field leads to the same sort of divergence in the $k_z$ momentum integration
we encountered in the $k$ sum in the de Sitter case, which can be handled by the replacement
(\ref{dkdt}) based on similar considerations of a fixed physical momentum cutoff. The reason
that the calculation  leading to (\ref{vacprobE}) together with a physical argument for the $k_z$ cutoff
gives the identical answer to Schwinger's proper time method \cite{Schw} is of course due to the fact
that the definition of particles by the positive frequency solutions of the time dependent mode
Eq. (\ref{yEmode}) is the same one selected by the covariant analyticity requirement
of the $m^2 - i 0^+$ prescription. For this correspondence to be unambiguous it is important
that the adiabatic frequency function $\Theta_{\lam}(u)$ in (\ref{Ephase}) have well-defined
asymptotic behavior at large $|u| \gg \sqrt\lam$, so that the {\it in} and {\it out} positive
frequency mode functions may be identified by the asymptotic behaviors of
the appropriate exact solutions of (\ref{yEmode}), even though the electric field
does not vanish in these asymptotic regions at very early or very late times.
Indeed exactly the same result (\ref{Erate}) is obtained if the electric field is switched
on and off smoothly \cite{Nar,NarNik,GavGit} in a finite time $T$. Then the Bogoliubov
coefficients have a non-trivial $k_z$ dependence and the integral over $k_z$
for finite $T$ is finite. Dividing by $T$ and taking the limit $T \rightarrow \infty$ one
recovers exactly the decay rate (\ref{Erate}) according to the replacement (\ref{dkdt})
above.

A technical deficiency of the asymptotic $| in\rag$ and $|out\rag$ states related to the
divergence of the $k_z$ integral or $k$ sum is that they are not Hadamard UV allowed states.
That is to say, the Wightman correlation functions computed in these states do not have the
same short distance singularities as those of flat space and mild deformations therefrom.
This has been known for some time in the de Sitter case, where in $|in\rag$ and $|out\rag$
states are two members of the `$\alpha$ vacuum' family of states with particular non-zero
values of the parameter $\alpha$ \cite{PartCreatdS,Allen}. Although this entire family of
states are formally de Sitter invariant under the $SO(4,1)$ subgroup of $O(4,1)$ continuously
connected to the identity, the two-point Wightman correlation function in all such states
other than the CTBD $\alpha=0$ state have short distance singularities as $x\rightarrow x'$
that differ than those in flat space. This implies a sensitivity of local short distance physics
to global state properties, at odds with usual expectations of renormalization and effective field theory.

The reason for this unphysical UV behavior of the {\it in} and {\it out} states
is the non-commutivity of the infinite time $|u| \rightarrow \infty$ and infinite momentum
$k \rightarrow \infty$ limit in the de Sitter case, or $(k_z, k_{\perp}) \rightarrow \infty$
limit in the $\bf E$ field case. The short distance or UV properties of the state rely
in Fourier space on the vacuum matching the flat space or zero field vacuum
to sufficiently high order at sufficiently high momentum or short distances, whereas
these large $k$ properties are lost if the infinite time limit is taken first. In the electric field
case the physical cutoff on $|k_z|$ is of order $eET$, so that the very high $k_z$ modes are
undisturbed from the ordinary vacuum and the unphysical ultraviolet behavior
of matrix elements and Green's functions in the initial state is removed when $T$ is
finite. The finite $T$ regulator thus eliminates any UV problem, and transfers the
divergence instead to the question of the long time or {\it infrared} secular evolution
of the system. Then time translational as well as time reversal symmetry is lost.

All idealized calculations in background fields that persist for infinite times do not give much
physical insight into the particle creation process itself in real time. In formulating a well-posed
time dependent problem one needs to define states in which the momentum dependent
particle creation process can be followed at any finite time. This leads to the introduction of adiabatic
vacuum and particle states defined at arbitrary times, not just in the asymptotic past or future.

\section{Adiabatic Vacuum States and Particle Creation in Real Time}
\label{Sec:Adb}

The $\it in$ and $\it out$ mode functions $f_ {(+)}$ and $f^{(+)}$ are pure positive
frequency particle modes in the asymptotic past and asymptotic future respectively,
while the time symmetric $\ups$ or $F$ is `halfway between' them and a positive frequency
mode at $u=0$. This suggests that it would be useful to introduce WKB mode functions
\be
\tilde f_{\bf k} = \frac{1}{\sqrt{2W_{\bf k}}} \, \exp\left(-i \int^t dt\, W_{\bf k}\right)
\label{adbmode}
\ee
that are {\it approximate} adiabatic positive frequency modes at any intermediate
time $t$, to interpolate between these limits. These approximate modes are related
to any of the {\it exact} mode function solutions $f_ {(+)}$ and $f^{(+)}$ or $\ups$
of the oscillator equation (\ref{oscmode}) or (\ref{Emodeq}) in the de Sitter of electric field
backgrounds (which we denote generically by $f_{\bf k}$) by a {\it time dependent}
Bogoliubov transformation
\be
\left( \begin{array}{c} f_{\bf k} \\ f_{\bf k}^* \end{array}\right) =
\left( \begin{array}{cc} \alpha_{\bf k}(t)& \beta_{\bf k}(t)\\
\beta^*_{\bf k}(t) & \alpha_{\bf k}^*(t)\end{array}\right)\
\left( \begin{array}{c} \tilde f_{\bf k}\\  \tilde f_{\bf k}^* \end{array}\right)
\label{Bogadb}
\ee
where we require that
\be
|\alpha_{\bf k}(t)|^2 - |\beta_{\bf k}(t)|^2 = 1
\label{alpbetnorm}
\ee
be satisfied at all times. The time dependent {\it real} frequency function $W_{\bf k}$ in (\ref{adbmode})
is to be chosen to match the exact frequency function $\Omega_k$  or $\omega_{\bf k}$ of the time
dependent harmonic oscillator equation (\ref{oscmode}) or (\ref{Emodeq}), {\it i.e.}  (\ref{defOmega}) or
(\ref{OmegaE}), to some order in the adiabatic expansion
\be
W_{\bf k}^2 \simeq \omega_{\bf k}^2 -\frac{1}{2} \frac{\ddot \omega_{\bf k}}{\omega_{\bf k}}
+ \frac{3}{4} \frac{\dot \omega_{\bf k}^2}{\omega_{\bf k}^2} + \dots
\label{adbfreq}
\ee
obtained by substituting (\ref{adbmode})  into the oscillator equation and expanding in
time derivatives of the frequency.

The expansion (\ref{adbfreq}) is adiabatic in the usual sense of slowly varying, in that it is clear
that the approximate positive frequency mode (\ref{adbmode}) more and more accurately approaches
an exact mode solution of the oscillator equation (\ref{oscmode}) or (\ref{Emodeq}), as (\ref{defOmega}) or
(\ref{OmegaE}) becomes a more slowly varying function of time, which is controlled by the strength
of the background gravitational or electric field. 

A second important property of the expansion (\ref{adbfreq}) is that it is an {\it asymptotic}
expansion (rather than a convergent series) which is {\it non-uniform} in $\bf k$. The higher
order terms fall off more and more rapidly at large $|{\bf k}|$, for {\it any} value of the background
field $H$ or $E$, or no matter how rapidly the background varies. This guarantees that the
adiabatic vacuum defined by (\ref{adbmode}) will match the usual Minkowski vacuum
at sufficiently short time and distance scales as $|{\bf k}| \rightarrow \infty$, for {\it any}
smoothly varying background field. This property of the adiabatic vacuum is essential
to the renormalization program for currents and stress-tensors, necessary to formulate
the backreaction problem for time varying background fields \cite{ParFul,BirBun,BirDav}.
In the literature the term {\it adiabatic} is often used in this second sense of the large
$|\bf k|$ behavior of vacuum modes and Green's functions for arbitrary (smooth)
backgrounds, independently of whether or not they are slowly varying in time \cite{BirDav}.

Because of the Wronskian normalization conditions (\ref{Wronups}) or (\ref{normE}), the
coefficients of the time dependent Bogoliubov transformation (\ref{Bogadb}) are
completely defined only if the first time derivatives of the exact mode functions in
terms of $\alpha_{\bf k}$ and $\beta_{\bf k}$ are also specified. The general form
of $\dot f_{\bf k}$ in terms of the adiabatic modes $\tilde f_{\bf k}$ that preserves
both the Wronskian condition (\ref{Wronups}) and (\ref{alpbetnorm}) is \cite{EMomTen}
\be
\frac{d}{dt} f_{\bf k} = \left(-iW_{\bf k} + \frac{V_{\bf k}}{2}\right)\alpha_{\bf k} \tilde f_{\bf k}
+ \left(iW_{\bf k} + \frac{V_{\bf k}}{2}\right)\beta_{\bf k} \tilde f_{\bf k}^*
\label{fmodedt}
\ee
where $V_{\bf k}$ is a second time dependent real function, with its own adiabatic expansion
given by the time derivative of $W_{\bf k}$ from (\ref{adbfreq}). Then the transformation
of bases (\ref{Bogadb}) may be viewed as a time dependent canonical transformation
in the phase space of the coordinates $f_{\bf k}$ and their conjugate momenta $\dot f_{\bf k}$.
The corresponding adiabatic particle and anti-particle creation and destruction operators
may be defined by setting the Fourier components
\be
\varphi_{\bf k}(t) \equiv a_{\bf k} f_{\bf k}(t) + b^{\dag}_{\bf k}f^*_{-\bf k} (t)
= \tilde a_{\bf k}(t) \tilde f_{\bf k} (t)+ \tilde b^{\dag}_{-\bf k}(t) \tilde f^*_{-\bf k}(t)
\label{Fourier}
\ee
equal so that the canonical transformation in the Fock space
(of a charged scalar field) is
\be
\left( \begin{array}{c} \tilde a_{\bf k}(t) \\ \tilde b_{-\bf k}^{\dag}(t) \end{array}\right) =
\left( \begin{array}{cc} \alpha_{\bf k}(t)& \beta_{\bf k}^*(t)\\
\beta_{\bf k}(t) & \alpha_{\bf k}^*(t)\end{array}\right)\
\left( \begin{array}{c} a_{\bf k}\\  b^{\dag}_{-\bf k}\end{array}\right)
\label{BogadbFock}
\ee
when referred to the time independent basis $(a_{\bf k}, b^{\dag}_{-\bf k})$.  For an uncharged
Hermitian scalar field, $b_{\bf k}$ and $b^{\dag}_{-\bf k}$ are replaced by $a_{\bf k}$ and  $a^{\dag}_{-\bf k}$
respectively. The time dependent instantaneous mean adiabatic particle number in the mode $\bf k$ is defined 
in the $(\tilde a_{\bf k}, \tilde b^{\dag}_{-\bf k})$ basis as
\vspace{-1mm}
\bea
{\cal N}_{\bf k}(t) &\equiv& \lag \tilde a^{\dag}_{\bf k}(t) \tilde a_{\bf k}(t) \rag = \lag \tilde b^{\dag}_{-\bf k}(t) \tilde b_{-\bf k}(t) \rag\nn
&=& |\alpha_{\bf k}(t)|^2 \lag a^{\dag}_{\bf k} a_{\bf k} \rag
+ |\beta_{\bf k}(t)|^2  \lag b_{-\bf k} b_{-\bf k}^{\dag}\rag\nn
&=& N_{\bf k} + (1 + 2 N_{\bf k}) |\beta_{\bf k}(t)|^2
\label{adbpart}
\eea
where
\vspace{-3mm}
\be
N_{\bf k} \equiv \lag a^{\dag}_{\bf k} a_{\bf k} \rag =  \lag b^{\dag}_{-\bf k} b_{-\bf k}\rag
\label{initN}
\ee
is the number of particles (assumed equal to the number of anti-particles) referred to the time
independent basis. This may be taken to be the particle number at the initial time $t=t_0$
provided that we initialize so that $|\beta_{\bf k}(t_0)|^2 = 0$.

With the definition of $V_{\bf k}$ in (\ref{fmodedt}), the time dependent Bogoliubov
coefficients may be found explicitly:
\bes
\bea
\alpha_{\bf k}&=& i \tilde f_{\bf k}^*
\left[\dot f_{\bf k} - \left( iW_{\bf k}+ \frac{V_{\bf k}}{2} \right) f_{\bf k}
\right]\\
\beta_{\bf k} &=& -i \tilde f_{\bf k}\left[ \dot f_{\bf k} +
\left( iW_{\bf k} - \frac{V_{\bf k}}{2}\right) f_{\bf k} \right]  \label{genbet}
\eea
\label{genalpbet}\ees
and in particular
\be
|\beta_{\bf k}(t)|^2 = \frac{1}{2W_{\bf k}} \left\vert\dot f_{\bf k} +
\left( iW_{\bf k} - \frac{V_{\bf k}}{2}\right) f_{\bf k} \right\vert^2
\label{betsq}
\ee
is determined in terms of the adiabatic frequency functions $(W_{\bf k}$, $V_{\bf k})$
and the exact mode function solution $f_{\bf k}$ of the oscillator equation (\ref{oscmode})
or (\ref{Emodeq}), which is specified by initial data $(f_{\bf k}, \dot f_{\bf k})$
at $t= t_0$. Although the choice of $(W_{\bf k}, V_{\bf k})$ is not unique, it is fairly
tightly constrained by the requirements of matching the adiabatic behavior of the
asymptotic expansion (\ref{adbfreq}) to sufficiently high order, but not higher than
is necessary to isolate the divergences of the current $\lag {\bf j} \rag$ or stress
tensor $\lag T^a_{\ b}\rag$ operators in their `vacuum-like' contributions. We shall
see that with these requirements, although the detailed time dependence of
${\cal N}_{\bf k}(t)$ depends on the precise choice of $(W_{\bf k}, V_{\bf k})$,
the main features of the adiabatic particle number are largely independent of that choice.

\begin{figure}[htp]
\vspace{-5mm}
\includegraphics[angle=90,height=9cm,viewport= 60 0 560 760, clip]{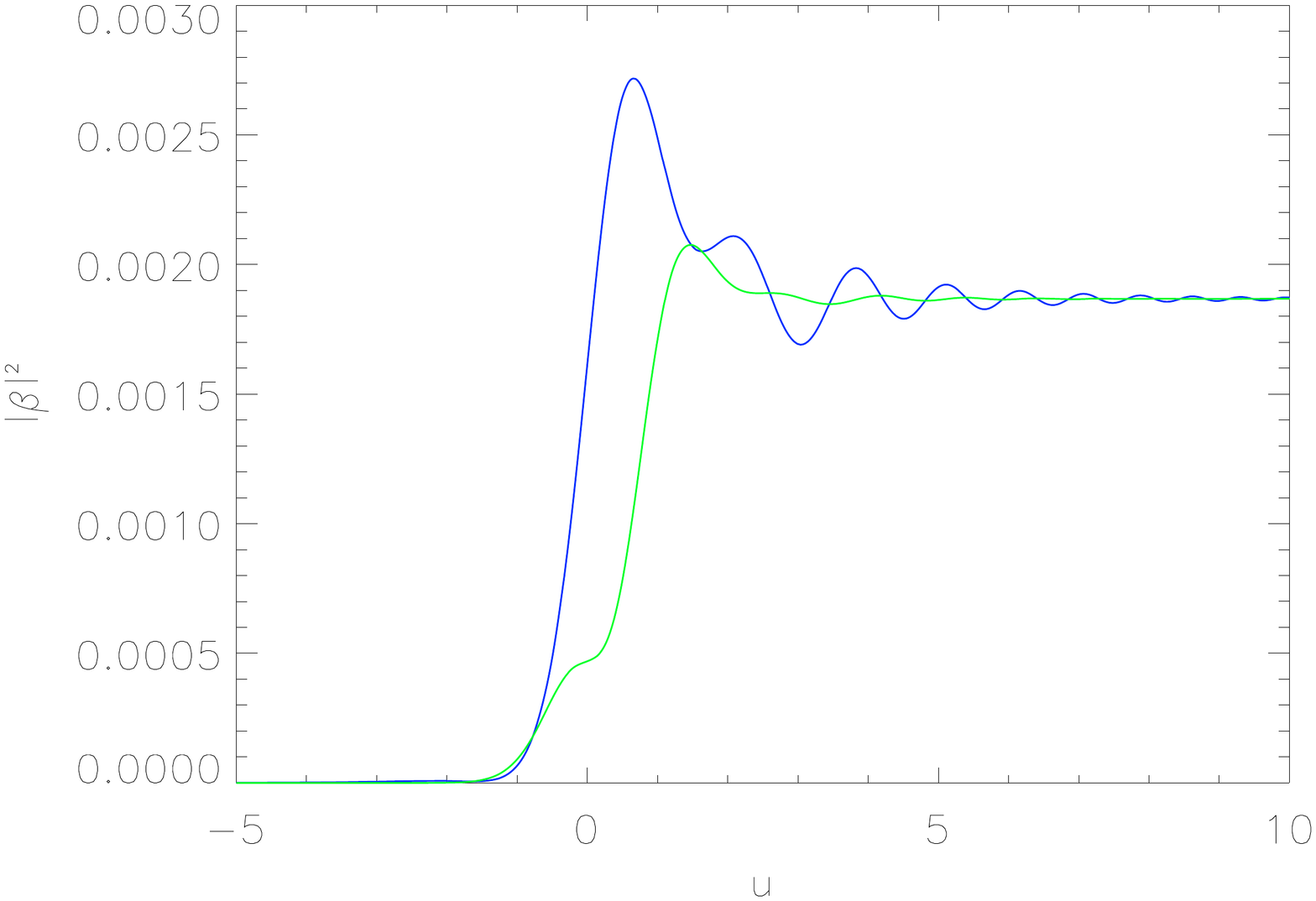}
\vspace{-5mm}
\caption{The mean number of particles created from the vacuum {\it in} state, given by (\ref{betsq})
with $f_{\bf k} =  f_{\lam \,(+)}(u)$ of (\ref{inoutyEmodes}), and with $\lam =1$. The blue curve with larger
oscillations is for the choice of ($W_{\bf k}, V_{\bf k}$) given by (\ref{WVE}), while the green curve is for
the second order adiabatic frequency of (\ref{W2}) and the same $V_{\bf k}$. Both change rapidly around 
$u=0$, and both tend to the same asymptotic value $e^{-2 \pi}= 0.001867$ of (\ref{Delbet}) as $u\rightarrow \infty$.}
\vspace{-3mm}
\label{Fig:EW0W2}
\end{figure}

Let us first apply this general adiabatic framework to the constant, uniform electric field
example. Although it is sufficient to choose the lowest order adiabatic frequency functions
\bes
\bea
&&W_{\bf k}^{(0)} = \omega_{\bf k} = \sqrt{\frac{eE}{2}}\ (u^2 + 4\lam)^{\frac{1}{2}} \label{WforE}\\
&&V_{\bf k}^{(1)} = -\frac{\dot \omega_{\bf k}}{\omega_{\bf k}} = -\sqrt{2eE}\  \frac{u}{u^2 + 4 \lam}
\eea
\label{WVE}\ees
in this case, we shall also study the second order choice
\be
W_{\bf k}^{(2)} =  \omega_{\bf k} - \frac{1}{4} \frac{\ddot \omega_{\bf k}}{  \omega_{\bf k}^2}
+ \frac{3}{8} \frac{\dot   \omega_{\bf k}^2}{  \omega_{\bf k}^4} = \sqrt{\frac{eE}{2}}\ (u^2 + 4\lam)^{\frac{1}{2}}
\left[1 - \frac{1}{(u^2 + 4 \lam)^2}  +  \frac{5}{2}\frac{u^2}{(u^2 + 4 \lam)^3} \right]
\label{W2}
\ee
for comparison purposes. The Bogoliubov coefficient  $|\beta_{\bf k}|^2$ and the adiabatic mean particle
number were studied in a constant electric field background with the choice $W_{\bf k}^{(0)}$ and $V_{\bf k} =0$
in \cite{QVlas}. In Fig. \ref{Fig:EW0W2} we plot $|\beta_{\bf k}|^2$ defined by (\ref{betsq}) with $f_{\bf k}$
the {\it in} vacuum mode function $f_{\lam \,(+)}(u)$ of (\ref{inoutyEmodes}) and for both the lowest
order and second order choices of adiabatic frequency $W_{\bf k}$, given by (\ref{WVE}) and
(\ref{W2}) respectively.

\begin{figure}[hbp]
\vspace{-5mm}
\includegraphics[height=6cm,viewport= 20 10 150 135, clip]{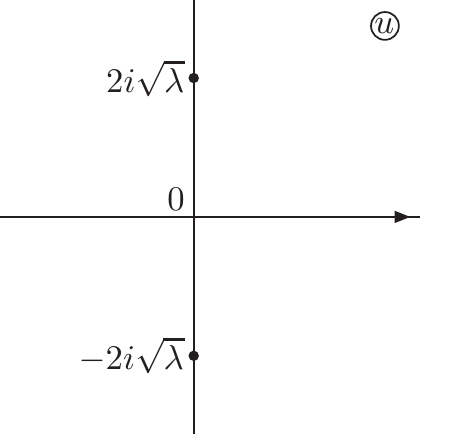}
\vspace{-3mm}
\caption{Location of the zeroes of the frequency function $\omega_{\bf k}$ in the complex
$u$ plane. Particle creation occurs as the real time $u$ contour passes the imaginary
axis at $u=0$ between the pair of these zeroes.}
\vspace{-3mm}
\label{Fig:EZeroes}
\end{figure}

A continuous but sharp rise in $|\beta_{\bf k}|^2$ is observed in each $k_z$ mode
around its `creation event,'  at $u=0$, {\it i.e.} at the time when the kinetic momentum
$p=k_z+ eEt =0$. Since the adiabatic mode functions are essentially WKB approximations
to the time dependent harmonic oscillator equation (\ref{oscmode}) or (\ref{Emodeq}),  the particle
creation process in real time and this rapid rise may be understood from a consideration of
the WKB turning points in the complex $u$ plane  \cite{PokKhal}.
These are defined by the values of $u$ where the frequency function
$\omega_{\bf k}$ vanishes. Since the solutions are oscillatory on the real time axis, those
turning points are located off the real line, and in the case of (\ref{Emodeq})-(\ref{OmegaE})
these zeroes of the frequency are at
\be
u = \pm u_{\lam} \equiv \pm \,2i\, \sqrt{\lam}
\label{Ezero}
\ee
illustrated in Fig. \ref{Fig:EZeroes}.

\begin{figure}[htp]
\vspace{-5mm}
\includegraphics[angle=90,height=9cm,viewport= 60 0 560 760, clip]{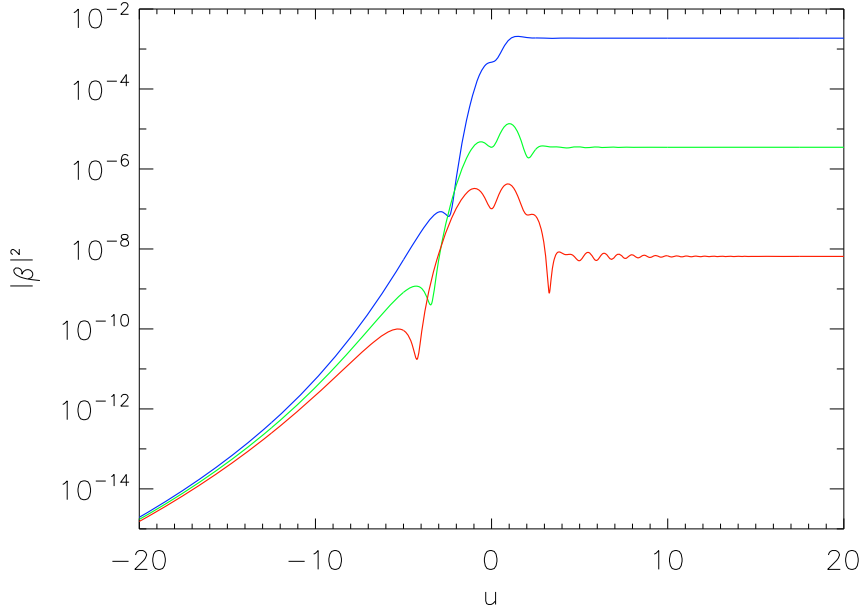}
\vspace{-5mm}
\caption{The mean number of particles created from the vacuum {\it in} state, given by (\ref{betsq})
with $f_{\bf k} =  f_{\lam \,(+)}(u)$ of (\ref{inoutyEmodes}), and the second order adiabatic frequency of (\ref{W2})
for $\lam =1, 2, 3$ (blue, green, red curves respectively). Note the logarithmic scale. The asymptotic values
for large $u$ are $1.87 \times 10^{-3}$, $3.49 \times 10^{-6}$, and $6.51 \times 10^{-9}$ for $\lam =1,2,3$ respectively,
in agreement with (\ref{Delbet}).}
\vspace{-3mm}
\label{Fig:Evarlam}
\end{figure}

The qualitative behavior of the Bogoliubov coefficient $\beta_{\bf k}(t)$ along the real $u$
axis can be found by finding the lines of steepest descent of the adiabatic phase function
$\Theta_{\lam}(u)$ of (\ref{Ephase}) in the complex $u$ plane, as they emanate from
$u_{\lam}$ \cite{PokKhal}. Far from the turning points, for $|u| \gg 2\sqrt\lam$, the exact
mode functions are well approximated by the adiabatic WKB mode function (\ref{adbmode})
and hence $|\beta_{\bf k}(t)|^2$ defined by (\ref{betsq}) will be approximately constant.
For $u \ll -|u_{\lam}| <0$ the adiabatic vacuum is approximately the {\it in} vacuum discussed
previously and $|\beta_{\bf k}(t)|^2$ is nearly zero if it is initialized so that $|\beta_{\bf k}(t_0)|^2 = 0$.
For $u \gg |u_{\lam}| > 0$, the adiabatic vacuum is approximately the {\it out}
vacuum. Again $|\beta_{\bf k}(t)|^2$ will be approximately constant in this region
and given approximately by the {\it total} Bogoliubov coefficient $B^{tot}_{\lam}$ from {\it in}
to {\it out}. In the region $u \in [-u_{\lam}, u_{\lam}]$, as $u$ passes nearest the
complex turning points (\ref{Ezero}), the exact mode function $f_{\bf k}(t)$ receives
an increasing admixture of the negative frequency component, and $|\beta_{\bf k}|^2$
changes rapidly from its {\it in} to {\it out} value. This change in $|\beta_{\bf k}(t)|^2$
in this region of $\Delta u \sim 4 \sqrt\lam$ or $\Delta t \sim 2 \sqrt{k_{\perp}^2 + m^2}/eE$
around $u=0$ (closest to the complex zero of $\omega_{\bf k}$) is given by (\ref{Blampart}) or
\be
\Delta |\beta_{\bf k}|^2 = |B^{tot}_{\lam}|^2 = e^{-2 \pi \lam}
\label{Delbet}
\ee
and can also be found from (\ref{Ephase}) by evaluating
$\exp [-4 \,{\rm Im}\, \Theta_{\lam} (u_{\lam})]$ \cite{PokKhal}.
We use the total Bogoliubov coefficient $B^{tot}_{\lam}$ since the rise
in $|\beta_{\bf k}|^2$ changes continuously in this region $u \in [-u_{\lam}, u_{\lam}]$
between the two complex zeroes (\ref{Ezero}) with no constant value halfway between.
The behavior of $|\beta_{\bf k}(u)|^2$ for various $\lam$ is plotted in Fig. \ref{Fig:Evarlam}
showing the asymptotic value of the jump in particle number consistent with (\ref{Delbet}).
Since this jump $|\beta_{\bf k}|^2$ occurs around $u=0$, the particle creation
`event' occurs at a different time $t = -k_z/eE$ for modes with different $k_z$.

Consider now the adiabatic initial data at some finite time $t_0$,
\bes
\bea
&&f_{\bf k} (t_0) = \frac{1}{\sqrt{2 \omega_{\bf k}(t_0)}}\\
&&  \frac{df_{\bf k}}{dt}\bigg\vert_{t=t_0} =   -\left(i\omega_{\bf k} + \frac{\dot \omega_{\bf k}}{2 \omega_{\bf k}}\right)
f_{\bf k}\Big\vert_{t=t_0}
\eea
\label{phiadbinitE}\ees
with $\omega_{\bf k}(t)$ given by (\ref{OmegaE}). This matches the adiabatic vacuum
with (\ref{WVE}) so that $\beta_{\bf k} (t_0) = 0$. Since the creation event occurs around
$u=0= k_z + eEt$, with a finite starting time only those modes for which the initial kinetic
momentum $p(t_0) = k_z + eEt_0<0$ can experience this creation event. They do so
at the time when their kinetic momentum $p(t) = k_z + eEt =0$, {\it i.e.} when the
particle initially moving in the opposite direction to the electric field is brought
to instantaneous rest $p(t) =0$ by the constant positive acceleration of the field and
begins to move in the direction of the electric field. On the other hand those modes
for which $p(t_0) = k_z + eEt_0>0$ are already moving in the same direction of the
electric field at the initial time and undergo no particle creation event at later times,
being already approximately in the $\it out$ vacuum state at the initial time $t_0$. Crudely
approximating the creation event as a step function at $u=0$ with step size (\ref{Delbet}),
the number of particles in mode $\bf k$ at time $t > t_0$ is
\bea
{\cal N}_{\bf k}(t) &=&(1 + 2N_{\bf k}) |\beta_{\bf k}(t)|^2 \approx \theta (p(t)) \theta (-p(t_0)) \, e^{-2 \pi \lam}\nn
&\approx& (1+ 2N_{\bf k})\, \theta (k_z +eEt)\, \theta (-k_z - eEt_0) \, e^{-2 \pi \lam}
\label{Epart}
\eea
where the factor of $1 + 2N_{\bf k}$ accounts for the induced creation rate of particles if there
are already particles $N_{\bf k} > 0$ in the initial state. From (\ref{Epart}) there is a `window function' in
$k_z$ for modes going through particle creation given by
\be
-eEt < k_z < -eEt_0
\label{Ewin}
\ee
which grows linearly with elapsed time $t-t_0$. The behavior of $|\beta_{\bf k}|^2$ is shown in
detail in Figs. \ref{Fig:EW0W2} and \ref{Fig:Evarlam}, and Figures 2-4 of Ref. \cite{QVlas}, with
the $\theta$ functions actually smooth functions of $u$ that rise on the time scale of
$\Delta u \sim 4 \sqrt{\lam}$, which can be accurately captured by the uniform asymptotic
approximation of the parabolic cylinder functions even for moderately small $\lam$ \cite{QVlas}.
Replacing this smooth rise of the average particle number by a step function already gives a
qualitatively correct picture of the semi-classical particle creation process mode by mode
in real time, with the correct asymptotic density of particles. It is the window function (\ref{Ewin})
which justifies the replacement of the integral over $k_z$ in (\ref{Erate}) by $eE$ times
the total elapsed time $T=t-t_0$, which can then be divided out to obtain the decay rate.
The window function (\ref{Ewin}) of the real time particle creation process also agrees
with the analysis of adiabatically switching on and off of the background electric
field, so that it acts only for a finite time \cite{Nar,NarNik,GavGit,AndMotSan}. It is this
definition of particles created by the electric field in the adiabatic basis that forms
the starting point in quantum theory for a kinetic description \cite{QVlas}.

The adiabatic basis also furnishes a simple physically well-motivated method
for defining renormalized expectation values of current and energy-momentum
bilinears in the quantum field. In the approximation in which the electric field background
is treated classically while the charged scalar matter field is quantized, the renormalized $j_z$
current expectation value is
\be
\lag t_0| j_z(t)| t_0\rag_{_R} = 2e \int \frac{d^3{\bf k}}{(2\pi)^3}\, (k_z + eEt) \left[ (1 + 2 N_{\bf k})
\vert f_{\bf k}(t)\vert^2 - \frac{1}{2 \omega_{\bf k}(t)}\right]
\label{jzexact}
\ee
where the leading divergence has been subtracted by the adiabatic vacuum term in which $|f_{\bf k}|^2$
has been replaced by $|\tilde f_{\bf k}|^2$ with (\ref{WVE}) and $N_{\bf k}$ replaced by zero. It can
be shown that this one subtraction removes all the UV divergences in the momentum
integral for a constant $E$ field \cite{KESCM}. A logarithmic divergence proportional to $\ddot E$
can be removed by using the second adiabatic order approximation for $W_{\bf k}$ in the expansion
(\ref{adbfreq}). As this term can easily be reabsorbed into coupling renormalization in backreaction
calculations and vanishes in any case for a constant $E$ field, the lowest order subtraction in
(\ref{jzexact}) is sufficient for our present purposes.

Substituting (\ref{Bogadb}) we obtain from (\ref{jzexact})
\be
\lag t_0| j_z(t)| t_0\rag_{_R} =  2e \int \frac{d^3{\bf k}}{(2\pi)^3}\, \frac{(k_z + eEt)}{\omega_{\bf k}}\,
\Big[\,{\cal N}_{\bf k}  + {\rm Re} \, (\alpha_{\bf k} \beta^*_{\bf k} e^{-2i\Theta_{\bf k}})\,\Big]
\label{qcurrent}
\ee
where
\be
\Theta_{\bf k} \equiv \int_{t_0}^t\omega_{\bf k}\, dt = \Theta_{\lam}(u(t)) - \Theta_{\lam}(u(t_0))
\label{defTheta}
\ee
is the adiabatic phase in (\ref{adbmode}), related to the function $\Theta_{\lam}(u)$ defined in (\ref{Ephase}).
Since $(k_z + eEt)/\omega_{\bf k} = p/\omega_{\bf k}$ is the $z$ component of the velocity of a classical particle
in the electric field, the first term in the integral of (\ref{qcurrent}) has a self-evident classical interpretation
as the contribution to the electric current of the positive plus negatively charged particles with phase space
number density ${\cal N}_{\bf k}$. The second term is a quantum interference term which has no classical
analog. This term is both rapidly oscillating in time and rapidly oscillating in $|\bf k|$ for fixed time, so
one would expect it to average out in the integral and give a relatively small contribution to the total current
compared to the first term. For the semi-classical particle interpretation based on the adiabatic modes
(\ref{adbmode}) to be most useful, this should be the case. If it is, one can also substitute the step
approximation (\ref{Epart}) for the particle density (assuming $N_{\bf k} =0$, {\it i.e.} no particles in the
initial state) and arrive at the simple result
\bea
\lag t_0| j_z(t)| t_0\rag_{_R} &\approx & 2e \int \frac{d^3{\bf k}}{(2\pi)^3}\,
\frac{(k_z + eEt)}{\omega_{\bf k} (t)} \,\theta(k_z +eEt)
\,\theta(-k_z - eEt_0)\, \exp\left[- \frac{\pi (k_{\perp}^2 + m^2)}{eE}\right]\nn
&& = \frac{e}{\pi}  \, \Big[\sqrt{e^2E^2(t-t_0)^2 + m^2} - m\Big]\,
\int_0^{\infty} \frac{dk_{\perp}^2}{4 \pi}\, \exp\left[- \frac{\pi (k_{\perp}^2 + m^2)}{eE}\right]\nn
&& \rightarrow \frac{e^3 E^2}{4\pi^3}  \, (t-t_0)\, e^{- \frac{\pi m^2}{eE\ }}
\label{jzapprox}
\eea
for the linear growth with time of the mean electric current of the created particles.
This exhibits the secular effect coming from the window function (\ref{Ewin}) opening
linearly with time so that more and more modes go through their particle
creation event as time goes on, each becoming accelerated very rapidly to
the speed of light, and making a constant contribution to the current.

\begin{figure}[htp]
\vspace{-1.1cm}
\includegraphics[height=10cm,viewport= 100 0 1600 1200, clip]{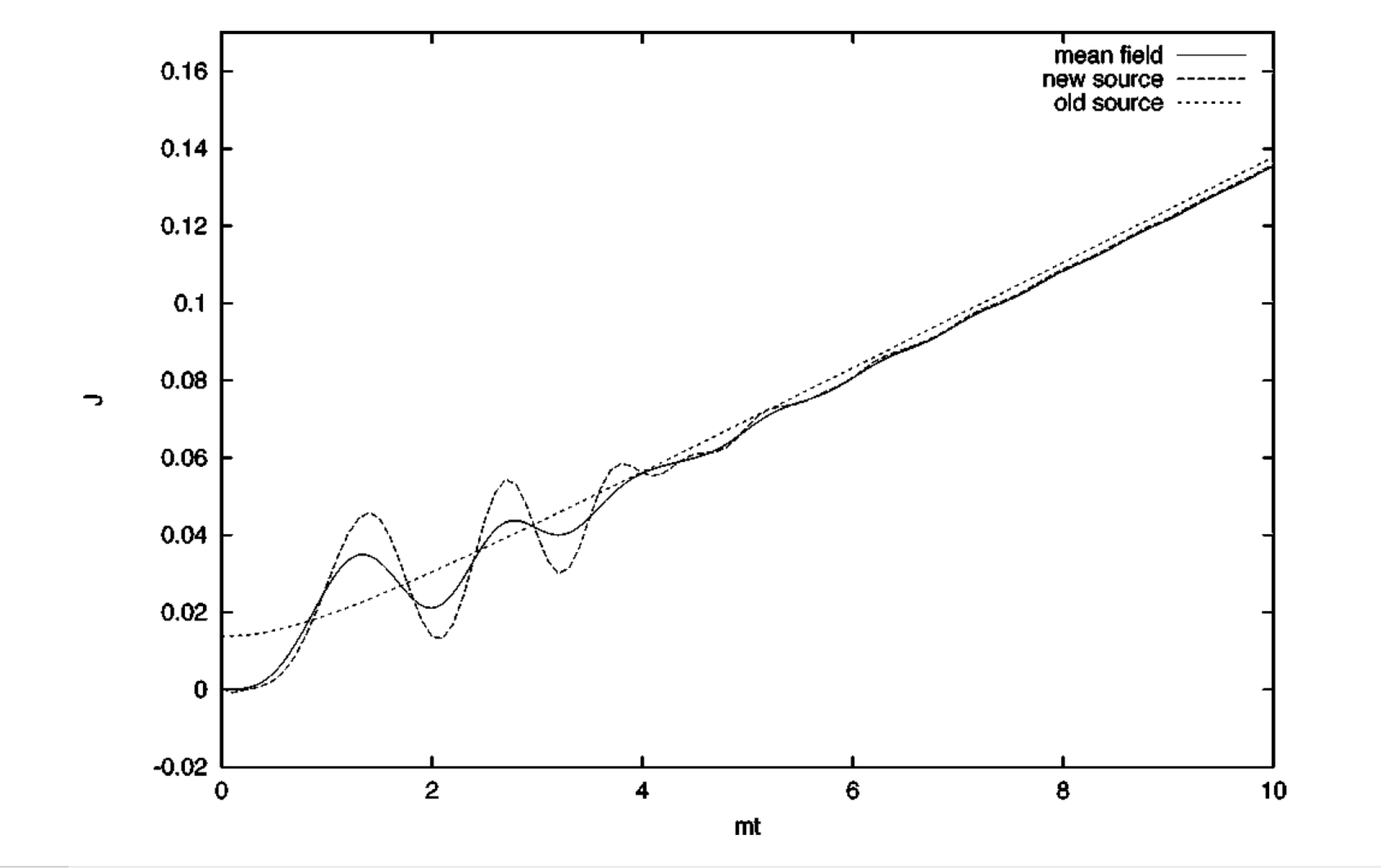}
\vspace{-3mm}
\caption{The linear growth of the electric current $J= \lag j_z\rag$ with time  in the case of
fixed constant background electric field in $1+1$ dimensions  in units of $e^2E$ for $eE/m^2 = 1$.
The three curves shown are the current of the exact renormalized current expectation value (\ref{jzexact}),
a uniform approximation described in Ref. \cite{QVlas} (new source) and with the simple window step
function of particle creation (\ref{Epart}), (\ref{jzapprox}), labelled old source here.}
\vspace{-3mm}
\label{Fig:CurrentE}
\end{figure}

One can also evaluate the exact expectation value (\ref{jzexact}) for a constant uniform electric
field background starting with the initial adiabatic data (\ref{phiadbinitE}) and compare it to the
simple step function approximation (\ref{jzapprox}). This comparison is shown in Fig. \ref{Fig:CurrentE} \cite{QVlas}.
The transient oscillations are the effect of the second quantum interference term in (\ref{qcurrent})
while the dominant {\it secular} effect of linear growth at late times is correctly captured
by the simple approximation (\ref{jzapprox}) based on the particle creation picture, labelled
as `old source' in Fig. \ref{Fig:CurrentE}. The curve labeled `new source' is the uniform
approximation of \cite{QVlas} that gives a slightly better approximation than the crude
step function approximation of (\ref{Epart}). Either gives correctly the coefficient of the linear
secular growth with time, which implies that backreaction must eventually be taken into account,
no matter how small $eE/m^2$ is, provided only that it is non-zero. This secular growth is a
non-perturbative infrared `memory effect' in the sense of depending upon the time elapsed
since the initial vacuum state is prepared at $t=t_0$. Note that this time dependence
due to particle creation is a spontaneous breaking of the time translational and time reversal
symmetry of the background constant $\bf E$ field \cite{Fluc}. The exponentially small tunneling
factor associated with the spontaneous Schwinger particle creation rate from
the vacuum shows that the effect is non-perturbative, but that however small,
it can  be overcome by a large initial state density of particles $N_{\bf k} >>1$
for which the induced particle creation and current is much larger.  Even in the initial
adiabatic vacuum case for $N_{\bf k} =0$, particle creation eventually overcomes the 
small tunneling factor at late times.

\section{Adiabatic States and Initial Data in de Sitter Space}
\label{Sec:AdbdS}

As in the electric field case, we introduce instantaneous adiabatic vacuum states in de Sitter space,
defined by the adiabatic mode functions
\be
\tilde f_k = \frac{1}{\sqrt{2W_k}} \, \exp\left(-i \int^{\tau} d\tau\, W_k\right)
\label{adbmodS}
\ee
analogous to (\ref{adbmode}). Due to spatial homogeneity and isotropy in the cosmological
case, these modes depend only upon the magnitude $k= |{\bf k}|$ which is the principal
quantum number of the spherical harmonic on ${\mathbb S}^3$. The time dependent
coefficients $\alpha_k(u)$ and $\beta_k(u)$ of the Bogoliubov transformation are defined by
\bes
\bea
f_k &=& \alpha_k\tilde f_k + \beta_k\tilde f_k^*\\
H\frac{d}{du} f_k &=& \left( -iW_k+ \frac{V_k}{2} \right) \alpha_k \tilde f_k
+ \left( iW_k + \frac{V_k}{2} \right) \beta_k\tilde f_k^* \label{dbog}
\eea
\label{ffdot}\ees
where $f_k$ is an exact mode function solution of (\ref{oscmode}).  They are given again 
by (\ref{genalpbet}), {\it viz.}
\be
|\beta_{\bf k}(t)|^2 = \frac{1}{2W_{\bf k}} \left\vert\dot f_{\bf k} +
\left( iW_{\bf k} - \frac{V_{\bf k}}{2}\right) f_{\bf k} \right\vert^2
\label{betsqgrav}
\ee
and (\ref{alpbetnorm}) is satisfied, provided only that both $W_k$ and $V_k$ are arbitrary real functions of time.

The analog of (\ref{Bogadb}) is now
\be
\left( \begin{array}{c} \tilde a_{klm_l}(u) \\ \tilde a_{kl\,-m_l}^{\dag}(u) \end{array}\right) =
\left( \begin{array}{cc} \alpha_k(u)& \beta_k^*(u)\\
\beta_k(u) & \alpha_k^*(u)\end{array}\right)\
\left( \begin{array}{c} a_{klm_l}\\  a^{\dag}_{kl\, -m_l}\end{array}\right)
\label{BogadbdS}
\ee
when referred to any time independent basis $(a_{klm_l}, a^{\dag}_{kl\,-m_l})$ for the Hermitian
scalar field $\Phi$ (not necessarily the CTBD basis). The time dependent mean adiabatic particle
number in the mode $(klm_l)$ is independent of $(lm_l)$ for $O(4)$ invariant adiabatic states and
may be defined by the analog of (\ref{adbpart}) in de Sitter space to be
\be
{\cal N}_k(u) = \lag \tilde a^{\dag}_{klm_l}(u) \tilde a_{klm_l}(u) \rag = N_k + (1 + 2 N_k) |\beta_k(u)|^2
\label{adbpartdS}
\ee
where
\be
N_k \equiv \lag a^{\dag}_{klm_l}a_{klm_l} \rag
\label{initNdS}
\ee
is the number of particles at the initial time $u=u_0$, provided $|\beta_{\bf k}(u_0)|^2 = 0$ is
initialized to zero at the initial time $u=u_0$. Note that with this initialization, the exact mode function
solution of (\ref{oscmode}) satisfies the initial conditions
\be
f_{k\gamma}(u_0) = \frac{1}{\sqrt{2W_k}}\bigg\vert_{u=u_0}\,,\qquad
\dot f_{k\gamma}(u_0) = \frac{1}{\sqrt{2W_k}}\left(-iW_k + \frac{V_k}{2}\right)\bigg\vert_{u=u_0}
\label{initdS}
\ee
and hence is a certain linear combination (time independent Bogoliubov transformation) of the CTBD mode
function $F_{k\gamma}$ and its complex conjugate $F_{k\gamma}^*$. Correspondingly, the time independent
basis operators $a_{klm_l}, a_{klm_l}^{\dag}$ in Fock space are certain linear combinations of the
$a^{\ups}_{klm_l}, a^{\ups\,\dag}_{klm_l}$ operators that define the de Sitter invariant state (\ref{dS}),
which can be expressed in terms of each other by time independent Bogoliubov
coefficients dependent upon the initial data (\ref{initdS}).

\begin{figure}[htp]
\includegraphics[height=9cm,viewport= 0 12 260 238 ,clip]{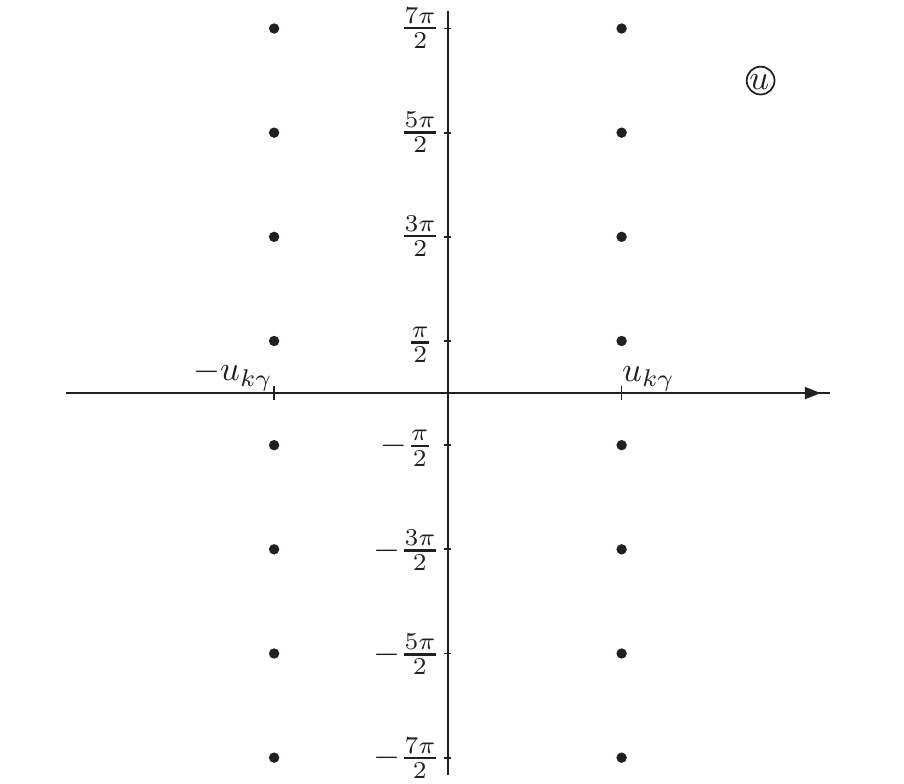}
\vspace{-2mm}
\caption{Location of the zeroes (\ref{ucomplex}) of the frequency function $\Omega_k$ (\ref{defOmega}) in the
complex $u$ plane for $\gamma^2 > 0$. Particle creation occurs as the real time $u$ contour passes through
the lines of these zeroes at  $u = \mp u_{k\gamma}$.}
\vspace{-3mm}
\label{Fig:ZeroesdS}
\end{figure}

As in the electric field example, the behavior of the solutions of the mode equation (\ref{oscmode}) can
be analyzed by general WKB methods in the complex $u$ plane. The zeroes of the frequency function,
$\Omega_k = 0$ in (\ref{OmegdS}) occur at the complex values $\cosh \bar u = \pm i\, \gamma^{-1}\sqrt{k^2 - 1/4}$,
or
\bes
\bea
&&\hspace{7mm} \bar u = u_R + i u_I= \pm u_{k\gamma} + i\pi \left(n + \tfrac{1}{2}\right) \quad {\rm with}
\quad n \in \mathbb{Z}\quad {\rm and}\\
&&\sinh u_{k\gamma} = \frac{\sqrt{k^2 - \frac{1}{4}}}{\gamma}\,, \qquad
u_{k\gamma} = \ln \left[ \frac{\sqrt{k^2 - \frac{1}{4}} + \sqrt{\gamma^2 + k^2 - \frac{1}{4}}}{\gamma}\right] \;.
\label{ukgam}\eea
\label{ucomplex}\ees
Thus there is an infinite line of zeroes of $\Omega_k$ in the complex $u$ plane
along the two vertical axes at $u= \pm u_{k\gamma}$ for $\gamma^2 >0$, {\it c.f.} Fig. \ref{Fig:ZeroesdS}.
The largest effect on the Bogoliubov coefficient $\beta_k (u)$ will occur when the real time contour
passes closest to these lines of complex turning points at $u= \pm u_{k\gamma}$. Hence there
are two `creation events' in global de Sitter space, one in the contracting and one in
the expanding phase symmetric around $u=0$.

We may consider the two limits
\bes
\bea
&& \hspace{1cm} u_{k\gamma} \rightarrow \frac{\sqrt{k^2 - \frac{1}{4}}}{\gamma} \rightarrow 0\quad {\rm for} \quad \gamma \gg k\,;
\qquad {\rm or}\label{gammalarge}\\
&& u_{k\gamma} \rightarrow \ln \left( \frac{2\sqrt{k^2 - \frac{1}{4}}}{\gamma}\right) \rightarrow
\ln \left( \frac{2k}{\gamma}\right)\rightarrow \infty
\quad {\rm for} \quad k \gg \gamma\,. \label{klarge}
\eea
\label{upmlimits}\ees
The first limit (\ref{gammalarge}) is the non-relativistic limit of very heavy particles whose rest mass is much larger than their physical
momentum $k/a$ at all times. These non-relativistic particles are created nearly at rest close to the symmetric point $u=0$
between the contracting and expanding de Sitter phases, so that the two events merge into one. The second limit (\ref{klarge})
is the relativistic limit of particles whose physical momentum is much larger than their rest mass for most of their history. These
particles are created in two bursts, at $u =\mp u_{k\gamma}$, when their physical momentum $k H\,\sech\, u_{k\gamma}$
is of the same order as their rest mass, so that they are moderately relativistic at creation. In the contracting phase
of de Sitter space $u<0$ these particles, created around $u=-u_{k\gamma}$, are blueshifted exponentially rapidly in $u$,
and thus become ultrarelativistic. This contracting phase with the created particles becoming ultrarelativistic
is therefore most analogous to the previous electric field example, and is the phase where we can expect the largest
backreaction effects. Conversely, in the expanding phase, $u >0$, the particles created around $u=+u_{k\gamma}$
will be subsequently exponentially redshifted in $u$, and therefore have a much smaller backreaction effect.
We emphasize that the time $\pm u_{k\gamma}$ is of the order of the horizon crossing of the mode
at $u \sim \pm \ln (2k)$ {\it only} for $\gamma \sim 1$. For large $\gamma \gg k$ the particle creation events occur
when the wavelength of the mode is much smaller than the horizon, while for  $\gamma \rightarrow 0$ the particle
creation events occur when the wavelength of the mode is much greater than the horizon.
Due to the different disposition of zeroes of the adiabatic frequency in the electric field and de Sitter cases,
{\it c.f.} Figs. \ref{Fig:EZeroes} and \ref{Fig:ZeroesdS}, there is no analog of this second burst of particle creation
in the electric field case.

For moderate values of $\gamma$, most of the $k$ modes fall into the second case (\ref{klarge}), and
experience two well-separated creation events at large $u_{k\gamma} \gg 1$ in both the contracting and expanding
phases of de Sitter space.  In contrast to the electric field case considered previously we may therefore distinguish
three distinct regions:
\bes
\bea
&& I :\qquad  u < - u_{k\gamma}\,,\qquad\qquad\quad in \\
&& II: \quad -u_{k\gamma} < u < u_{k\gamma}\,,\qquad CTBD\\
&& III: \quad u_{k\gamma} < u\,,\qquad\qquad\quad\ \ out
\eea
\label{regions}\ees
where we have indicated the character of the adiabatic vacuum in each region.
If one takes the infinite time limits $u \rightarrow \mp \infty$ with $k$ and $\gamma$ and hence
$u_{k\gamma}$ fixed, one is automatically in the first {\it in} region or the third
{\it out} region respectively. This corresponds to the {\it in/out} scattering problem considered
in Sec. \ref{Sec:Scat}. If on the other hand one takes the $k \rightarrow \infty$ limit for fixed $u, \gamma$
 then Eq.\ (\ref{klarge}) shows that one is always in region II, where the CTBD state is the adiabatic vacuum.
This shows explicitly the non-commutivity of the infinite $u$ and infinite $k$ limits, with the
transition between the two limits occurring at $u = \pm u_{k\gamma}$.

Next we consider the mode function(s) and adiabatic vacuum state specified by the initial
values (\ref{initdS}) at an arbitrary finite time $u_0 < 0$. The modes for a given value of $k$ fall into 
two possible classes:
\bes
\bea
(i)&&\  -u_{k\gamma} < u_0 < 0 \\
(ii)&&\  u_0 < -u_{k\gamma} <0 \,.
\eea
\ees
For modes in the first class {\it (i)} the initial time $u_0$ is already later than the first creation event.
For these modes in region II, the adiabatic initial condition is close to the CTBD state in the high $k$
limit, $\beta_k \approx 0$ and nothing further happens in the contracting phase, as they
remain in region II for all $u_0 < u \le 0$. In contrast, the $k$ modes in class {\it (ii)} are approximately
in the {\it in} vacuum state initially. These modes have yet to go through
their particle creation event which occurs at the later time $u=  -u_{k\gamma} > u_0$ in the
contracting phase. At that time, the adiabatic particle vacuum switches rapidly to
approximately the CTBD state as $u$ increases past $-u_{k\gamma}$. Thus this mode sees
its time dependent Bogoliubov coefficient change rapidly in a few expansion times
($\Delta u \sim 1$ since the imaginary part of the nearest complex zero of $\Omega_k$ is $\pi/2$
and independent of $k, \gamma$) from approximately zero to a  non-zero plateau determined
by the Bogoliubov coefficient  (\ref{Boutvalue}). Approximating the jump in particle number
at these creation events by a step function as before, we have
\bes
\bea
&& \Delta |\beta_k|^2 = |B^{in}_{k\gamma}|^2 = \frac{e^{-\pi \gamma}}{2 \sinh (\pi \gamma)}
= \frac{1}{e^{2\pi\gamma} - 1}  \label{delbeta1} \\
&&{\cal N}_k (u) \approx
\theta (u + u_{k\gamma})\ \theta(-u_{k\gamma} - u_0)\,\Delta {\cal N}_{1,k\gamma}
\quad {\rm for} \quad u <0\,,\quad u_0<0  \\
&&\quad \Delta {\cal N}_ {1,k\gamma}=  (1+ 2N_k)\, \Delta|\beta_k|^2 = \frac{1+ 2N_k}{e^{2\pi\gamma} - 1}
\label{DelN1}
\eea
\label{Ncontract}\ees
in the contracting phase. The first $\theta$ function in (\ref{Ncontract}) specifies the time of
the creation event when the step occurs, while the second $\theta$ function restricts the
modes to class {\it (ii)} for which the step occurs at a later time $u =  -u_{k\gamma}> u_0$
in the contracting phase. These two $\theta$ functions give the `window function'
which is similar to that found in the electric field case (\ref{Ewin}), namely
\be
K_{\gamma}(u) \equiv \sqrt{\gamma^2 \sinh^2 u + \tfrac{1}{4}} < k < \sqrt{\gamma^2 \sinh^2 u_0 + \tfrac{1}{4}}
= K_{\gamma}(u_0)
\label{windS}
\ee
in the contracting phase of de Sitter space for which $u_0 < u \le 0$. Like (\ref{Ewin}) this window function
has an upper limit fixed by the initial time and a lower limit which decreases as time evolves (for $u<0$).

If we continue the evolution past the symmetric point $u=0$ into the expanding de Sitter phase,
all of the modes of class {\it (ii)} have experienced the first particle creation event, and
then begin (with the smallest value of $k$ first) to experience a second creation event at
$u=+u_{k\gamma}$. Thus the modes of class {\it (ii)} which started in region I undergo
two creation events with a total Bogoliubov transformation of (\ref{ABtotval}),
while the modes of class {\it (i)} which started in region II undergo only the second creation
event in the expanding phase for which the single Bogoliubov transformation $B_{k\gamma}^{out}$
applies. Again approximating these creation events by step functions we obtain
\bes
\bea
&&{\cal N}_k (u) \approx \left[\theta( u_{k\gamma}-u) \, \theta(-u_{k\gamma} - u_0)
+ \theta (u - u_{k\gamma}) \theta(u_0 + u_{k\gamma}) \right]\,\Delta {\cal N}_{1,k\gamma}\nn
&& \qquad +\,  \theta (u - u_{k\gamma})\,\theta(-u_{k\gamma} - u_0)\, \Delta {\cal N}_{2,k\gamma}
\quad {\rm for}\quad u>0\,,\quad u_0 <0\label{Nexpand}\\
&&\hspace{1.5cm} \Delta {\cal N}_{2,k\gamma} = (1+ 2N_k) \,|B_{k\gamma}^{tot}|^2
= (1+ 2N_k)\,{\rm csch}^2(\pi\gamma)\label{DelN2}
\eea
\label{Nexpand}\ees
in the expanding phase of de Sitter space. The window function for this
second creation event in the expanding phase is now
\be
k <  \sqrt{\gamma^2 \sinh^2 u + \frac{1}{4}} = K_{\gamma}(u)
\label{windSexp}
\ee
for the modes undergoing the second creation event at $u=+ u_{k\gamma}$. Those with
$k < K_{\gamma}(u_0)$ undergo both the first and second creation events with
$\Delta{\cal N} =\Delta {\cal N}_{2,k\gamma}$, while those with $k>  K_{\gamma}(u_0)$ 
experience only the second creation event with $\Delta {\cal N}= \Delta {\cal N}_{1,k\gamma}$.

This analysis may be repeated if the initial time $u_0 > 0$ is in the expanding phase. In this case all
modes initially in region II, with $u_0 < u_{k\gamma}$ undergo a single creation event at $u=+u_{k\gamma}$.
Hence we have
\be
{\cal N}_k (\tau) \approx \theta (u - u_{k\gamma})\, \theta(u_{k\gamma} - u_0)\, \Delta {\cal N}_{1,k\gamma}
 \quad {\rm for}\quad  u, \ u_0>0
\label{u0pos}
\ee
replacing (\ref{Nexpand}). The window function in $k$ is now the reverse of (\ref{windS}), namely
\be
K_{\gamma}(u_0) < k < K_{\gamma}(u)
\label{windSpos}
\ee
which like (\ref{windSexp}) shows an upper limit that increases with time.

The various cases (\ref{Ncontract}), (\ref{Nexpand}) and  (\ref{u0pos}) may be collected into one result
\bea
&&{\cal N}_k (\tau) \approx \Big[\theta (u_{k\gamma}- |u|)\ \theta(-u_{k\gamma} - u_0)
+ \theta (u - u_{k\gamma})\ \theta(u_{k\gamma} - |u_0|)\Big]\, \left(\frac{1+ 2N_k}{e^{2\pi\gamma} - 1}\right)\nn
&& \hspace{2cm}  +\  \theta (u - u_{k\gamma})\,\theta(-u_{k\gamma} - u_0)\, (1+ 2N_k)\,{\rm csch}^2(\pi\gamma)
\label{totN}
\eea
valid for all values of $u$ and initial times $u_0$. From this or (\ref{Nexpand}) it is clear that for fixed $k$,
with $u_0 \rightarrow - \infty, u \rightarrow + \infty$, the mode experiences both particle creation
events and we recover (\ref {Boutpart}), while for a finite interval of time only those modes for which
$u_0 < -u_{k\gamma}$ and $u_{k\gamma} < u$ experience both creation events. Thus taking
the symmetric limit with $u = -u_0 >0$, the values of $k$ satisfying both these conditions are cut off at
the maximum value $K_{\gamma}(u)$, {\it i.e.}
\be
k \le K_{\gamma} (u) \rightarrow \frac{\gamma}{2} \, e^{|u|}\qquad {\rm or} \qquad
\ln K_{\gamma}  (u) \rightarrow |u| + \ln \left(\frac{\gamma}{2}\right)
\label{Kcutoff}
\ee
as $|u| \rightarrow \infty$, which is exactly the cutoff (\ref{lnKu}) that we argued on physical grounds earlier
in Sec. \ref{Sec:Scat} (and in Ref. \cite{PartCreatdS}) should be used in the $k$ sum of (\ref{vacprob})
to calculate the finite decay rate per unit volume (\ref{Decayrate}) of de Sitter space to massive particle creation
in the limit $V_4 \rightarrow \infty$. The constant in (\ref{lnKu}) has been determined to be $\ln(\gamma/2)$ by
our detailed analysis of the particle creation process in real time. The non-Hadamard short distance behavior
of the {\it in} and {\it out} states found in \cite{PartCreatdS} has also been removed by regulating the large $k$
behavior with a finite initial and final time, since the modes for which $k > K_{\gamma}(u)$ remain in the CTBD
state in region II for all $-|u_0| < u < |u_0|$ and the CTBD state is known to have the correct short distance
behavior \cite{BunDav}.

\begin{figure}[htp]
\vspace{-5mm}
  \centering
  \includegraphics[angle=90,height=9cm,viewport= 60 0 560 720, clip]{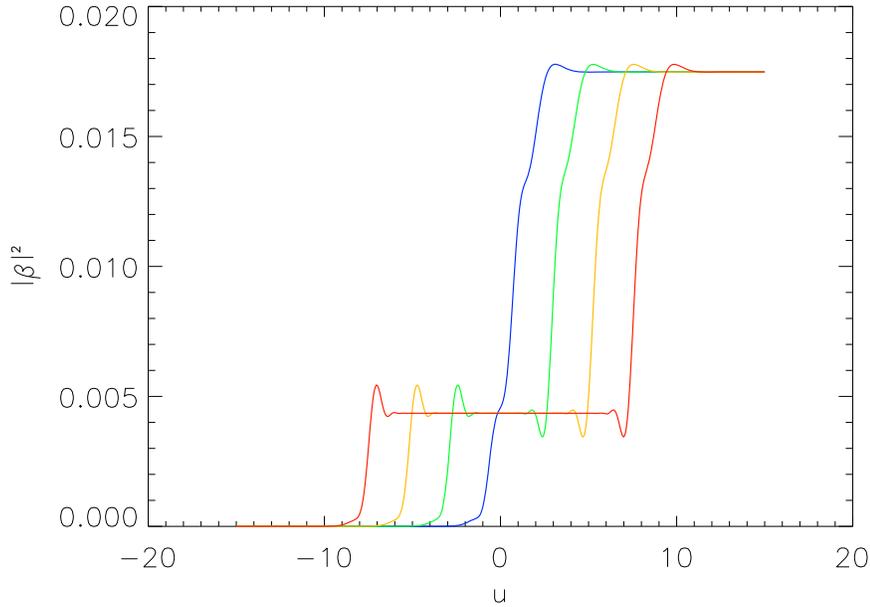}
  \vspace{-7mm}
  \caption{Pllotted is $|\beta_k(u)|^2$ for $m = H$ defined by (\ref{betsqgrav}) with the initial
  adiabatic matching time $u_0 = -15$ and the second order matching defined by (\ref{Wk2Vk1}). The innermost blue
  curve is for $k = 1$, the green for $k = 10$,  the yellow for $k = 100$ and the outermost red for $k = 1000$, the latter $3$
  values showing two clearly separated particle creation events. The values of $u_{k\gamma}$  given by (\ref{ucomplex}) 
  are $0.35$, $2.31$, $5.44$, and $10.0$ respectively for these values of $k$ and $\gamma$. The asymptotic value of 
  $|\beta_k|^2$ of all the curves for large $u$ is $0.01748$ in agreement with (\ref{DelN2}) for $N_k =0$. 
  The intermediate plateau is at $0.00435$ in agreement with (\ref{DelN1}).}
  \vspace{-3mm}
  \label{Fig:bui15}
\end{figure}

\begin{figure}[hbp]
  \centering
  \includegraphics[angle=90,height=9cm,viewport= 60 0 560 720, clip]
  {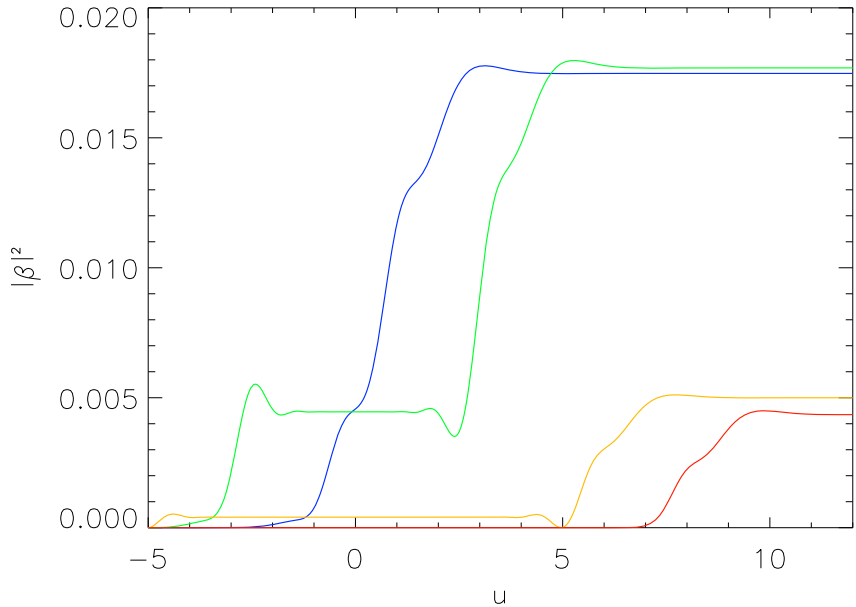}
  \vspace{-7mm}
  \caption{Plotted is $|\beta_k(u)|^2$ for the same values of $\gamma$ and $k$ as in Fig. \ref{Fig:bui5}, but with the initial
  adiabatic matching time $u_0 = -5$.  Note that for the two highest values of $k$ at $100$ and $1000$ (the lower yellow
  and red curves), a marked first particle creation event does not occur since $u_0 > -u_{k\gamma}$ for these modes.
  The asymptotic value of the lower red curve at large $u$ is $0.00435$ in agreement with the first term of (\ref{totN}).
  The yellow curve for $k=100$ has a small contribution from the first creation event since $u_0$ and $-u_{k\gamma}$
  are comparable.}
    \label{Fig:bui5}
\end{figure}

\begin{figure}[htp]
  \centering
  \includegraphics[angle=90,height=9cm,viewport= 60 0 560 720, clip]
  {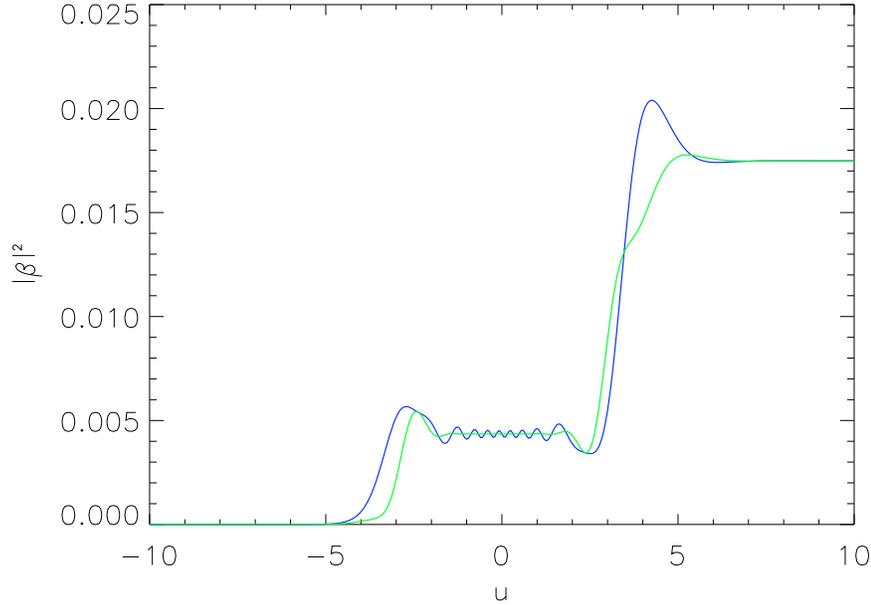}
  \caption{Plotted is $|\beta_k(u)|^2$ for $k = 10$ when the matching time is $u_0= -15$.  The blue curve corresponds to
  the zeroth order adiabatic vacuum state specified by $W_k = \Omega_k$ with $V_k$ given by \eqref{WkVksimple}.
  The green curve corresponds to the second order adiabatic vacuum state specified by \eqref{Wk2Vk1}.}
  \label{Fig:wkbcompare}
\end{figure}

The actual smooth behaviors of $|\beta_k(u)|^2$ defined by (\ref{betsqgrav}) for various $k$ and $u_0=-15$ and $u_0=-5$
are shown in Figs. \ref{Fig:bui15} and \ref{Fig:bui5} respectively. The increases in $|\beta_k(u)|^2$ occur on a
time scale $\Delta u \sim 1$ for all the modes. The values chosen for the adiabatic frequency functions
$(W_k, V_k)$ are
\bes
\bea
&&W_k^{(2)} = \Omega_k + \frac{3}{8} \frac{\dot \omega_k^2}{\omega_k^3} -\frac{1}{4} \frac{\ddot \omega_k}{\omega_k^2} \nn
&& \qquad = \Omega_k + \frac{h^2}{8\,\omega_k}\left(1 - \frac{6m^2}{\omega_k^2} + \frac{5m^4}{\omega_k^4}\right)
+ \frac{\dot h}{4\, \omega_k} \left(1 - \frac{m^2}{\omega_k^2}\right)\,,\\
&&V_k^{(1)}= - \frac{\dot \omega_k}{\omega_k}= h  \left(1 - \frac{m^2}{\omega_k^2}\right)
\eea
\label{Wk2Vk1}\ees
correct up to second order in the adiabatic expansion. A comparison of $|\beta_k(u)|^2$ for this choice and the simpler choice
\bes
\bea
&&W_k^{(0)} = \Omega_k =  H\left[ \left(k^2 - \tfrac{1}{4}\right) \sech^2 u + \gamma^2\right]^{\frac{1}{2}}\\
\vspace{2mm}
&&V_k^{(1)}= -\frac{\dot \omega_k}{\omega_k} = h\left(1 - \frac{m^2}{\omega_k^2}\right)
\eea
\label{WkVksimple}\ees
for the $k=10$ mode and $u_0 =-15$ is shown in Fig. \ref{Fig:wkbcompare}.

\begin{figure}[hbp]
  \centering
  \vspace{-9mm}
  \includegraphics[angle=90,height=9cm,viewport= 60 0 560 720, clip]
  {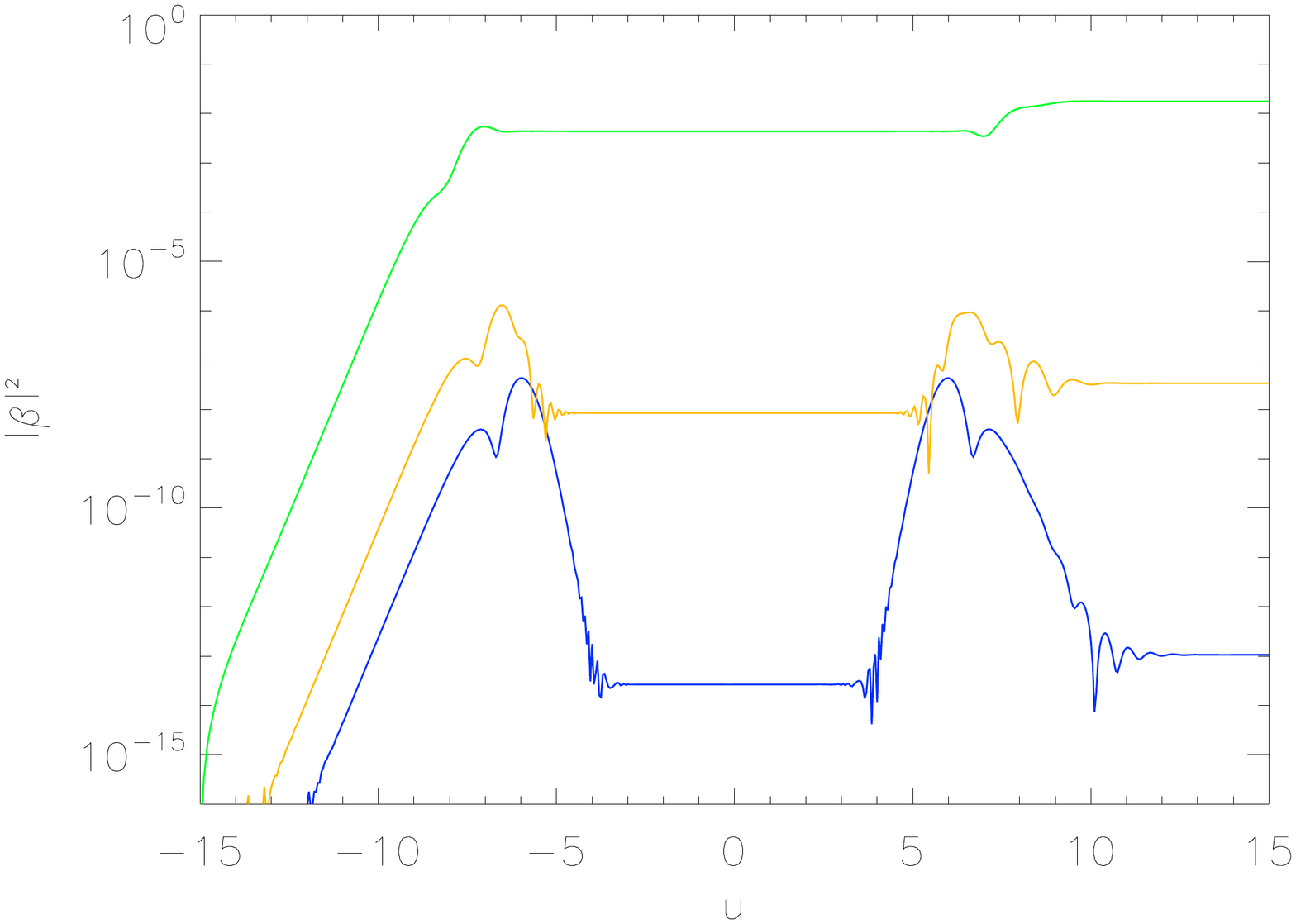}
  \vspace{-5mm}
  \caption{Plotted is $|\beta_k(u)|^2$ for fixed $k = 1000$ and adiabatic matching time $u_0= -15$, for three values of the mass:
  $m=H$ (upper, green),  $m=3H$ (middle, yellow),  $m=5H$ (lower, blue). Note the logarithmic scale.
  The asymptotic values of $|\beta_k|^2$ for large $u$ are $1.748 \times 10^{-2}, 3.391 \times 10^{-8}$ and $1.062 \times 10^{-13}$
  respectively, in agreement with (\ref{DelN2}) for $N_k =0$. The intermediate plateaux at $u=0$ are at $4.35 \times 10^{-3}, 8.48 \times 10^{-9}$,
  and $2.66 \times 10^{-14}$ respectively, in agreement with (\ref{DelN1}). The particle creation events occur at $\mp u_{k\gamma}$
  with $u_{k\gamma} = 7.74, 6.52$ and $6.00$ respectively for the three values of $m$.}
  \label{Fig:mass123}
\end{figure}

As in the electric field case ({\it c.f.} Fig. \ref{Fig:EW0W2}) the detailed time structure of the creation event is different
with different choices of $(W_k, V_k)$, but the qualitative features and asymptotic values (and intermediate plateau value)
are independent of the choice. The second order WKB choice (\ref{Wk2Vk1}) suppresses the oscillations observed with
the choice (\ref{WkVksimple}) and comes closer to the approximate step function description. As predicted by the previous
WKB analysis and (\ref{totN}), the modes with $k = 1$ and $k =10$ in Fig. \ref{Fig:bui5} go through both creation events
with a rapid increase in $|\beta_k(u)|^2$ occurring for each at the appropriate value of $\mp u_{k\gamma}$. The modes with 
$k = 100$ and $k = 1000$ for which $u_{k\gamma} > |u_0|$ only go through a marked second creation event, although the yellow
curve for $k=100$ has a small contribution from the first creation event, since $u_0= -5$ and  $-u_{k\gamma} = -5.44$
are comparable for this mode. The value of $|\beta_k(u)|^2$ after the first creation event is well approximated by (\ref{delbeta1})
or (\ref{DelN1}) with $N_k =0$ for an initial vacuum, while for those modes undergoing two creation events the second
plateau of $|\beta_k(u)|^2$ for $u>u_{k\gamma}$ is given by (\ref{DelN2}).  In Fig. \ref{Fig:mass123} we also compare
$|\beta_k(u)|^2$ for fixed $k=1000$ and $u_0=-15$, for three different values of the mass, showing the dependence
of the time of the creation events on $\gamma$ given by (\ref{ukgam}).

\begin{figure}[htp]
  \centering
  \vspace{-5mm}
  \includegraphics[angle=90,height=9cm,viewport= 60 0 560 720, clip]
  {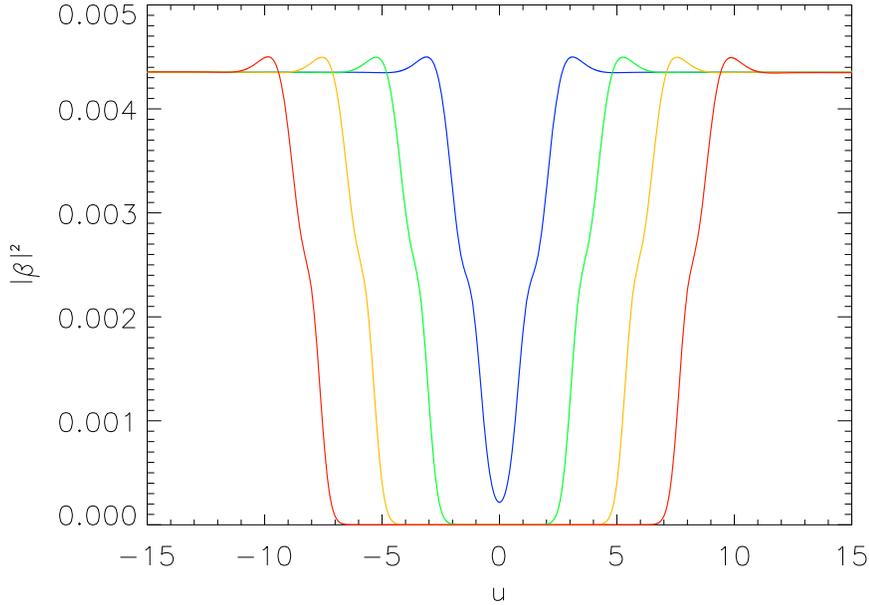}
  \vspace{-5mm}
  \caption{Plotted is $|\beta_k(u)|^2$ for the CTBD state for various values of $k$: $k = 1$ (blue), $k=10$ (green).
  $k = 100$ (orange), $k = 1000$ (red). Note that the first event at $u=-u_{k\gamma}$ is a particle {\it annihilation}
  event in which the particle number decreases from $0.004352$ to zero in each $k$ mode,
  rising again to the same value at $u=+u_{k\gamma}$ in a completely time symmetric manner.}
  \vspace{-7mm}
  \label{Fig:CTBD}
\end{figure}

For comparison we also plot the adiabatic particle number as a function of $u$ for the CTBD state in Fig. \ref{Fig:CTBD}.
This figure shows that the CTBD state contains particles in its initial condition and the first event at $u=-u_{k\gamma}$ is
actually a particle {\it destruction} event. The phase coherent initial particles in modes with principal quantum number
$k$ find each other and annihilate at the time $u=-u_{k\gamma}$, canceling each other precisely in region II. At the
later time $u=u_{k\gamma}$ these particles are created again in a completely time symmetric manner. This is 
clearly a delicately balanced coherent process that is artificially arranged by initial conditions in the CTBD state. 
In an accompanying paper we show that a small perturbation of the CTBD state upsets this balance and leads
again to instability \cite{AndMotDSVacua}.

\section{Particle Creation in Spatially Flat FLRW Poincar\'e Coordinates}
\label{Sec:Flat}

The analysis of particle creation in the spatially closed ${\mathbb S}^3$ coordinates of de Sitter space
of the previous section can just as well be carried out in the spatially flat Poincar\'e coordinates
of (\ref{FLRW}), more commonly used in cosmology. The wave equation (\ref{waveq}) again separates
in the usual Fourier basis $\Phi \sim \phi_k(\tau) e^{i {\bf k \cdot x}}$. Removing the scale
factor by defining the mode function $f_k = a^{\frac{3}{2}} \phi_k$ as in (\ref{fdef})
but with $a = \exp(H\tau)$ in this case gives the mode equation
\be
\left[\frac{d^2}{d\tau^2} + k^2 e^{-2H\tau} + m^2 - \frac{H^2}{4\,}\right] f_k(\tau) = 0
\label{oscflat}
\ee
with $k \equiv |\bf k|$. This equation again has the form of an harmonic oscillator equation with a time dependent
frequency which is given by
\be
\omega_k^2 (\tau) = k^2 e^{-2H\tau} + m^2 - \frac{H^2}{4\,} \;.
\label{omflat}
\ee
Thus with this change all of the methods employed in the spatially closed sections or the electric field background 
may be utilized again. In particular, for $\gamma^2 >0$ we have over the barrier scattering in a non-trivial one 
dimensional potential, and we should expect the stationary waves incident from the left as $\tau \rightarrow -\infty$
to be partially reflected and partially transmitted to the right as $\tau \rightarrow \infty$. This
scattering will result again in a non-trivial Bogoliubov transformation between the positive
frequency particle solutions at early times in the {\it in} vacuum and those at late times in the
{\it out} vacuum, {\it i.e.} particle creation, just as in the electric field case.

By making the change of variables
\be
z \equiv \frac{k}{H} \,e^{-H\tau}
\label{defz}
\ee
Eq.\,(\ref{oscflat}) may be transformed into Bessel's equation with imaginary index $\nu = \pm i \gamma$. Thus
the exact solutions are Bessel or Hankel functions with this index. The particular solution
\be
\bar \ups_{\gamma} (z) \equiv \frac{1}{2} \sqrt{\frac{\pi}{H}}\, e^{- \frac{\pi \gamma}{2}} e^{\frac{i \pi}{4}}
H^{(1)}_{i \gamma} (z) =  \sqrt{\frac{\pi}{H}} \,\frac{e^{\frac{\pi \gamma}{2}}e^{\frac{i \pi}{4}}}{e^{2 \pi \gamma} - 1}
\,\Big[ e^{\pi \gamma} J_{i \gamma}(z) - J_{-i\gamma}(z)\Big]
\label{BDflat}
\ee
is the CTBD solution in flat coordinates, with the asymptotic behavior
\be
\bar \ups_{\gamma} (z) \rightarrow \frac{1}{\sqrt{2Hz}}\, e^{iz}\qquad {\rm as}\qquad z \rightarrow \infty\,.
\label{barupslarz}
\ee
Since $\omega_k = H \sqrt{z^2 + \gamma^2} \rightarrow H z$ and
\bea
\Theta_{\gamma}(z) &\equiv &\int^{\tau(z)}\, d\tau \, \omega_k (\tau) = - \int^z \frac{dz}{z} \sqrt{z^2 + \gamma^2}\nn
&=& - \sqrt{z^2 + \gamma^2} - \frac{\gamma}{2} \ln \left[ \frac{\sqrt{z^2 + \gamma^2} - \gamma}{\sqrt{z^2 + \gamma^2} + \gamma}\right]
\rightarrow  -z  \qquad {\rm as} \qquad  z\rightarrow \infty
\eea
the solution (\ref{BDflat}) is also the correctly normalized adiabatic {\it in} vacuum solution
\be
f_{k (+)} (\tau) = \bar \ups_{\gamma} (z) \rightarrow \frac{1}{\sqrt{2 \omega_k}} \, e^{-i \Theta_{\gamma}}
\qquad {\rm as} \qquad  \tau \rightarrow -\infty
\label{flatin}
\ee
in the Poincar\'e coordinates.

This much is standard and may be found in standard references \cite{BirDav}. However, in the opposite limit
of late times $\tau \rightarrow \infty$
\be
\omega_k \rightarrow H \gamma \qquad {\rm and}\qquad
\Theta_{\gamma}(z) \rightarrow -\gamma \ln z\qquad {\rm as} \qquad z\rightarrow 0
\label{z0lim}
\ee
and therefore
\be
f_k^{(+)} (\tau) = \left(\frac{2H}{k}\right)^{i\gamma} \frac{ \Gamma(1 + i \gamma)}{\sqrt{2H\gamma}}\,
J_{i \gamma}(z) \rightarrow \left(\frac{H}{k}\right)^{i\gamma}\frac{z^{i\gamma}}{\sqrt{2H\gamma}}
= \frac{1}{\sqrt{2 H \gamma}} \, e^{-i\gamma H \tau} \quad {\rm as} \quad \tau \rightarrow \infty
\ee
is the properly normalized positive frequency {\it out} solution, which agrees with the adiabatic form
$e^{-i \Theta_{\gamma}}/\sqrt{2 \omega_k}$ at late times. Comparison with the last form
of (\ref{BDflat}) shows that indeed there is a nontrivial mixing of positive and negative frequencies
at late times in the CTBD state.  The Bogoliubov coefficients are
\bes
\bea
&&A_{\gamma} =   \frac{\sqrt{2\pi\gamma}\ e^{\frac{i \pi}{4}}}{2^{i\gamma}\,\Gamma(1 + i \gamma)}
\ \frac{e^{ \frac{3\pi \gamma}{2}}}{e^{2 \pi \gamma} - 1} \\
&&B_{\gamma} = -\frac{\sqrt{2\pi\gamma}\ e^{\frac{i \pi}{4}}}{2^{i\gamma}\,\Gamma(1 + i \gamma)}
\ \frac{e^{\frac{\pi \gamma}{2}}}{e^{2 \pi \gamma} - 1}
\eea
\label{Bogflat}
\ees
with $|A_{\gamma}|^2 - |B_{\gamma}|^2 = 1$.  Note that
\be
|B_{\gamma}|^2 = \frac{1}{e^{2 \pi \gamma} - 1} = |B^{out}_{k\gamma}|^2 \;.
\label{Bsqflat}
\ee
Thus $B_k$ has exactly the same magnitude as the corresponding Bogoliubov coefficient 
Eq.\, (\ref{Boutvalue}) obtained previously in the closed ${\mathbb S}^3$ spatial sections.
This equality is to be expected since in the asymptotic future the closed spatial sections have
negligible spatial curvature and there is no local difference with the flat sections. This
mode mixing and particle creation effect in the flat Poincar\'e coordinates, which follows from
the second form of (\ref{BDflat}), seems to have been overlooked in \cite{BirDav}, which states
that ``there is no particle production."

The particle creation process may be analyzed in real time in the flat Poincar\'e coordinates
by using the methods of Sec. \ref{Sec:Adb}. Indeed the zeroes of (\ref{omflat}) in
the complex $z$ plane occur at
\be
z_{\gamma} = \pm i \gamma
\ee
which represents an infinite series of zeroes at $H\tau = \ln (k/H\gamma) + i \pi (n + \frac{1}{2})$
similar to those in the complex $u$ plane in Eq.\ (\ref{ucomplex}) and Fig. \ref{Fig:ZeroesdS}.
Since the Poincar\'e coordinates cover only one half of the de Sitter manifold, where it
is only expanding (or in the other half where it is only contracting), there is only one
line of complex zeroes in Poincar\'e coordinates and only one creation event occurring at
\be
\tau_{k\gamma} = \frac{1}{H} \ln \left(\frac{k}{H \gamma}\right)
\label{tauk}
\ee
for the mode with Fourier component $k= |{\bf k}|$. If one starts the evolution at some finite
initial time $\tau_0$ then only those modes with $k$ in the range determined by
\be
\tau_0 < \tau_{k\gamma} < \tau
\ee
will experience their creation event at a later time $\tau$. The number density of
particles in modes with $k= |{\bf k}|$ at time $\tau$ is therefore
\be
{\cal N}_k(\tau) \approx \theta (\tau_{k\gamma} - \tau_0)\, \theta(\tau - \tau_{k\gamma}) \,(1 + 2 N_k)
\,|B_{\gamma}|^2
\label{npartflat}
\ee
in the approximation that the particles are created instantaneously when $\tau$ passes through
$\tau_{k\gamma}$.

In the expanding phase of de Sitter space, whether described by closed ${\mathbb S}^3$
or flat ${\mathbb R}^3$ spatial sections, these particles will be redshifted in energy
and make a decreasing contribution to the energy density and pressure at later times.
In the next section we compute the energy density of the created particles
which grow exponentially in the contracting part of de Sitter space due to their blueshifting
toward the extreme ultrarelativistic limit. This does not occur in the Poincar\'e sections with
only monotonic expansion, {\it for spatially homogeneous states}. In \cite{Attract} we showed 
that the energy density and pressure relax to the values of the de Sitter invariant 
CTBD state for all such UV allowed spatially homogeneous states and for all $M^2 > 0$ 
in the expanding phase. Nevertheless the same particle creation  and vacuum instability 
effect (or more precisely one half of it) is present in the Poincar\'e coordinates as in the 
closed section coordinates of the full hyperboloid. Spatially inhomogeneous states have
a different behavior and are studied in \cite{AndMotDSVacua}.

\section{Stress-Energy Tensor of Created Particles}
\label{Sec:SET}

In this section we consider the stress-energy tensor of the created particles, and their ability to affect
the background de Sitter spacetime by backreaction. The energy-momentum tensor of the scalar field
with arbitrary curvature coupling $\xi$ is
\be
T_{ab} =(\nabla_a {\bf \Phi}) (\nabla_b {\bf \Phi}) - \frac{g_{ab}}{2}
 (\nabla^c {\bf \Phi}) (\nabla_c{\bf \Phi}) - \frac{m^2}{2} g_{ab} {\bf \Phi}^2
+  \xi \big[g_{ab} \sq - \nabla_a\nabla_b + G_{ab}\big]{\bf \Phi}^2
\label{SET}
\ee
where $G_{ab}$ is the Einstein tensor. Assuming a metric of the form (\ref{RW}) and spatial
homogeneity and isotropy of the state on the ${\mathbb S}^3$ sections, the only non-vanishing
components of the expectation value of $T_{ab}$ are the energy density $\varepsilon = \lag T_{\tau\tau}\rag$
and the isotropic pressure $p = \frac{1}{3} \lag T^i_{\ i}\rag$. The scalar field operator $\bf \Phi$ can be expressed
in terms of the exact mode function solutions of (\ref{oscmode}) such that
\be
{\bf \Phi} (u,\hat N) = a^{-\frac{3}{2}}\sum_{k=1}^{\infty}\sum_{l=0}^{k-1}\sum_{m_l=-l}^{l}
\left\{ a_{klm_l}\, f_k(u)\, Y_{klm_l} (\hat N) +  a_{klm_l}^{\dagger} f^*_k(u)
Y_{klm_l}^*\, (\hat N)\right\}  \;.
\label{Phiopf}
\ee
Assuming the mode functions satisfy the arbitrary initial conditions (\ref{initdS}), and specializing to conformal
coupling $\xi = \frac{1}{6}$, we find
\bes
\bea
\varepsilon\big\vert_{\xi = \frac{1}{6}} &=&
\frac{1}{4 \pi^2 a^3} \sum_{k=1}^{\infty}\, (1 + 2N_k) \left[ |\dot f_k|^2
- h\, {\rm Re}(f_k^* \dot f_k)
+ \left(\frac{k^2}{a^2} + m^2 + \frac{h^2}{4\,} \right) |f_k|^2 \right]\\
p\big\vert_{\xi = \frac{1}{6}}  &=& \frac{1}{12 \pi^2 a^3} \sum_{k=1}^{\infty}\, (1 + 2N_k)\left[ |\dot f_k|^2
- h\, {\rm Re}(f_k^* \dot f_k)
+ \left(\frac{k^2}{a^2} - m^2 + \frac{h^2}{4\,} \right) |f_k|^2 \right]\,.
\eea
\label{epmode}\ees
The exact mode functions $f_k$ and their time derivatives can be expressed in terms of
the adiabatic functions $\tilde f_k$ and the time dependent Bogoliubov coefficients
$(\alpha_k, \beta_k)$ by (\ref{genalpbet}) and (\ref{ffdot}). Thus (\ref{epmode}) may
be expressed in the general adiabatic basis as
\bes
\bea
\varepsilon\big\vert_{\xi = \frac{1}{6}} &=&  \frac{1}{2 \pi^2 a^3} \sum_{k=1}^{\infty} k^2
\Big[\varepsilon^{\cal N}_{k} \left({\cal N}_k + \tfrac{1}{2}\right)
+ \varepsilon^{\cal R}_{k} {\cal R}_k  +  \varepsilon^{\cal I}_{k} {\cal I}_k \Big] \\
p\big\vert_{\xi = \frac{1}{6}} &=& \frac {1}{2 \pi^2 a^3}\sum_{k=1}^{\infty} k^2
\Big[p^{\cal N}_{k} \left({\cal N}_k + \tfrac{1}{2}\right)
+ p^{\cal R}_{k} {\cal R}_k  +  p^{\cal I}_{k} {\cal I}_k\Big]
\eea
\label{epdecomp}
\ees
where the three terms labelled by $\cal N$, $\cal R$, and $\cal I$ are
\bes
\bea
\varepsilon^{\cal N}_k\Big\vert_{\xi =\frac{1}{6}}&=& \frac{1}{2W_k}
\left[\omega^2_k + W_k^2 + \frac{(V_k - h)^2}{4}  \right]\\
\varepsilon^{\cal R}_k\Big\vert_{\xi =\frac{1}{6}}&=& \frac{1}{2W_k}
\left[\omega^2_k - W_k^2 + \frac{(V_k - h)^2}{4} \right] \\
\varepsilon^{\cal I}_k\Big\vert_{\xi =\frac{1}{6}}&=& \frac{V_k - h}{2}
\eea
\label{epsNRI}
\ees
in the energy density, and
\bes
\bea
p^{\cal N}_k\Big\vert_{\xi =\frac{1}{6}}&=& \frac{1}{6W_k} \left[\omega^2_k -2m^2 + W^2_k
+ \frac{(V_k - h)^2}{4} \right]\\
p^{\cal R}_k\Big\vert_{\xi =\frac{1}{6}}&=& \frac{1}{6W_k}
\left[ \omega^2_k -2m^2 - W^2_k + \frac{(V_k - h)^2}{4}  \right] \\
p^{\cal I}_k\Big\vert_{\xi =\frac{1}{6}}&=& \frac{V_k - h}{6}\,,
\eea
\label{pNRI}\ees
in the pressure, with ${\cal N}_k$ given by (\ref{adbpartdS}), and ${\cal R}_k, {\cal I}_k$ given by
\bes
\bea
{\cal R}_k &=& (1 + 2N_k)\,{\rm Re} (\alpha_k\beta_k^* e^{-2i\Theta_k})\\
{\cal I}_k&=& (1 + 2N_k)\,{\rm Im} (\alpha_k\beta_k^* e^{-2i\Theta_k})
\eea
\label{calRI}\ees
in terms of the adiabatic phase
\be
\Theta_k \equiv  \int_{\tau_0}^{\tau}d\tau\,W_k\,.
\label{adbTheta}
\ee
The ${\cal N}_k$ terms have a quasi-classical interpretation as the energy density and pressure
of particles with single particle energies $\varepsilon_k^{\cal N}$. The $\frac{1}{2}$
in these terms has the natural interpretation of the quantum zero point energy in the adiabatic
vacuum specified by $(W_k, V_k)$. The ${\cal R}_k$ and ${\cal I}_k$ terms are oscillatory quantum
interference terms that have no classical analog, analogous to the last term of (\ref{qcurrent}).

The mode sums over $k$ in (\ref{epdecomp}) are generally quartically divergent in four dimensions.
It is in handling and removing these divergent contributions in the mode sums that the adiabatic method
is most useful \cite{Parker,ParFul,BirBun,BirDav,AndPark,EMomTen,AndEak}.  Although a fourth
order adiabatic subtraction is needed in general, when $\xi = \frac{1}{6}$ it is sufficient
to subtract only the second order adiabatic expressions
\bes
\bea
&&\varepsilon^{(2)} = \frac{1}{4 \pi^2 a^3} \sum_{k=1}^{\infty}k^2 \varepsilon_k^{(2)}\,\\
&&p^{(2)} = \frac{1}{4 \pi^2 a^3} \sum_{k=1}^{\infty}k^2 p_k^{(2)}
\eea
\label{sumadb2}\ees
with
\bes
\bea
&&\varepsilon_k^{(2)}\Big\vert_{\xi =\frac{1}{6}} =
\omega_k + \frac{h^2 m^4}{8 \omega_k^5} \\
&&
p_k^{(2)}\Big\vert_{\xi =\frac{1}{6}} = \frac{1}{3}\left[ \omega_k - \frac{m^2}{\omega_k}
- \frac{m^4}{8 \omega_k^5}\, (2\dot h + 5 h^2)
+ \frac{5 m^6 h^2}{ 8 \omega_k^7} \right]
\eea
\label{adb2}\ees
to arrive at a finite, renormalized and conserved stress tensor. The reason for this is that
it may be shown that the only possible remaining divergence is logarithmic and proportional
to $(\xi - \frac{1}{6})^2$, and correspondingly there are no $\omega_k^{-3}$ terms in either of
the expressions (\ref{adb2}) for conformal coupling $\xi = \frac{1}{6}$. Moreover the logarithmic
divergence is proportional to the tensor $^{(1)}H_{ab}$ \cite{BirDav,EMomTen} which vanishes
in de Sitter space (similar to the vanishing of the counterterm proportionl to $\ddot E$ when $E$
is a constant).

The difference of the $\frac{1}{2}$ vacuum-like $\cal N$ terms in (\ref{epdecomp})
and the subtraction terms are
\bes
\bea
&&\varepsilon_{vac}= \frac{1}{4 \pi^2 a^3} \sum_{k=1}^{\infty} k^2 \left(\varepsilon^{\cal N}_{k}- \varepsilon_k^{(2)}\right)
\label{epvac}\\
&&p_{vac} = \frac{1}{4 \pi^2 a^3} \sum_{k=1}^{\infty} k^2 \left(p^{\cal N}_{k}- p_k^{(2)}\right)
\eea
\label{epspvac}\ees
with the summands
\bes
\bea
&&\varepsilon^{\cal N}_{k}- \varepsilon_k^{(2)} = \frac{1}{2W_k}
\left[(W_k- \omega_k)^2 + \frac{(V_k - h)^2}{4}\right]  - \frac{h^2m^4}{8\, \omega_k^5} \label{epsub}\\
&&p^{\cal N}_{k}- p_k^{(2)} = \frac{1}{3} (\varepsilon^{\cal N}_{k}- \varepsilon_k^{(2)} )
+ \frac{\ m^2}{3W_k\omega_k\!\!}\ (W_k-\omega_k) + \frac{m^4}{12\omega_k^5}\,(\dot h + 3 h^2)
- \frac{5 m^6h^2}{24\,\omega_k^7}\,. \label{psub}
\eea
\label{epspsub}\ees
Thus in order for the sums in the renormalized energy-momentum tensor expectation value,
subtracted as in (\ref{epspvac}) to converge, it is sufficient for the summands (\ref{epspsub}) to fall off
as $k^{-5}$ or faster at large $k$. This is the important physical condition on the
definition of the adiabatic mode functions $(W_k, V_k)$, which restricts the choice of the
adiabatic vacuum state. The last term of (\ref{epsub}) and the last two terms of (\ref{psub})
already satisfy this condition. Inspection of the other terms in (\ref{epspsub}) shows that in
order to satisfy this condition it is sufficient for
\be
W_k- \omega_k = {\cal O}\left(k^{-3}\right) \quad {\rm and} \quad V_k -h =  {\cal O}\left(k^{-2}\right)
\quad {\rm as} \quad  k \rightarrow \infty
\ee
and then the sums in (\ref{epspvac}) will converge quadratically. Either of the choices
(\ref{Wk2Vk1}) or (\ref{WkVksimple}) of the last section satisfy this condition. Let us
emphasize that the choice of $(W_k, V_k)$ only affects how the individual ${\cal N}, {\cal R}$
and $\cal I$ terms contribute to the stress tensor expectation value in (\ref{epdecomp})
while the sum of all the  contributions and the subtraction terms (\ref{sumadb2})-(\ref{adb2})
are independent of that choice.

Thus with the vacuum contributions subtracted (\ref{epdecomp}) becomes
\bes
\bea
\varepsilon_{_R} &=&  \frac{1}{2 \pi^2 a^3} \sum_{k=1}^{\infty} k^2
\Big[\varepsilon^{\cal N}_{k} {\cal N}_k
+ \varepsilon^{\cal R}_{k} {\cal R}_k  +  \varepsilon^{\cal I}_{k} {\cal I}_k \Big]
+ \varepsilon_{vac}
\label{erenorm}\\
p_{_R}&=& \frac {1}{2 \pi^2 a^3}\sum_{k=1}^{\infty} k^2
\Big[p^{\cal N}_{k}{\cal N}_k
+ p^{\cal R}_{k} {\cal R}_k  +  p^{\cal I}_{k} {\cal I}_k\Big] + p_{vac}
\eea
\label{eprenorm}\ees
It should make very little difference which of the choices for $(W_k, V_k)$ one uses to define the
instantaneous adiabatic vacuum and time dependent Bogoliubov coefficients, since they
all fall off at large $k$, and will give qualitatively the same behavior of the particle
creation effects while passing through the lines of complex zeroes in Fig. \ref{Fig:ZeroesdS}.
The change in the plateau values $\Delta {\cal N}_{1,k\gamma}$ or $\Delta {\cal N}_{2,k\gamma}$
obtained after one or two creation events are the same for all definitions and the only
difference is in the detailed time dependence during the creation `event' itself, and only
for the lower $k$ modes. 

In our actual numerical calculations we have used the full fourth order adiabatic subtraction
as described in Ref. \cite{AndEak} in which the renormalization counterterms are separated
into terms which are divergent and terms which are finite. The latter can be integrated to form an
analytic contribution to the stress-energy tensor that is separately conserved. The full renormalized
stress-energy tensor is then given by (\ref{eprenorm}) with
\vspace{-3mm}
\bes
\bea
&&\varepsilon_{vac}= \frac{1}{4 \pi^2 a^3}
\sum_{k=1}^{\infty} k^2 \left(\varepsilon^{\cal N}_{k}- \frac{k}{a} -
\frac{m^2 a}{2 k} + \frac{m^4 a^3}{8 k^3} \right) + \varepsilon_{\rm an}
\label{epvac4}
\eea
\bea
p_{vac} = \frac{1}{4 \pi^2 a^3} \sum_{k=1}^{\infty} k^2 \left(p^{\cal N}_{k}-
\frac{ k a}{3}  + \frac{m^2 a^2 k}{6} - \frac{m^4 a^3}{8 k} \right)  + p_{\rm an}  \;.
\eea
\label{epspvac4}\ees
The analytic terms are given by
\bes
\bea
\varepsilon_{\rm an} &=& \frac{1}{2880 \pi^2} \, \left(\frac{6 \,\dddot{a}
   \dot{a}}{a^2} +\frac{6 \,\ddot{a}\dot{a}^2}{a^3}  -\frac{3 \, \ddot{a}^2}{a^2}
  -\frac{6 \,\dot{a}^4}{a^4}+\frac{6}{a^4}  \right)  - \frac{m^2}{96 \pi^2} \left( \frac{\dot{a}^2}{a^2} + \frac{1}{a^2} \right) \nn
 && - \frac{m^4}{64 \pi^2} \left[ \frac{1}{2} + \log\left(\frac{m^2 a^2}{4} \right)  + 2 C \right]
 \eea
 \vspace{-7mm}
 \bea
p_{\rm an} &=&  \frac{1}{2880 \pi^2} \, \left( -\frac{2\, \ddddot{a}}{a}  -\frac{4\, \dddot{a}\dot{a}}{a^2}  +\frac{8\, \ddot{a}
   \dot{a}^2}{a^3}  -\frac{3 \,\ddot{a}^2}{a^2}
-\frac{2 \,\dot{a}^4}{a^4}+\frac{2}{a^4} \right)
   + \frac{m^2}{288 \pi^2} \left(\frac{2\, \ddot{a}}{a} + \frac{\dot{a}^2}{a^2}  + \frac{1}{a^2}  \right) \nn
    & & +\frac{m^4}{64 \pi^2} \left[ \frac{7}{6} +   \log\left(\frac{m^2 a^2}{4} \right)  + 2 C \right]
\eea
\ees
with $C$ Euler's constant. This differs from the vacuum subtraction in (\ref{epspsub}) by finite terms,
which one can check remain small for all times in de Sitter space.

\begin{figure}[h]
  \centering
  \vspace{-3mm}
  \includegraphics[angle=90,height=9cm,viewport= 60 0 560 720, clip]
  {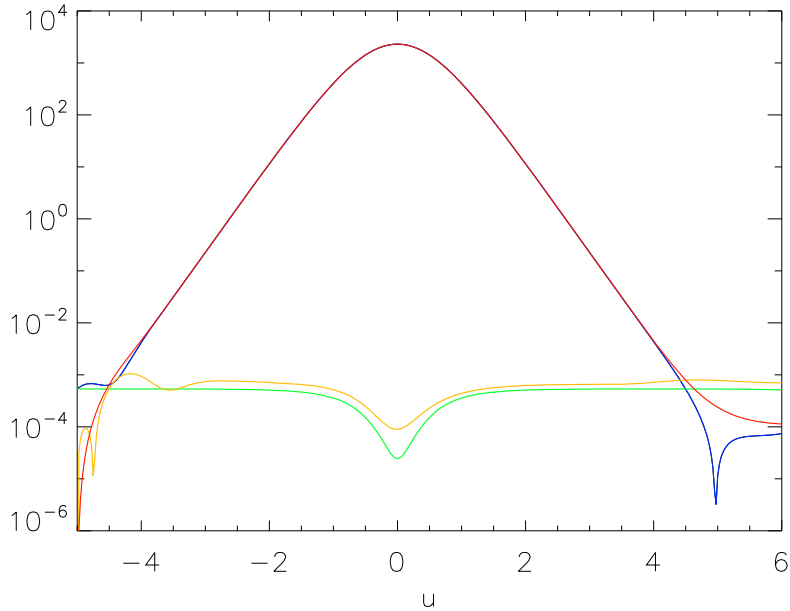}
  \vspace{-4mm}
  \caption{The absolute values of various contributions in (\ref{erenorm}) to the energy density for an adiabatic state
  when  $m= H$ are shown for a matching time of $u_0 = -5$. The blue curve is the total energy density, the red curve
  is the contribution from the ${\cal N}_k$ term, the green curve is the contribution from $\varepsilon_{\rm vac}$, and the
  orange curve is the contribution from the sum of the ${\cal R}_k$ and the ${\cal I}_k$ terms. Note the logarithmic scale.}
  \vspace{-3mm}
  \label{Fig:comparison}
\end{figure}

One way to assess the usefulness of the particle description is to analyze its contribution
to the stress-energy tensor.  This is done in Fig. \ref{Fig:comparison} where the full energy density
and that due to the various terms in the energy density (\ref{eprenorm}) are plotted.  It is clear from the
plot that near $u = 0$ the $\varepsilon_k^{\cal N}{\cal N}_k$  term provides by far the dominant contribution 
to the stress-energy tensor. At very early times and very late times this is not the case.  At early times this is expected
since the particle definition is designed to give a vacuum state at the matching time.  At late times
it is also expected since the energy density of the particles redshifts away.

\begin{figure}[htp]
  \centering
   \vspace{-5mm}
   \includegraphics[angle=90,height=9cm,viewport= 60 0 560 720, clip]{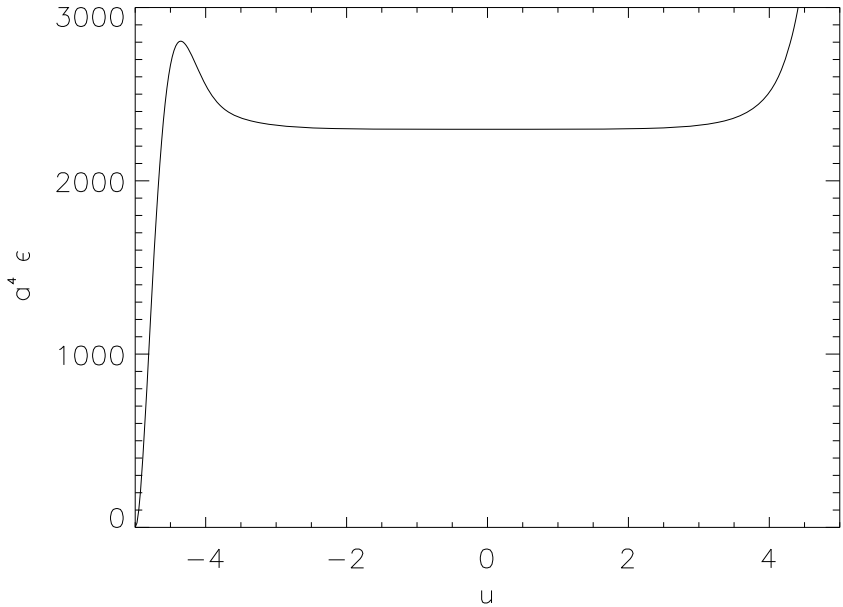}
   \vspace{-5mm}
  \caption{The product of the fourth power of the scale factor and the energy density for an adiabatic
  state when $m= H$ is shown.  The matching time for the adiabatic state is $u_0=-5$. }
  \vspace{-3mm}
  \label{Fig:scale}
  \end{figure}
  
From the window function (\ref{windS}) in the contracting phase of de Sitter space the energy density
for the initial adiabatic vacuum matched by (\ref{initdS}) at $u=u_0$ should behave like
\be
\varepsilon_{_R} \simeq \frac{1}{2 \pi^2 a^3} \sum_{k = K_{\gamma}(u)}^{K_{\gamma}(u_0)}\,
k^2 \varepsilon_k^{\cal N}  \,\Delta {\cal N}_{1\,k\gamma}\simeq
\frac{1}{8\pi^2} \frac{K^4_{\gamma}(u_0)}{a^4(u)} \,  \frac{1}{e^{8 \pi \gamma} - 1} \simeq
\frac{\gamma^4}{8\pi^2} \frac{1}{e^{2 \pi \gamma} - 1} \,\frac{\sinh^4 u_0}{a^4(u)}
\label{expandTpart}
\ee
and therefore grow exponentially with the fourth power of the scale factor as $a(u)$ decreases,
consistent  with our discussion above of the highest $k$ modes contributing in the window
$K(u) < k < K(u_0)$ in the contracting phase, where their effects on the
stress tensor are blueshifted, becoming highly relativistic. This is what accounts for the enormous
growth of the particle contribution to the energy density $\varepsilon_k^{\cal N}$ in Fig. \ref{Fig:comparison}
as it rapidly dominates the quantum $\varepsilon_k^{\cal R}, \varepsilon_k^{\cal I}$ and vacuum terms
in (\ref{eprenorm}). This is completely analogous to the secular
growth of the current in a constant uniform electric field starting from the adiabatic vacuum plotted
in Fig.  \ref{Fig:CurrentE}. To check the $a^{-4}$ relativistic dependence predicted by (\ref{expandTpart})
we plot the energy density due to the particles multiplied by $a^4(u)$ in
Fig. \ref{Fig:scale}. As expected the resulting quantity is approximately constant for a large
range of $u$, deviating from this behavior only for initial values of $u$ of order $u_0< 0$ and again as the
particles are redshifted away at $u$ of order $|u_0|$.

\begin{figure}[htp]
  \vspace{-5mm}
  \centering
  \includegraphics[angle=90,height=9cm,viewport= 60 0 560 720, clip]{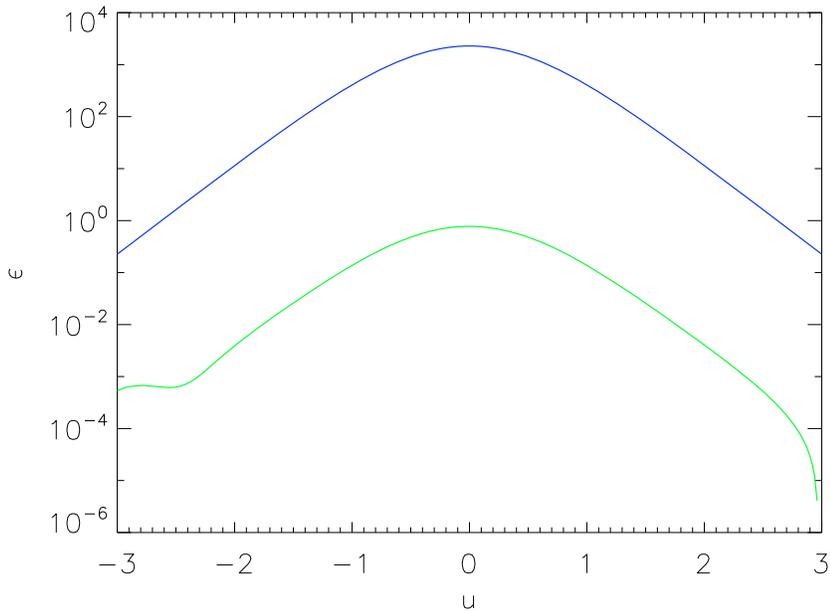}
   \vspace{-3mm}
  \caption{The energy density for an adiabatic state when $m= H$ is shown for an adiabatic matching time
  of $u_0 = -3$ (green curve) and a matching time of $u = -5$ (blue curve).}
  \vspace{-3mm}
  \label{Fig:enden1}
\end{figure}

The estimate (\ref{expandTpart}) also predicts that the maximum value of the energy density at the
symmetric point $u=0$ varies with the fourth power of $\sinh u_0$. In Fig. \ref{Fig:enden1} the energy
density is plotted for two different adiabatic matching times. It is also clear from the plots that the maximum
energy density at $u=0$ is substantially larger for the earlier adiabatic matching time, consistent
with (\ref{expandTpart}). This is expected since an earlier matching time allows more modes to go through the first
particle creation phase and increases the upper limit $K_{\gamma}(u_0)$ in (\ref{expandTpart}).

These results show that the adiabatic particle definition is a very useful one, since its
contribution to the energy density dominates when the particles become ultrarelativistic, that
the energy density of the created particles grows exponentially in the contracting phase
of de Sitter space, and most importantly that the maximum of the energy density also
grows exponentially with the initial time as $u_0 \rightarrow -\infty$. This shows
conclusively that global or `eternal' de Siter space is unstable to particle creation,
as their arbitrarily large energy densities will necessarily lead to a very
large backreaction on the classical spacetime when used as a source for
the semi-classical Einstein equations, especially as $u_0 \rightarrow -\infty$.

That the effect grows in the contracting phase of de Sitter, when modes are being
blueshifted to the ultrarelativistic regime, and the large $k$ UV part of the mode sum
dominates, as opposed to the expanding phase when the particles are
redshifted and the low $k$ modes dominate is not surprising. To see this effect
properly one needs to start with the adiabatic basis at a finite (early) initial time.
We observe here one major difference between the electric field and de Sitter cases
in the basic kinematics. The electric field is a vector field and uniformly
accelerates all charged particles of a given charge in one direction. Particles
(virtual or real) with initially negative kinetic momenta along the direction of
the electric field are eventually brought momentarily to rest, and then turn around
with continually increasing positive kinetic momentum ever after.  It is these
modes in the quantum theory that undergo particle creation at the turnaround
time, and make the secular contribution to the current as their
kinetic momentum and energy grow without bound and the corresponding
particles approach the speed of light. Thus this late time contribution
is clearly relativistic and UV dominated. There is only one creation event
for each wavenumber mode.

On the other hand in the de Sitter case the physical or kinetic momentum is $p = k/a$
which is isotropic, depending only on the magnitude of $\bf k$ and not its direction
in the spatially homogeneous states we are considering.
There are two creation events for each $k$ mode, one in the contracting phase
of de Sitter space, the second in the expanding phase. The first creation event
is quite analogous to the electric field case in that once created the particles
are blueshifted (exponentially in this case), rapidly becoming ultrarelativistic
and making a contribution to the energy density and pressure that grows
like $a^{-4}$, typical of ultrarelativistic particles in the contracting
phase of de Sitter space. In the expanding phase of de Sitter space
the situation is reversed, the created particles are redshifted and make
a decreasing contribution to the stress tensor, at least for spatially homogeneous
states, so that even a steady rate of particle creation cannot produce an effect 
in the stress tensor secularly growing in time. Instead the vacuum and
other $\cal R$ and $\cal I$ interference terms in (\ref{eprenorm}) remain
of the same order as the particle creation term at late times and together
their sum approaches the de Sitter invariant CTBD value \cite{Attract}.
There is no exact analog of this second behavior in the electric field case.

\section{Adiabatic Switching On of de Sitter Space and the In Vacuum}
\label{Sec:Switch}

In the case of the spatially uniform electric field there is an exactly soluble problem
in which the electric field is adiabatically switched on and off according to the profile
\be
A_z(t) = - E\, T \, \tanh \left(t/T\right)\,,\qquad  {\bf E} (t) = E \,{\bf \hat z}\ \sech^2 \left(t/T\right)  \;.
\label{Aztanht}
\ee
By taking the limit $T \rightarrow \infty$ at the end of the calculation, the constant
uniform field (\ref{Egauge}) is recovered \cite{Nik,NarNik,GavGit}. On the other hand if $T$
is finite and $t \rightarrow \pm \infty$ the electric field goes to zero exponentially rapidly,
and the particle and anti-particle solutions are the standard Minkowski ones (with $p = k_z \pm eET$),
so there is no doubt that ${\it in}$ and ${\it out}$ states may be identified with the Minkowski vacuum
and the excitations above that state correspond to the usual definition of particles.

No such analytically soluble model for de Sitter space is known. However there is no difficulty
in studying the time profile of the scale factor
\be
a_T(u) = \frac{1}{H} \,\cosh \bigg[ HT \tanh\left(\frac{u}{HT}\right)\bigg]
\label{adbswitch}
\ee
in the line element (\ref{RW}) and the associated solutions of the scalar field mode
equation (\ref{oscmode}) by numerical methods. With this profile at fixed $T$, in the infinite time limits
$u\rightarrow \mp \infty$, $a(u)$ approaches the constant $H^{-1} \cosh (HT)$ so that the spacetime
is static and the particle number is uniquely defined by the asymptotic constant positive and negative
frequency functions. On the other hand for $HT \rightarrow \infty$ at fixed $u$, $a_T(u)$ approaches the de Sitter
scale factor $H^{-1} \cosh u$. Thus (\ref{adbswitch}) interpolates between static spacetimes
in both the remote past and remote future with a symmetrical de Sitter contracting and expanding
phase in between.

\begin{figure}[htp]
  \centering
   \vspace{-3mm}
   \includegraphics[angle=90,height=9cm,viewport= 60 0 560 720, clip]{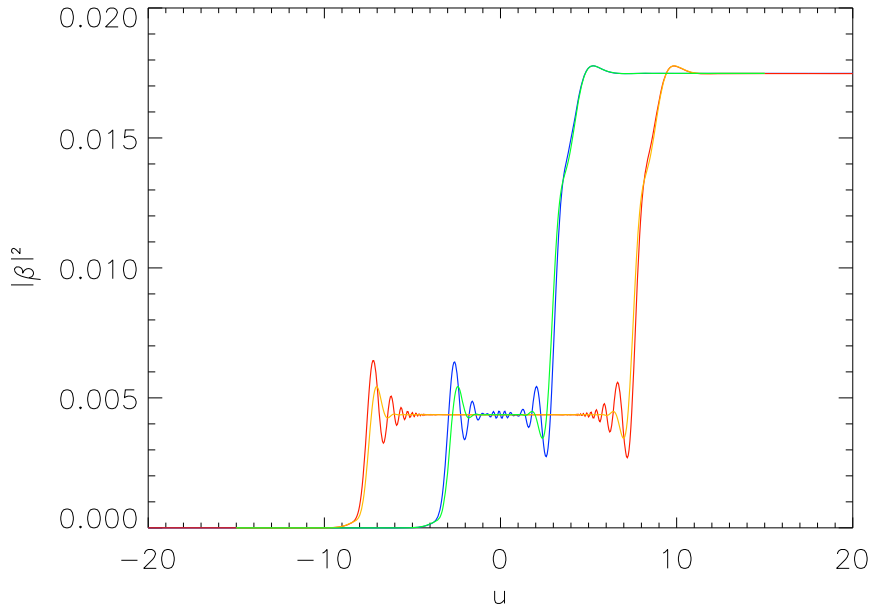}
    \vspace{-5mm}
  \caption{Comparison of $|\beta_k(u)|^2$ for the {\it in} vacuum in exact de Sitter space for matching
  time $u_0 = -15$, and that for the smooth adiabatic switching on of de Sitter space according to the
  time profile (\ref{adbswitch}) for $HT= 1000$.  Plotted is $|\beta_k(u)|^2$ for $k = 10$, $u_0=-15$ 
  (green, exact and blue, $HT=1000$),  and $|\beta_k(u)|^2$ for $k = 1000$, $u_0=-15$
  $HT=1000$ (red, exact and orange, $HT=1000$ ). }
  \vspace{-3mm}
  \label{Fig:finiteT}
\end{figure}

In Fig. \ref{Fig:finiteT} we show a comparison of the time dependent adiabatic particle number
$|\beta_k(u)|^2$ for the fixed de Sitter and adiabatic switched metric of the form (\ref{adbswitch}),
for two representative values of $k$. The similarity of the curves for the two models with the same
value of $k$ shows that essentially the {\it same} particle production process takes place in the fixed
de Sitter space background or with the adiabatic switching on of the background from a static
initial metric according to (\ref{adbswitch}), at least for a large range of $k$ when $T$ is large enough.
This supports the choice of the {\it in} state and positive frequency mode function (\ref{modein})
in de Sitter space, as that one selected by turning on the de Sitter background adiabatically.
Certainly the state produced in this way is very different from the maximally symmetric CTBD
state defined by analytic continuation from Euclidean ${\mathbb S}^4$, in its low
momentum modes, as expected by our analysis of adiabatic vacua and particle creation:
compare Fig. \ref{Fig:CTBD}. A detailed study of the state and subsequent evolution produced
by (\ref{adbswitch}) and other adiabatic profile functions will be presented in a subsequent
publication \cite{AndMotSan}.

\section{Summary and Discussion}
\label{Sec:Conc}

In this paper we have presented a detailed study of the spontaneous particle production of a massive
free field theory in geodesically complete de Sitter space in real time. It is this spontaneous production
of particles from the vacuum that is the basis for the instability of global de Sitter space.
The formulation of particle creation as an harmonic oscillator with a time dependent frequency,
or equivalently, as a one-dimensional stationary state scattering problem, determines the {\it in}
and {\it out} positive energy particle states for massive fields in de Sitter space. This emphasizes
the very close analogy with the spontaneous creation of charged particle/anti-particle pairs in a uniform,
constant electric field. In each case the background gravitational or electric field configuration is
symmetric under time reversal. Hence in each case it is possible to find a time reversal symmetric
state in which no net particle creation occurs, and for which the imaginary part of the one-loop
effective action and the decay rate vanishes identically. Such a maximally symmetric state 
allows a `self-consistent' solution of the semi-classical Maxwell or Einstein equations, with vanishing
electric current or de Sitter invariant stress-tensor. The artificiality of such a time symmetric state
is apparent in the stationary scattering formulation, since it corresponds to choosing a very special
coherent superposition of positive and negative frequency scattering solutions globally, which
exactly cancels each spontaneous particle creation event by a time reversed particle annihilation
event, {\it c.f.} Fig. {Fig:CTBD}. This corresponds to adjusting the state of the quantum field to contain 
just as many pairs coming in from infinity and with the precisely right phase relations between them, 
so as to exactly cancel the electric currents or stress-energies of the pairs being spontaneously 
produced by the electric or de Sitter backgrounds. This is clearly not a true vacuum state in either case.

In situations such as these, the extension of the concepts of `particle' and `vacuum'  from
flat Minkowski space with no background fields must be reconsidered carefully. The
essential generalization of the Feynman prescription of particles propagating forward in time and 
anti-particles propagating backward in time is to define {\it in} and {\it out} `vacuum' states
corresponding to the choice of pure positive frequency modes (\ref{adbmode}) at intermediate
times which are asymptotic to the exact particle {\it in} and {\it out} solutions~\eqref{inouttoBD} 
of the oscillator equation (\ref{oscmode}) in the remote past and remote future.  Mathematically 
this is the condition that the positive frequency particle modes are analytic functions of $m^2$ 
in the upper or lower complex plane which are regular as $t \rightarrow \mp \infty$ respectively, 
and which corresponds also to the Schwinger-DeWitt method and choice of proper time contour. 
This should settle the question of whther the effective action in de Sitter space is real or imaginary
due to particle creation effects \cite{Akhm}. 

We have provided evidence that the {\it in} state is also the state obtained by turning the background 
fields on and off again according to a finite time $T$ parameter which may be taken to infinity 
at the end of all calculations. By any of these equivalent methods one obtains the standard Schwinger 
decay rate (\ref{Erate}) for scalar charged particle creation in a constant, uniform electric field. 
By applying these same two methods to the background gravitational field of de Sitter space, 
one obtains the vacuum persistence amplitude and decay rate (\ref{Decayrate}). Hence global 
de Sitter space is unstable to particle creation for the same reason as a constant, uniform electric 
field is in electrodynamics. This provides a mechanism for the relaxation of vacuum energy
into matter or radiation and at least one possible route to the solution of the vacuum
energy problem relying only upon known physics \cite{Fluc,NJP}.

Although the definition of the adiabatic particle number (\ref{adbpart})
necessarily comes with some ambiguity in a time dependent background, and depends
upon two frequency functions $(W_k, V_k)$ in (\ref{betsq}) that are not unique, their choice
is highly constrained by the requirements of rendering the vacuum zero-point contributions
to physical currents, and the energy density and pressure (\ref{epspvac}) ultraviolet finite.
The detailed time profile of the particle number depends upon the particular choice of
adiabatic particle number through $(W_k, V_k)$, but the qualitative features and the asymptotic
values in either the constant electric field or de Sitter cases do not, {\it c.f.} Figs.
\ref{Fig:EW0W2} and \ref{Fig:wkbcompare}. The rapid change in the adiabatic particle
number around `creation events' can be understood from the location of the
zeroes of the adiabatic frequency function in the complex time plane,
{\it c.f.} (\ref{Ezero}) and Figs. \ref{Fig:EZeroes}-\ref{Fig:Evarlam} in the electric field
case and (\ref{ucomplex}) and Figs. \ref{Fig:ZeroesdS}-\ref{Fig:mass123} for de Sitter
space. Some adiabatic particle number definition of this kind is certainly necessary
to make the transition from QFT to kinetic theory, since the Boltzmann equation assumes
that particle numbers and densities can be defined.

The usefulness of the particle concept is seen in the evaluation of expectation values
of currents and stress tensors, and particularly in their secular terms, which are most
important for backreaction. The `window function'  (\ref{Ewin}) of pair creation in the
electric field background accounts very well for the linear secular growth in time of the
current of the produced pairs in Fig. \ref{Fig:CurrentE} \cite{QVlas}. In the de Sitter case
the corresponding window function (\ref{windS}) accounts very well for the exponential
growth of the energy density and pressure of the created particles in the contracting phase, 
{\it c.f.} Eq.\ (\ref{expandTpart}) and Fig. \ref{Fig:comparison}. The linear growth in time in the first 
case and exponential $a^{-4}$ growth in time in the contracting de Sitter case, Fig. \ref{Fig:scale}
are just what should be expected as the created particles are accelerated and rapidly become
ultrarelativistic. Because of this secular growth backreaction effects on the background electric field 
must be taken into account through the semi-classical Maxwell equations.  Likewise for the de Sitter 
geometry for early enough matching times $u_0$, {\it c.f.} \ref{expandTpart}), backreaction effects 
similarly must be taken into account through the semi-classical Einstein equations.

This detailed study of particle creation removes a possible objection to the use
of the strictly asymptotic {\it in} and {\it out} states in \cite{PartCreatdS} to calculate the decay
rate of de Sitter space, namely that these states are not Hadamard UV allowed
states as defined {\it e.g.} in \cite{ShortDistDecohere}. Instead they are members of the
one parameter family of `$\alpha$ vacua' invariant under the $SO(4,1)$
subgroup of the de Sitter group continuously connected to the identity \cite{PartCreatdS,Allen},
although not the discrete ${\mathbb Z}_2$ inversion symmetry (\ref{inversion}).
The difficulty is removed by the recognition that the large $|u|$ time and large $k$
limits do not commute. If one starts with UV allowed states such as the adiabatic initial
state (\ref{initdS}) at a finite initial time, and evolves forward for a finite
time, only a finite number of modes experience particle creation events, according
to the appropriate window function. The non-Hadamard $\alpha$ vacua are
produced only in the improper limit of $|u| \rightarrow \infty$ in eternal de Sitter space,
never in any finite time starting with UV finite initial data. Although the result for the
decay rate (\ref{Decayrate}) with an appropriate physical cutoff is the same, only
a detailed analysis of the $k$ and time dependence allows a description free
of any spurious UV problems and focuses attention on the resulting necessary breaking
of de Sitter invariance instead \cite{Fluc,deSMM}.

In the electric field case it is generally accepted that particle creation will lead to
eventual `shorting' of the electric field, although to date this process has only been
studied in a large $N$ semiclassical approximation \cite{KESCM}, which is not adequate
to show the true long time behavior of the system even in QED. This depends upon
self-interactions, and the long time behavior of correlation functions that are not accessible
to the standard weak coupling approximations. Such processes involving multiple
interactions in a medium, possibly very far from equilibrium, are generally described in many-body
physics in the kinetic Boltzmann equation approximation, where all time reversal invariance properties
of the underlying QFT are lost, and irreversible behavior is expected. What is perhaps less widely
appreciated is that this breaking of time reversal invariance has its roots in the definition of the
vacuum itself and the distinction between particles and anti-particles in QFT by the
$m^2 - i0^+$ prescription for the Feynman propagator and Schwinger-DeWitt
proper time method.  When interactions are turned on, the bare mass becomes a dressed self-energy
function $\Sigma - i {\Gamma}/2$  and the pole moves away from the real axis. The imaginary part
is now finite and gives the quasi-particle lifetime in the medium. Causality fixes the sign
of this imaginary part, and that same causal prescription is already present in the {\it free}
propagator in the limit $\Gamma \rightarrow 0^+$ that the interactions are turned off.
It is this causal boundary condition anticipating the inclusion of interactions, rather
than the interactions themselves, which breaks time reversal symmetry.

Spontaneous particle creation {\it vs.} the exact annihilation of particles in the CTBD state,
\!{\it c.f.} \hspace{-3mm} Fig. \hspace{-3mm}  \ref{Fig:CTBD} raises another interesting point
about the origins of time irreversibility, entropy and the second law in QFT. That particle creation
is in some sense an irreversible process in which entropy increases \cite{HuKanPav} can
be made precise by means of the quantum density matrix expressed in the adiabatic particle basis.
Since adiabatic particle number is by construction an adiabatic invariant of the evolution,
the diagonal elements of the density matrix are slowly varying in this basis. In contrast, the
off-diagonal elements are rapidly varying functions both of time and of momentum
at a fixed time. Then it is reasonable to average over those rapidly varying phases
and construct the {\it reduced} density matrix which shows general (though not strictly
monotonic) increase in time as particles are created \cite{CKHM}, much as Figs. \ref{Fig:Evarlam},
\ref{Fig:bui15} and \ref{Fig:bui5} for the particle number itself do. This is equivalent
to the approximation of neglecting the oscillatory term(s) in the current (\ref{qcurrent})
or stress-tensor (\ref{eprenorm}) expectation values, which as we have seen is
a very good approximation over long times when the secular effects of particle
creation dominate. Clearly no such interpretation is possible for the time
symmetric CTBD state in which phase correlations are exactly preserved, and particle 
annihilations represent a {\it decrease} in the effective entropy. One would 
expect such processes to be statistically disfavored.

The fact that particles can achieve arbitrarily high energies for persistent constant field
backgrounds producing a secular effect in both the electric field and de Sitter cases
underscores the interesting interplay of UV and IR aspects. Since anomalies perform exactly 
this function of connecting the UV to the IR, they can play an important role \cite{DSAnom,Zak}, 
a connection we explore in detail in \cite{AndMotDSVacua}. In de Sitter space with ${\mathbb S}^3$ 
spatial sections the blueshifting toward ultrahigh energies is clear in the contracting phase. 
In the expanding phase of de Sitter space the created particles defined in the ${\mathbb S}^3$ 
sections are redshifted and do not produce any growing secular effect in spatially homogeneous
states. In fact, one can prove  that the energy density and pressure always tend to the de Sitter invariant 
Bunch-Davies value for fields with positive effective masses, $m^2 + \xi R >0$, produced in any UV
allowed $O(4)$ invariant state \cite{Attract}. It is this redshifting and damping effect of the expansion 
upon perturbations in FRW metrics, which because of inflation almost all attention has been focused,
and consideration only of spatially homogenous states that leads to the impression that de Sitter space 
is stable. 

This is clearly incorrect for global ('eternal') de Sitter space which has both a
contracting and an expanding phase. In fact, apart from initial conditions or boundary conditions,
de Sitter space is a homogeneous manifold in which all points are equivalent, so the very notions
of expansion or contraction have no meaning unless a preferred set of de Sitter breaking set of
time slicings is chosen. With the ${\mathbb S}^3$ time slicing chosen to cover all of de Sitter
space, for early matching times (as discussed above), the exponentially growing energy density 
and pressure of the created particles will necessarily produce an enormous backreaction on the 
geometry if taken into account even semi-classically, and even without self-interactions. Hence 
one may never arrive at the expanding or inflationary phase of de Sitter space at all.
This emphasizes the importance of and sensitivity to the initial conditions of
inflation \cite{Vil}.

Moreover although the description of adiabatic vacua and particle definition given
is dependent upon spatial homogeneity and isotropy in the ${\mathbb S}^3$
time slicings, the tendency to create particles and the possibility for these particles to
be kinematically blueshifted and produce large backreaction effects in de Sitter space, is not
dependent upon coordinate choices. We have shown in Sec. \ref{Sec:Flat} that even in the
Poincar\'e coordinates with flat spatial sections which cover only the expanding half of de Sitter
space, particle production also takes place. If only the expanding half of de Sitter space is
considered, then the particles are redshifted, so that their energy density does not grow,
{\it if} the initial state is given and {\it if} the particle modes are defined with respect the usual
FRW spatially flat sections. Once it is granted that global de Sitter space with its exact  $O(4,1)$
symmetry is unstable to particle creation, then there is no reason to restrict one's attention
{\it a priori} even to spatially homogeneous or isotropic sections. Fluctuations in the mean
stress tensor $\lag T_{ab}(x) T_{cd}(x')\rag$ due to particle creation are surely spatially inhomogeneous.
A previous study of linear response has indicated the importance of spatially
inhomgeneous perturbations on the de Sitter horizon, in the static coordinates  of de
Sitter space \cite{DSAnom}.

Here again the electric field example may be helpful. One can describe a constant, uniform
electric field in either a time dependent or time independent but spatially dependent gauge.
Both are equally good for describing the idealized situation without boundaries in
either space or time. However, what actually happens depends sensitively on boundary
or initial conditions. Just as one can consider relaxing the constancy in time of the background
to study the dependence upon vacuum initial conditions, adiabatically switching it on and
then off, one could also consider the arguably more physical situation of relaxing strict spatial
homogeneity, allowing the electric field to be established by some charge distribution
at a large but finite distance away from the local region of interest. Sensitivity of the vacuum
to initial conditions will likely then be accompanied by sensitivity to the spatial boundary conditions,
and the final evolution may be quite different globally. The distance scale over which
the particle creation takes place is of order $2mc^2/eE$ in an electric field background, so that
is the scale at which one might expect spatial inhomogeneities to develop in a random
particle creation process. The natural scale for such inhomogenities to develop
in de Sitter space is the horizon scale $H^{-1}$. Therefore one should allow for perturbations
on the horizon scale which can be sensitive to the blueshifting kinematics in a way
that perturbations based on the spatially homogeneous ${\mathbb S}^3$ or flat
Poincar\'e sections are not. We address spatially inhomogeneous perturbations of the
CTBD state in an accompanying paper \cite{AndMotDSVacua}, depending upon the direction 
of $\bf k$ as well as its magnitude, and show that these effects may be even more significant that the
average particle creation rate in spatially homogeneous states studied in this paper.

\vspace{3mm}
\centerline{\large{\bf Acknowledgements}}
\vspace{3mm}
P. R. A. would like to thank Dillon Sanders for help with the early stages of this project.
This work was supported in part by the National Science Foundation under Grant Nos. PHY-0856050
and PHY-1308325.  The numerical computations herein were performed on the WFU DEAC cluster;
we thank WFU's Provost Office and Information Systems Department for their generous support.

\appendix

\section{Geometry and Coordinates of de Sitter Space}
\label{App:GeometrydS}

The de Sitter manifold is most conveniently defined as the single sheeted hyperboloid
\be
\eta_{AB}X^A X^B =  -(X^0)^2 + \sum_{i=1}^3 X^iX^i + (X^4)^2 = \frac{1}{H^2}
\label{dSman}
\ee
embedded in five dimensional flat Minkowski spacetime,
\be
ds^2 = \eta_{AB}\, dX^A\,dX^B = - (dX^0)^2 + (dX^1)^2 + (dX^2)^2 + (dX^3)^2 + (dX^4)^2\,.
\label{flat5}
\ee
This manifold has the isometry group $O(4,1)$ with the maximal number of continuous
symmetry generators ($10$) for any solution of the vacuum Einstein field equations,
\be
R^{a}_{\ b} - \frac{R}{2} \,\delta^{a}_{\ b} + \Lambda \,\delta^{a}_{\ b} = 0
\label{Ein}
\ee
in four dimensions. It also has the discrete inversion symmetry
\be
X^A \rightarrow - X^A
\label{inversion}
\ee
which is not continuously connected to the identity, making the isometry group of the full de Sitter
manifold $O(4,1) = {\mathbb Z}_2 \otimes SO(4,1)$. The Riemann tensor, Ricci tensor, and scalar curvature are
\bes
\bea
R^{ab}_{\ \ cd} &=& H^2 \left(\delta^a_{\ c}\,\delta^b_{\ d} - \delta^a_{\ d}\,\delta^b_{\ c}\right)\\
R^a_{\ b} &=& 3 H^2\, \delta^a_{\ b}\\
R &=& 12 H^2
\eea
\ees
with the Hubble constant $H$ related to $\Lambda$ by
\be
H = \sqrt{\frac{\Lambda}{3}}\,.
\ee

\begin{figure}[h]
\begin{center}
\vspace{-5mm}
\includegraphics[height=6cm,width=6cm]{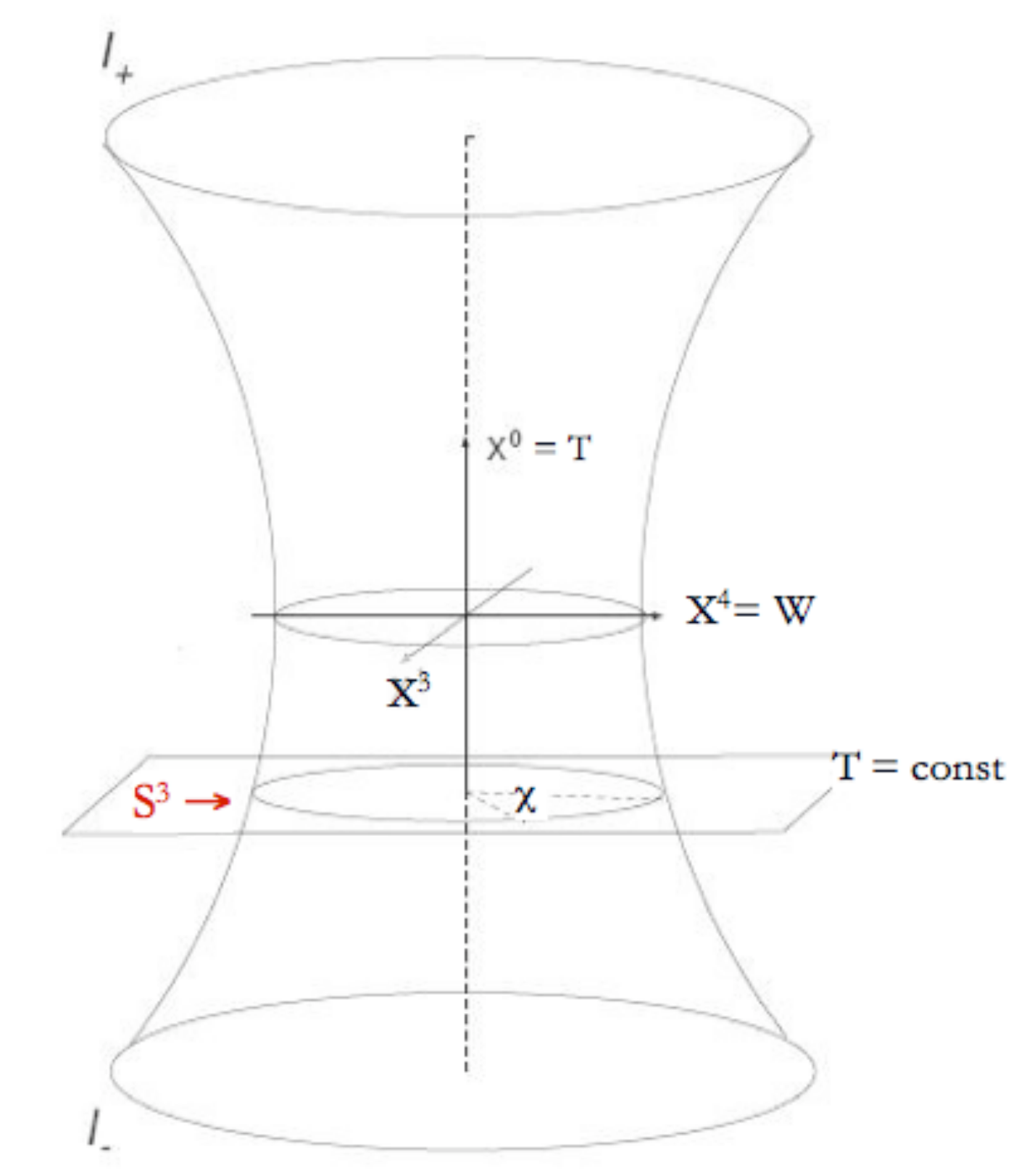}
\caption{The de Sitter manifold represented as a single sheeted hyperboloid of revolution
about the $X^0$ axis, embedded in five dimensional flat spacetime $(X^0, X^a), \,a= 1,\dots 4$, in which the
$X^1$, $X^2$ coordinates are suppressed. The hypersurfaces at constant $X^0=H^{-1} \sinh u$ are
three-spheres, $\mathbb{S}^3$. The $\mathbb{S}^3$ at $X^0= \pm \infty$ are denoted by $I_{\pm}$.}
\label{Fig:deShyper}
\end{center}
\vspace{-7mm}
\end{figure}

The globally complete hyperbolic coordinates $(u, \chi, \theta, \phi)$ of de Sitter space are defined by
\bes
\bea
X^0 &=& \frac{1}{H}\, \sinh u\,,\\
X^i &=& \frac{1}{H}\, \cosh u\,\sin\chi\, \hat n^i\,,\qquad i =1, 2, 3\,.\\
X^4&=&  \frac{1}{H}\, \cosh u\,\cos\chi\,
\eea
\label{hypercoor}\ees
where $\hat n = (\sin\theta\,\cos\phi\,,\sin\theta\,\sin\phi\,,\cos\theta)$
is the unit vector on $\mathbb{S}^2$. In these coordinates the de Sitter line element takes the form
\be
ds^2 = \frac{1}{H^2} \big( -du^2 + \cosh^2 u\, d\Sigma^2\big)
\label{hypermet}
\ee
where
\be
d\Sigma^2 \equiv d \hat N \cdot d \hat N = d\chi^2 + \sin^2\chi\,  (d\theta^2 + \sin^2\theta\, d\phi^2)
\label{S3}
\ee
is the standard round metric on $\mathbb{S}^3$. Hence in the coordinates of (\ref{hypercoor})
which cover the entire de Sitter manifold the de Sitter line element (\ref{hypermet}) is a hyperboloid
of revolution whose constant $u$ sections are three-spheres, {\it c.f.} Fig. \ref{Fig:deShyper}.
In (\ref{S3}) and the following we make use of the shorthand notation,
\be
\hat N = (\sin\chi\ \hat n,\, \cos\chi ) \label{Ndef}
\ee
for the unit four-vector of $\mathbb{S}^3$ in the $(X^i, X^4)$ coordinates of the flat space embedding.

In cosmology it is more common to use instead the Friedmann-Lema\^itre-Robertson-Walker
(FLRW) line element with flat $\mathbb{R}^3$ spatial sections, {\it viz.}
\be
ds^2 = - d\tau^2 + e^{2H\tau}\, d {\bf x}^2 =  - d\tau^2 + e^{2H\tau}\,(d\varrho^2 + \varrho^2 d\Omega^2)
\label{FLRW}
\ee
with $\varrho \equiv |{\bf x}|$. De Sitter space (\ref{dSman})-(\ref{flat5}) can be brought into the
flat FLRW form (\ref{FLRW}) by setting
\bes
\bea
X^0 &=& \frac{1}{H} \sinh (H\tau) + \frac{H\varrho^2}{2}\, e^{H\tau}\\
X^i  &=& e^{H\tau} \,\varrho \,\hat n^i\,,\qquad i =1, 2, 3 \\
X^4 &=&  \frac{1}{H}\cosh (H\tau) - \frac{H\varrho^2}{2}\, e^{H\tau}\,.
\eea\label{RWcoor}\ees
From (\ref{RWcoor}) $T + W > 0$ in these coordinates for all $\tau \in (-\infty, \infty)$, with the
hypersurfaces  of constant FLRW time $\tau$ slicing the hyperboloid in Fig. \ref{Fig:deShyper}
at a $45^{\circ}$ angle. The null surface at $T + W = 0$ is approached in the limit
$\tau \rightarrow -\infty$.

Hence the flat FLRW coordinates (\ref{FLRW}) break the time inversion
symmetry of global de Sitter space and cover only one half of the full de Sitter hyperboloid
in which the spatial sections are always expanding as $\tau$ increases. The other half of
the full de Sitter hyperboloid with $T + W < 0$ is obtained if $H\tau$ is replaced by $-H\tau' + i \pi$,
so that $\exp(H\tau)$ is replaced by $-\exp(-H\tau')$, and the line element (\ref{FLRW}) now
takes the form
\be
ds^2 = - d\tau^{\prime\,2} + e^{-2H\tau'}\, d {\bf x}^2
\ee
with the spatial sections are always contracting as $\tau'$ increases. The null surface
$T+ W =0$ is approached in the limit $\tau' \rightarrow + \infty$.


\begin{thebibliography}{99}
\vfil\eject

\centerline{\large{\bf References}}
\vspace{4mm}
\bibitem{Pau} W. Pauli (unpublished); C. P. Enz and A. Thellung, {\it Helv. Phys. Acta} {\bf 33} 839 (1960).

\bibitem{Infl} See {\it e. g.} A. R. Liddle and D. H. Lyth, {\it Cosmological
Inflation and Large-Scale Structure}, Cambridge Univ. Press, Cambridge (2000); or\hfil\break
V. Mukhanov, {\it Physical Foundations of Cosmology}, Cambridge Univ. Press, Cambridge (2005).

\bibitem{SNI} A. G. Riess {\it et. al.}, {\it Astron. J.} {\bf 116}, 1009 (1998);\hfil\break
S. Perlmutter {\it et. al.}, {\it Astrophys. J.} {\bf 517}, 565 (1999);\hfil\break
J. L. Tonry {\it et. al.}, {\it Astrophys. J.} {\bf 594}, 1 (2003).

\bibitem{Parker} L. Parker, {\it Phys. Rev. Lett.} {\bf 21}, 562 (1968); {\it Phys. Rev.} {\bf 183}, 1057 (1969);
{\it Phys. Rev. D} {\bf 3}, 346 (1971), Erratum {\it ibid.} 2546 (1971); {\it Phys. Rev. Lett.} {\bf 28}, 705 (172),
Erratum {\it ibid.} 1497 (1972).

\bibitem{Zel}   Ya.B. Zel'dovich, {\it Pis. Zh. Eksp. Teor. Fiz.} {\bf 12}, 443 (1970) [JETP Lett. {\bf 12}, 307 (1970)].

\bibitem{ParFul} L. Parker and S. A. Fulling, {\it Phys. Rev. D} {\bf 9}, 341 (1974);\hfil\break
S. A. Fulling and L. Parker, {\it Ann. Phys.} {\bf 87}, 176 (1974);\hfil\break
S. A. Fulling, L. Parker, and B. L. Hu, {\it Phys. Rev. D} {\bf 10}, 3905 (1974).

\bibitem{BirBun} N. D. Birrell, {\it Proc. R. Soc. Lon. B} {\bf 361}, 513 (1978);\hfil\break
T. S. Bunch, {\it J. Phys. A: Math. Gen.} {\bf 13}, 1297  (1980).

\bibitem{BirDav} N. D. Birrell and P. C. W. Davies, {\it Quantum Fields
in Curved Space} Cambridge Univ. Press, Cambridge (1982).

\bibitem{PartCreatdS} E. Mottola, {\it Phys. Rev. D} {\bf 31}, 754 (1985).

\bibitem{Fluc} E. Mottola, {\it Phys. Rev. D} {\bf 33}, 2136 (1986);
{\it Physical Origins of Time Asymmetry}, J. J. Halliwell {\it et al.} eds., pp. 504-515,
Cambridge,  Cambridge Univ. Press (1993).

\bibitem{NJP} I. Antoniadis, P. O. Mazur and E. Mottola, {\it New J. Phys.} {\bf 9}, 11 (2007).

\bibitem{Poly} A. M. Polyakov, {\it Nucl. Phys. B} {\bf 797}, 199 (2008);
{\it ibid.} {\bf 834}, 316 (2010); e-print arXiv:1209.4135;\hfil\break
D. Krotov and A. M. Polyakov, {\it Nucl. Phys. B} {\bf 849}, 410 (2011).

\bibitem{AkhBui} E. T. Akhmedov and P.V. Buividovich, {\it Phys. Rev. D} {\bf 78}, 104005 (2008).

\bibitem{Nacht}  O. Nachtmann, {\it Commun. Math. Phys.} {\bf 6}, 1 (1967).

\bibitem{CherTag} N. A. Chernikov and E. A. Tagirov, {\it Ann. Inst. H. Poincar\'e Phys. Theor.}
{\bf A9}, 109 (1968);\hfil\break
E. A. Tagirov, {\it Ann. Phys.} {\bf 76}, 561 (1973).

\bibitem{BunDav} T. S. Bunch and P. C. W. Davies, {\it Proc. R. Soc. A} {\bf 360}, 117 (1978).

\bibitem{GibHaw} G. W. Gibbons and S. W. Hawking, {\it Phys. Rev. D} {\bf 15}, 2738
(1977);\hfil\break
A. S. Lapedes, {\it J. Math. Phys. }{\bf 19}, 2289 (1978).

\bibitem{Ford} L. H. Ford, {\it Phys. Rev. D}  {\bf 31}, 710 (1985).

\bibitem{AntMot} I. Antoniadis and E. Mottola, preprint CERN-TH-4605/86 (1986);
{\it Jour. Math. Phys.} {\bf 32}, 1037 (1991).

\bibitem{Wood} N. C. Tsamis and R. P. Woodard,  {\it Nuc. Phys. B} {\bf 474}, 235 (1996);
{\it Ann. Phys.} {\bf 253}, 1 (1997); {\bf 267},145 (1998).

\bibitem{Schw} J. Schwinger, {\it Phys. Rev.} {\bf 82}, 664 (1951).

\bibitem{DeWitt} B. S. DeWitt {\it Dynamical Theory of Groups and Fields}, Gordon and Breach, New York (1965);
{\it Phys. Rep.} {\bf 19}, 295 (1975).

\bibitem{Rumpf}  H. Rumpf, {\it Phys. Lett. B} {\bf 6}, 272 (1976); {\it Nuovo Cim.} {\bf B35}, 321 (1976);\hfil\break
H. Rumpf and H. K. Urbantke, {\it Ann. Phys.} {\bf 114}, 332 (1978).

\bibitem{Nar} N. B. Narozhnyi, {\it Zh. Eksp. Teor. Fiz.} {\bf 54}, 676 (1968)
[Sov. Phys. JETP {\bf 27}, 360 (1968)].

\bibitem{Nik} A. I. Nikishov, {\it Zh. Eksp. Teor. Fiz.} {\bf 57}, 1210 (1970)
[Sov. Phys. JETP {\bf 30}, 660 (1970)].

\bibitem{NarNik} N. B. Narozhnyi and A. I. Nikishov, {\it Yad. Fiz.} {\bf 11}, 1072 (1970)
[Sov. J. Nuc. Phys. {\bf 11}, 596 (1970)].

\bibitem{FradGitShv} E. S. Fradkin, D. M. Gitman, and Sh. M. Shvartsman,
{\it Quantum Electrodynamics with Unstable Vacuum}, Springer-Verlag, Berlin (1991).

\bibitem{KESCM} F. Cooper, E. Mottola, B. Rogers, and P. Anderson, ``Pair Production From an External Electric Field", 
in the Proceedings of the Santa Fe Workshop on Intermittency in High Energy Collisions, edited by 
F. Cooper, R. Hwa and I. Sarcavic (World Scientific, 1991) p. 399;\hfil\break
Y. Kluger, J. M. Eisenberg, B. Svetitsky, F. Cooper, and E. Mottola,
{\it Phys. Rev. Lett.}  {\bf 67}, 2427 (1991); {\it Phys. Rev. D} {\bf 45}, 4659 (1992);\hfil\break
F. Cooper, J. M. Eisenberg, Y. Kluger, E. Mottola, and B. Svetitsky, {\it Phys. Rev. D} {\bf 48}, 190 (1993).

\bibitem{GavGit} S. O. Gavrilov and D. M. Gitman, {\it Phys. Rev. D} {\bf 53}, 7162 (1996).

\bibitem{QVlas} Y. Kluger, E. Mottola, and J. M. Eisenberg, {\it Phys. Rev. D} {\bf 58},
125015 (1998).

\bibitem{AndMotDSVacua} P. R. Anderson and E. Mottola, LA-UR-13-26990, submitted to {\it Phys. Rev. D}.

\bibitem{EMomTen} S. Habib, C. Molina-Par\'is and E. Mottola, {\it Phys. Rev. D}
{\bf 61}, 024010 (1999).

\bibitem{Attract} P. R. Anderson, W. Eaker, S. Habib, C. Molina-Par\'is, and
E. Mottola, {\it Phys. Rev. D} {\bf 62}, 124019 (2000).

\bibitem{Feyn} R. P. Feynman, {\it Phys. Rev.}, {\bf 76}, 749 (1949).

\bibitem{FeynPath} R. P. Feynman, {\it Rev. Mod. Phys.}, {\bf 20}, 367 (1948); \hfil\break
R. P. Feynman and A. R. Hibbs, {\it Quantum Mechanics and Path Integrals}, McGraw-Hill, New York (1965).

\bibitem{Bate} {\it Higher Transcendental Functions,} Vols. I and II, Bateman Manuscript Project,
A. Erd\'elyi, ed. McGraw-Hill, New York (1953).

\bibitem{Gutz} M. Gutzwiller, {\it Helv. Phys. Acta} {\bf 29}, 213 (1956).

\bibitem{Rumpf81}  H. Rumpf, {\it Phys. Rev. D} {\bf 24}, 275 (1981).

\bibitem{Laser} A. R. Bell and J. G. Kirk,  {\it Phys. Rev. Lett.} {\bf 101}, 200403 (2008);\hfil\break
D. B. Blaschke, A. V. Prozorkevich, G. Ropke, C. D. Roberts, S. M. Schmidt, D. S. Shkirmanov, and
S. A. Smolyansky, {\it Eur. Phys. J. D} {\bf 55}, 341 (2009);\hfil\break
G. V. Dunne, H. Gies, and R. Sch\"utzhold, {\it Phys. Rev. D} {\bf 80}, 111301 (2009); \hfil\break
G. V. Dunne, {\it Int. J. Mod. Phys. A} {\bf 25}, 2373 (2010).

\bibitem{Allen} B. Allen,  {\it Phys. Rev. D} {\bf 32}, 3136 (1985).

\bibitem{PokKhal} V. L. Pokrovskii and I. M. Khalatnikov, {\it Zh. Eksp. Teor. Fiz.} {\bf 40}, 1713 (1961)
[Sov. Phys. JETP {\bf 13}, 1207 (1961)].

\bibitem{AndMotSan} P. R. Anderson, E. Mottola, and D. H. Sanders, in preparation.

\bibitem{Schwmod} J. Schwinger, {\it Phys. Rev.} {\bf 125}, 397 (1962); {\bf 128}, 2425 (1962).

\bibitem{Cole} S. Coleman, {\it Phys. Rev. D} {\bf 11}, 2088 (1975);\hfil\break
J. Fr\"ohlich and E. Seiler, {\it Helv. Phys. Acta} {\bf 49}, 889 (1976).

\bibitem{John} K. Johnson, {\it Phys. Lett.} {\bf 5}, 253 (1963).

\bibitem{AndPark} P. R. Anderson and L. Parker, {\it Phys. Rev. D} {\bf 36}, 2963 (1987).

\bibitem{AndEak} P. R. Anderson and W. Eaker, {\it Phys. Rev. D}{\bf 61}, 024003 (1999).

\bibitem{ShortDistDecohere} P. R. Anderson, C. Molina-Par\'is, and
E. Mottola, {\it Phys. Rev. D} {\bf 72}, 043515 (2005).

\bibitem{Akhm} E. T. Akhmedov, {\it Mod. Phys. Lett.} {\bf A25}, 2815 (2010).

\bibitem{deSMM} P. O. Mazur and E. Mottola, {\it Nucl. Phys. B} {\bf 278}, 694 (1986).

\bibitem{HuKanPav}  B. L. Hu and D. Pavon, {\it Phys. Lett. B} {\bf 180}, 329 (1986);\hfil\break
B. L. Hu and H. E. Kandrup, {\it Phys. Rev. D} {\bf 35}, 1776 (1987).

\bibitem{CKHM} S. Habib, Y. Kluger, E. Mottola, and J. P. Paz, {\it Phys. Rev. Lett.}
{\bf 76}, 4660 (1996); \hfil\break
F. Cooper, S. Habib, Y. Kluger, and E. Mottola, {\it Phys. Rev. D}
{\bf 55}, 6471 (1997).

\bibitem{DSAnom} P. R. Anderson, C. Molina-Par\'is, and E. Mottola,
 {\it Phys. Rev. D} {\bf 80}, 084005 (2009).

\bibitem{Zak} E. Mottola, {\it Acta Physica Polonica B} {\bf 41}, 2031 (2010).

\bibitem{Vil} A. Vilenkin, {\it Phys. Rev. D} {\bf 46}, 2355 (1992).

\end{thebibliography}
\end{document}